\documentclass[longauth]{aa}

\usepackage{pdflscape}
\usepackage{graphicx}
\usepackage{natbib}
\usepackage{scalerel}
\usepackage{tabularx} 
\usepackage{pdflscape}
\usepackage{rotating}
\usepackage{subcaption} 
\usepackage[font=small]{caption} 
\usepackage[labelfont=bf]{caption} 
\usepackage[singlelinecheck=false]{caption} 

\usepackage[table]{xcolor}

\usepackage[nolist,nohyperlinks]{acronym}

\usepackage{txfonts}
\usepackage[pdfencoding=auto,psdextra]{hyperref}
\hypersetup{
    colorlinks=true,
    linkcolor=blue,
    filecolor=magenta,      
    urlcolor=blue,
    citecolor=blue
}
\usepackage{orcidlink} 
\newcommand{\orcid}[1]{\orcidlink{#1}}

\usepackage[nameinlink,capitalise]{cleveref}
\crefname{section}{Sect.}{Sects.}
\Crefname{section}{Section}{Sections}
\crefname{figure}{Fig.}{Figs.}
\Crefname{figure}{Figure}{Figures}
\crefname{equation}{Eq.}{Eqs.}
\Crefname{equation}{Equation}{Equations}

\makeatletter
\renewcommand*\aa@pageof{, page \thepage{} of \pageref*{LastPage}}
\makeatother
\usepackage{lastpage}

\usepackage[utf8]{inputenc}

\usepackage[switch, modulo]{lineno}

\usepackage{float}

\usepackage{euclid}

\newcommand{\sbmag}{\ensuremath{\,\mathrm{mag}\,\mathrm{arcsec}^{-2}}}

\begin{document}

\acrodef{AGN}{active galactic nucleus}
\acrodef{ASIC}{application specific integrated circuit}
\acrodef{BFE}{brighter-fatter effect}
\acrodef{BCDs}{blue compact dwarfs}
\acrodef{BCGs}{blue compact galaxies}
\acrodef{BCG}{brightest cluster galaxy}
\acrodef{CaLA}{camera-lens assembly}
\acrodef{CCD}{charge-coupled device}
\acrodef{CEA}{Comité Energie Atomique}
\acrodef{CoLA}{corrector-lens assembly}
\acrodef{CDS}{Correlated Double Sampling}
\acrodef{CFC}{cryo-flex cable}
\acrodef{CFHT}{Canada-France-Hawaii Telescope}
\acrodef{CGH}{computer-generated hologram}
\acrodef{CNES}{Centre National d'Etude Spacial}
\acrodef{CPPM}{Centre de Physique des Particules de Marseille}
\acrodef{CPU}{central processing unit}
\acrodef{CTE}{coefficient of thermal expansion}
\acrodef{DCU}{Detector Control Unit}
\acrodef{DES}{Dark Energy Survey}
\acrodef{DGL}{diffuse galactic light}
\acrodef{DPU}{Data Processing Unit}
\acrodef{DS}{Detector System}
\acrodef{EDS}{Euclid Deep Survey}
\acrodef{EE}{encircled energy}
\acrodef{ERO}{Early Release Observations}
\acrodef{ESA}{European Space Agency}
\acrodef{EWS}{Euclid Wide Survey}
\acrodef{FDIR}{Fault Detection, Isolation and Recovery}
\acrodef{FGS}{fine guidance sensor}
\acrodef{FOM}[FoM]{figure of merit}
\acrodef{FOV}[FoV]{field of view}
\acrodef{FPA}{focal plane array}
\acrodef{FWA}{filter-wheel assembly}
\acrodef{FWC}{full-well capacity}
\acrodef{FWHM}{full width at half maximum}
\acrodef{GC}{globular cluster}
\acrodef{GWA}{grism-wheel assembly}
\acrodef{H2RG}{HAWAII-2RG}
\acrodef{HST}{{\it Hubble} Space Telescope}
\acrodef{HSC}{Subaru-Hyper Suprime-Cam}
\acrodef{ISM}{interstellar medium}
\acrodef{IP2I}{Institut de Physique des 2 Infinis de Lyon}
\acrodef{JWST}{{\em James Webb} Space Telescope}
\acrodef{IAD}{ion-assisted deposition}
\acrodef{ICGC}{intracluster globular cluster}
\acrodef{ICU}{instrument control unit}
\acrodef{ICL}{Intra-cluster light}
\acrodef{ICM}{intra-cluster medium}
\acrodef{IGM}{intragalactic medium}
\acrodef{IMF}{initial mass function}
\acrodef{IPC}{inter-pixel capacitance}
\acrodef{LAM}{Laboratoire d'Astrophysique de Marseille}
\acrodef{LED}{light-emitting diode}
\acrodef{LF}{Luminosity function}
\acrodef{LSB}{low surface brightness}
\acrodef{LSST}{Legacy Survey of Space and Time}
\acrodef{MACC}{Multiple Accumulated}
\acrodef{MLI}{multi-layer insulation}
\acrodef{MMU}{Mass Memory Unit}
\acrodef{MPE}{Max-Planck-Institut für extraterrestrische Physik}
\acrodef{MPIA}{Max-Planck-Institut für Astronomie}
\acrodef{MW}{Milky Way}
\acrodef{NA}{numerical aperture}
\acrodef{NASA}{National Aeronautic and Space Administration}
\acrodef{JPL}{NASA Jet Propulsion Laboratory}
\acrodef{MZ-CGH}{multi-zonal computer-generated hologram}
\acrodef{NGVS}{Next Generation Virgo Survey}
\acrodef{NI-CU}{NISP calibration unit}
\acrodef{NI-OA}{near-infrared optical assembly}
\acrodef{NI-GWA}{NISP Grism Wheel Assembly}
\acrodef{NIR}{near-infrared}
\acrodef{NISP}{Near-Infrared Spectrometer and Photometer}
\acrodef{NSCs}{Nuclear star clusters}
\acrodef{PARMS}{plasma-assisted reactive magnetron sputtering}
\acrodef{PLM}{payload module}
\acrodef{PTFE}{polytetrafluoroethylene}
\acrodef{PV}{performance verification}
\acrodef{PWM}{pulse-width modulation}
\acrodef{PSF}{point spread function}
\acrodef{QE}{quantum efficiency}
\acrodef{RGB}{red-green-blue}
\acrodef{RMS}{root mean square}
\acrodef{ROI}[RoI]{region of interest}
\acrodef{ROIC}{readout-integrated chip}
\acrodef{ROS}{reference observing sequence}
\acrodef{SBF}{surface brightness fluctuation}
\acrodef{SCA}{sensor chip array}
\acrodef{SCE}{sensor chip electronic}
\acrodef{SCS}{sensor chip system}
\acrodef{SED}{spectral energy distribution}
\acrodef{SDSS}{Sloan Digital Sky Survey}
\acrodef{SGS}{science ground segment}
\acrodef{SHS}{Shack-Hartmann sensor}
\acrodef{SMF}{stellar mass function}
\acrodef{SNR}[SNR]{signal-to-noise ratio}
\acrodef{SED}{spectral energy distribution}
\acrodef{SiC}{silicon carbide}
\acrodef{SVM}{service module}
\acrodef{UCDs}{ultra compact dwarfs}
\acrodef{UDGs}{ultra diffuse galaxies}
\acrodef{UNIONS}{Ultraviolet Near Infrared Optical Northern Survey}
\acrodef{VGC}{Virgo cluster catalogue}
\acrodef{VIS}{visible imager}
\acrodef{WD}{white dwarf}
\acrodef{WCS}{world coordinate system}
\acrodef{WFE}{wavefront error}
\acrodef{ZP}{zero point}
%
%

\title{\Euclid: Early Release Observations -- Overview of the Perseus cluster and analysis of its luminosity and stellar mass functions\thanks{This paper is published on behalf of the Euclid Consortium}}

\author{J.-C.~Cuillandre\orcid{0000-0002-3263-8645}\thanks{\email{jc.cuillandre@cea.fr}}\inst{\ref{aff1}}
\and M.~Bolzonella\orcid{0000-0003-3278-4607}\inst{\ref{aff2}}
\and A.~Boselli\orcid{0000-0002-9795-6433}\inst{\ref{aff3},\ref{aff4}}
\and F.~R.~Marleau\orcid{0000-0002-1442-2947}\inst{\ref{aff5}}
\and M.~Mondelin\orcid{0009-0004-5954-0930}\inst{\ref{aff1}}
\and J.~G.~Sorce\orcid{0000-0002-2307-2432}\inst{\ref{aff6},\ref{aff7},\ref{aff8}}
\and C.~Stone\orcid{0000-0002-9086-6398}\inst{\ref{aff9}}
\and F.~Buitrago\orcid{0000-0002-2861-9812}\inst{\ref{aff10},\ref{aff11}}
\and Michele~Cantiello\orcid{0000-0003-2072-384X}\inst{\ref{aff12}}
\and K.~George\orcid{0000-0002-1734-8455}\inst{\ref{aff13}}
\and N.~A.~Hatch\orcid{0000-0001-5600-0534}\inst{\ref{aff14}}
\and L.~Quilley\orcid{0009-0008-8375-8605}\inst{\ref{aff15}}
\and F.~Mannucci\orcid{0000-0002-4803-2381}\inst{\ref{aff16}}
\and T.~Saifollahi\orcid{0000-0002-9554-7660}\inst{\ref{aff17},\ref{aff18}}
\and R.~S\'anchez-Janssen\orcid{0000-0003-4945-0056}\inst{\ref{aff19}}
\and F.~Tarsitano\orcid{0000-0002-5919-0238}\inst{\ref{aff20}}
\and C.~Tortora\orcid{0000-0001-7958-6531}\inst{\ref{aff21}}
\and X.~Xu\orcid{0000-0001-7980-8202}\inst{\ref{aff18}}
\and H.~Bouy\orcid{0000-0002-7084-487X}\inst{\ref{aff22},\ref{aff23}}
\and S.~Gwyn\orcid{0000-0001-8221-8406}\inst{\ref{aff24}}
\and M.~Kluge\orcid{0000-0002-9618-2552}\inst{\ref{aff25}}
\and A.~Lan\c{c}on\orcid{0000-0002-7214-8296}\inst{\ref{aff17}}
\and R.~Laureijs\inst{\ref{aff26}}
\and M.~Schirmer\orcid{0000-0003-2568-9994}\inst{\ref{aff27}}
\and Abdurro'uf\orcid{0000-0002-5258-8761}\inst{\ref{aff28}}
\and P.~Awad\orcid{0000-0002-0428-849X}\inst{\ref{aff18}}
\and M.~Baes\orcid{0000-0002-3930-2757}\inst{\ref{aff29}}
\and F.~Bournaud\inst{\ref{aff1}}
\and D.~Carollo\orcid{0000-0002-0005-5787}\inst{\ref{aff30}}
\and S.~Codis\inst{\ref{aff1}}
\and C.~J.~Conselice\orcid{0000-0003-1949-7638}\inst{\ref{aff31}}
\and V.~De~Lapparent\orcid{0009-0007-1622-1974}\inst{\ref{aff32}}
\and P.-A.~Duc\orcid{0000-0003-3343-6284}\inst{\ref{aff33}}
\and A.~Ferr\'e-Mateu\orcid{0000-0002-6411-220X}\inst{\ref{aff34},\ref{aff35}}
\and W.~Gillard\orcid{0000-0003-4744-9748}\inst{\ref{aff36}}
\and J.~B.~Golden-Marx\orcid{0000-0002-6394-045X}\inst{\ref{aff14}}
\and P.~Jablonka\orcid{0000-0002-9655-1063}\inst{\ref{aff37}}
\and R.~Habas\orcid{0000-0002-4033-3841}\inst{\ref{aff12}}
\and L.~K.~Hunt\orcid{0000-0001-9162-2371}\inst{\ref{aff16}}
\and S.~Mei\orcid{0000-0002-2849-559X}\inst{\ref{aff38}}
\and M.-A.~Miville-Desch\^enes\orcid{0000-0002-7351-6062}\inst{\ref{aff39}}
\and M.~Montes\orcid{0000-0001-7847-0393}\inst{\ref{aff35},\ref{aff34}}
\and A.~Nersesian\orcid{0000-0001-6843-409X}\inst{\ref{aff40},\ref{aff29}}
\and R.~F.~Peletier\orcid{0000-0001-7621-947X}\inst{\ref{aff18}}
\and M.~Poulain\orcid{0000-0002-7664-4510}\inst{\ref{aff41}}
\and R.~Scaramella\orcid{0000-0003-2229-193X}\inst{\ref{aff42}}
\and M.~Scialpi\orcid{0009-0006-5100-4986}\inst{\ref{aff43},\ref{aff44},\ref{aff16}}
\and E.~Sola\orcid{0000-0002-2814-3578}\inst{\ref{aff45}}
\and J.~Stephan\inst{\ref{aff19}}
\and L.~Ulivi\orcid{0009-0001-3291-5382}\inst{\ref{aff44},\ref{aff43},\ref{aff16}}
\and M.~Urbano\orcid{0000-0001-5640-0650}\inst{\ref{aff17}}
\and R.~Z\"oller\orcid{0000-0002-0938-5686}\inst{\ref{aff13},\ref{aff25}}
\and N.~Aghanim\orcid{0000-0002-6688-8992}\inst{\ref{aff7}}
\and B.~Altieri\orcid{0000-0003-3936-0284}\inst{\ref{aff46}}
\and A.~Amara\inst{\ref{aff47}}
\and S.~Andreon\orcid{0000-0002-2041-8784}\inst{\ref{aff48}}
\and N.~Auricchio\orcid{0000-0003-4444-8651}\inst{\ref{aff2}}
\and M.~Baldi\orcid{0000-0003-4145-1943}\inst{\ref{aff49},\ref{aff2},\ref{aff50}}
\and A.~Balestra\orcid{0000-0002-6967-261X}\inst{\ref{aff51}}
\and S.~Bardelli\orcid{0000-0002-8900-0298}\inst{\ref{aff2}}
\and R.~Bender\orcid{0000-0001-7179-0626}\inst{\ref{aff25},\ref{aff13}}
\and C.~Bodendorf\inst{\ref{aff25}}
\and D.~Bonino\orcid{0000-0002-3336-9977}\inst{\ref{aff52}}
\and E.~Branchini\orcid{0000-0002-0808-6908}\inst{\ref{aff53},\ref{aff54},\ref{aff48}}
\and M.~Brescia\orcid{0000-0001-9506-5680}\inst{\ref{aff55},\ref{aff21},\ref{aff56}}
\and J.~Brinchmann\orcid{0000-0003-4359-8797}\inst{\ref{aff57},\ref{aff58}}
\and S.~Camera\orcid{0000-0003-3399-3574}\inst{\ref{aff59},\ref{aff60},\ref{aff52}}
\and V.~Capobianco\orcid{0000-0002-3309-7692}\inst{\ref{aff52}}
\and C.~Carbone\orcid{0000-0003-0125-3563}\inst{\ref{aff61}}
\and J.~Carretero\orcid{0000-0002-3130-0204}\inst{\ref{aff62},\ref{aff63}}
\and S.~Casas\orcid{0000-0002-4751-5138}\inst{\ref{aff64}}
\and F.~J.~Castander\orcid{0000-0001-7316-4573}\inst{\ref{aff65},\ref{aff66}}
\and M.~Castellano\orcid{0000-0001-9875-8263}\inst{\ref{aff42}}
\and S.~Cavuoti\orcid{0000-0002-3787-4196}\inst{\ref{aff21},\ref{aff56}}
\and A.~Cimatti\inst{\ref{aff67}}
\and G.~Congedo\orcid{0000-0003-2508-0046}\inst{\ref{aff68}}
\and L.~Conversi\orcid{0000-0002-6710-8476}\inst{\ref{aff69},\ref{aff46}}
\and Y.~Copin\orcid{0000-0002-5317-7518}\inst{\ref{aff70}}
\and F.~Courbin\orcid{0000-0003-0758-6510}\inst{\ref{aff37}}
\and H.~M.~Courtois\orcid{0000-0003-0509-1776}\inst{\ref{aff71}}
\and M.~Cropper\orcid{0000-0003-4571-9468}\inst{\ref{aff72}}
\and A.~Da~Silva\orcid{0000-0002-6385-1609}\inst{\ref{aff73},\ref{aff74}}
\and H.~Degaudenzi\orcid{0000-0002-5887-6799}\inst{\ref{aff20}}
\and A.~M.~Di~Giorgio\orcid{0000-0002-4767-2360}\inst{\ref{aff75}}
\and J.~Dinis\orcid{0000-0001-5075-1601}\inst{\ref{aff73},\ref{aff74}}
\and M.~Douspis\orcid{0000-0003-4203-3954}\inst{\ref{aff7}}
\and F.~Dubath\orcid{0000-0002-6533-2810}\inst{\ref{aff20}}
\and C.~A.~J.~Duncan\inst{\ref{aff31}}
\and X.~Dupac\inst{\ref{aff46}}
\and S.~Dusini\orcid{0000-0002-1128-0664}\inst{\ref{aff76}}
\and M.~Farina\orcid{0000-0002-3089-7846}\inst{\ref{aff75}}
\and S.~Farrens\orcid{0000-0002-9594-9387}\inst{\ref{aff1}}
\and S.~Ferriol\inst{\ref{aff70}}
\and S.~Fotopoulou\orcid{0000-0002-9686-254X}\inst{\ref{aff77}}
\and M.~Frailis\orcid{0000-0002-7400-2135}\inst{\ref{aff30}}
\and E.~Franceschi\orcid{0000-0002-0585-6591}\inst{\ref{aff2}}
\and S.~Galeotta\orcid{0000-0002-3748-5115}\inst{\ref{aff30}}
\and B.~Gillis\orcid{0000-0002-4478-1270}\inst{\ref{aff68}}
\and C.~Giocoli\orcid{0000-0002-9590-7961}\inst{\ref{aff2},\ref{aff78}}
\and P.~G\'omez-Alvarez\orcid{0000-0002-8594-5358}\inst{\ref{aff79},\ref{aff46}}
\and A.~Grazian\orcid{0000-0002-5688-0663}\inst{\ref{aff51}}
\and F.~Grupp\inst{\ref{aff25},\ref{aff13}}
\and L.~Guzzo\orcid{0000-0001-8264-5192}\inst{\ref{aff80},\ref{aff48}}
\and S.~V.~H.~Haugan\orcid{0000-0001-9648-7260}\inst{\ref{aff81}}
\and J.~Hoar\inst{\ref{aff46}}
\and H.~Hoekstra\orcid{0000-0002-0641-3231}\inst{\ref{aff82}}
\and W.~Holmes\inst{\ref{aff83}}
\and I.~Hook\orcid{0000-0002-2960-978X}\inst{\ref{aff84}}
\and F.~Hormuth\inst{\ref{aff85}}
\and A.~Hornstrup\orcid{0000-0002-3363-0936}\inst{\ref{aff86},\ref{aff87}}
\and P.~Hudelot\inst{\ref{aff32}}
\and K.~Jahnke\orcid{0000-0003-3804-2137}\inst{\ref{aff27}}
\and M.~Jhabvala\inst{\ref{aff88}}
\and E.~Keih\"anen\orcid{0000-0003-1804-7715}\inst{\ref{aff89}}
\and S.~Kermiche\orcid{0000-0002-0302-5735}\inst{\ref{aff36}}
\and A.~Kiessling\orcid{0000-0002-2590-1273}\inst{\ref{aff83}}
\and M.~Kilbinger\orcid{0000-0001-9513-7138}\inst{\ref{aff1}}
\and T.~Kitching\orcid{0000-0002-4061-4598}\inst{\ref{aff72}}
\and R.~Kohley\inst{\ref{aff46}}
\and B.~Kubik\orcid{0009-0006-5823-4880}\inst{\ref{aff70}}
\and K.~Kuijken\orcid{0000-0002-3827-0175}\inst{\ref{aff82}}
\and M.~K\"ummel\orcid{0000-0003-2791-2117}\inst{\ref{aff13}}
\and M.~Kunz\orcid{0000-0002-3052-7394}\inst{\ref{aff90}}
\and H.~Kurki-Suonio\orcid{0000-0002-4618-3063}\inst{\ref{aff91},\ref{aff92}}
\and O.~Lahav\orcid{0000-0002-1134-9035}\inst{\ref{aff93}}
\and D.~Le~Mignant\orcid{0000-0002-5339-5515}\inst{\ref{aff3}}
\and S.~Ligori\orcid{0000-0003-4172-4606}\inst{\ref{aff52}}
\and P.~B.~Lilje\orcid{0000-0003-4324-7794}\inst{\ref{aff81}}
\and V.~Lindholm\orcid{0000-0003-2317-5471}\inst{\ref{aff91},\ref{aff92}}
\and I.~Lloro\inst{\ref{aff94}}
\and D.~Maino\inst{\ref{aff80},\ref{aff61},\ref{aff95}}
\and E.~Maiorano\orcid{0000-0003-2593-4355}\inst{\ref{aff2}}
\and O.~Mansutti\orcid{0000-0001-5758-4658}\inst{\ref{aff30}}
\and O.~Marggraf\orcid{0000-0001-7242-3852}\inst{\ref{aff96}}
\and K.~Markovic\orcid{0000-0001-6764-073X}\inst{\ref{aff83}}
\and N.~Martinet\orcid{0000-0003-2786-7790}\inst{\ref{aff3}}
\and F.~Marulli\orcid{0000-0002-8850-0303}\inst{\ref{aff97},\ref{aff2},\ref{aff50}}
\and R.~Massey\orcid{0000-0002-6085-3780}\inst{\ref{aff98},\ref{aff99}}
\and S.~Maurogordato\inst{\ref{aff100}}
\and H.~J.~McCracken\orcid{0000-0002-9489-7765}\inst{\ref{aff32}}
\and E.~Medinaceli\orcid{0000-0002-4040-7783}\inst{\ref{aff2}}
\and M.~Melchior\inst{\ref{aff101}}
\and Y.~Mellier\inst{\ref{aff102},\ref{aff32}}
\and M.~Meneghetti\orcid{0000-0003-1225-7084}\inst{\ref{aff2},\ref{aff50}}
\and E.~Merlin\orcid{0000-0001-6870-8900}\inst{\ref{aff42}}
\and G.~Meylan\inst{\ref{aff37}}
\and J.~J.~Mohr\orcid{0000-0002-6875-2087}\inst{\ref{aff13},\ref{aff25}}
\and M.~Moresco\orcid{0000-0002-7616-7136}\inst{\ref{aff97},\ref{aff2}}
\and L.~Moscardini\orcid{0000-0002-3473-6716}\inst{\ref{aff97},\ref{aff2},\ref{aff50}}
\and R.~Nakajima\inst{\ref{aff96}}
\and R.~C.~Nichol\orcid{0000-0003-0939-6518}\inst{\ref{aff47}}
\and S.-M.~Niemi\inst{\ref{aff26}}
\and C.~Padilla\orcid{0000-0001-7951-0166}\inst{\ref{aff103}}
\and S.~Paltani\orcid{0000-0002-8108-9179}\inst{\ref{aff20}}
\and F.~Pasian\orcid{0000-0002-4869-3227}\inst{\ref{aff30}}
\and K.~Pedersen\inst{\ref{aff104}}
\and W.~J.~Percival\orcid{0000-0002-0644-5727}\inst{\ref{aff105},\ref{aff106},\ref{aff107}}
\and V.~Pettorino\inst{\ref{aff26}}
\and S.~Pires\orcid{0000-0002-0249-2104}\inst{\ref{aff1}}
\and G.~Polenta\orcid{0000-0003-4067-9196}\inst{\ref{aff108}}
\and M.~Poncet\inst{\ref{aff109}}
\and L.~A.~Popa\inst{\ref{aff110}}
\and L.~Pozzetti\orcid{0000-0001-7085-0412}\inst{\ref{aff2}}
\and F.~Raison\orcid{0000-0002-7819-6918}\inst{\ref{aff25}}
\and A.~Renzi\orcid{0000-0001-9856-1970}\inst{\ref{aff111},\ref{aff76}}
\and J.~Rhodes\orcid{0000-0002-4485-8549}\inst{\ref{aff83}}
\and G.~Riccio\inst{\ref{aff21}}
\and E.~Romelli\orcid{0000-0003-3069-9222}\inst{\ref{aff30}}
\and M.~Roncarelli\orcid{0000-0001-9587-7822}\inst{\ref{aff2}}
\and R.~Saglia\orcid{0000-0003-0378-7032}\inst{\ref{aff13},\ref{aff25}}
\and D.~Sapone\orcid{0000-0001-7089-4503}\inst{\ref{aff112}}
\and P.~Schneider\orcid{0000-0001-8561-2679}\inst{\ref{aff96}}
\and T.~Schrabback\orcid{0000-0002-6987-7834}\inst{\ref{aff5}}
\and A.~Secroun\orcid{0000-0003-0505-3710}\inst{\ref{aff36}}
\and G.~Seidel\orcid{0000-0003-2907-353X}\inst{\ref{aff27}}
\and S.~Serrano\orcid{0000-0002-0211-2861}\inst{\ref{aff66},\ref{aff113},\ref{aff65}}
\and C.~Sirignano\orcid{0000-0002-0995-7146}\inst{\ref{aff111},\ref{aff76}}
\and G.~Sirri\orcid{0000-0003-2626-2853}\inst{\ref{aff50}}
\and J.~Skottfelt\orcid{0000-0003-1310-8283}\inst{\ref{aff114}}
\and L.~Stanco\orcid{0000-0002-9706-5104}\inst{\ref{aff76}}
\and P.~Tallada-Cresp\'{i}\orcid{0000-0002-1336-8328}\inst{\ref{aff62},\ref{aff63}}
\and A.~N.~Taylor\inst{\ref{aff68}}
\and H.~I.~Teplitz\orcid{0000-0002-7064-5424}\inst{\ref{aff115}}
\and I.~Tereno\inst{\ref{aff73},\ref{aff11}}
\and R.~Toledo-Moreo\orcid{0000-0002-2997-4859}\inst{\ref{aff116}}
\and I.~Tutusaus\orcid{0000-0002-3199-0399}\inst{\ref{aff117}}
\and E.~A.~Valentijn\inst{\ref{aff18}}
\and L.~Valenziano\orcid{0000-0002-1170-0104}\inst{\ref{aff2},\ref{aff118}}
\and T.~Vassallo\orcid{0000-0001-6512-6358}\inst{\ref{aff13},\ref{aff30}}
\and G.~Verdoes~Kleijn\orcid{0000-0001-5803-2580}\inst{\ref{aff18}}
\and Y.~Wang\orcid{0000-0002-4749-2984}\inst{\ref{aff115}}
\and J.~Weller\orcid{0000-0002-8282-2010}\inst{\ref{aff13},\ref{aff25}}
\and E.~Zucca\orcid{0000-0002-5845-8132}\inst{\ref{aff2}}
\and A.~Biviano\orcid{0000-0002-0857-0732}\inst{\ref{aff30},\ref{aff119}}
\and C.~Burigana\orcid{0000-0002-3005-5796}\inst{\ref{aff120},\ref{aff118}}
\and G.~Castignani\orcid{0000-0001-6831-0687}\inst{\ref{aff2}}
\and G.~De~Lucia\orcid{0000-0002-6220-9104}\inst{\ref{aff30}}
\and V.~Scottez\inst{\ref{aff102},\ref{aff121}}
\and A.~Mora\orcid{0000-0002-1922-8529}\inst{\ref{aff122}}
\and P.~Simon\inst{\ref{aff96}}
\and J.~Mart\'{i}n-Fleitas\orcid{0000-0002-8594-569X}\inst{\ref{aff122}}
\and D.~Scott\orcid{0000-0002-6878-9840}\inst{\ref{aff123}}}
										   
\institute{Universit\'e Paris-Saclay, Universit\'e Paris Cit\'e, CEA, CNRS, AIM, 91191, Gif-sur-Yvette, France\label{aff1}
\and
INAF-Osservatorio di Astrofisica e Scienza dello Spazio di Bologna, Via Piero Gobetti 93/3, 40129 Bologna, Italy\label{aff2}
\and
Aix-Marseille Universit\'e, CNRS, CNES, LAM, Marseille, France\label{aff3}
\and
INAF - Osservatorio Astronomico di Cagliari, Via della Scienza 5, 09047 Selargius (CA), Italy\label{aff4}
\and
Universit\"at Innsbruck, Institut f\"ur Astro- und Teilchenphysik, Technikerstr. 25/8, 6020 Innsbruck, Austria\label{aff5}
\and
Univ. Lille, CNRS, Centrale Lille, UMR 9189 CRIStAL, 59000 Lille, France\label{aff6}
\and
Universit\'e Paris-Saclay, CNRS, Institut d'astrophysique spatiale, 91405, Orsay, France\label{aff7}
\and
Leibniz-Institut f\"{u}r Astrophysik (AIP), An der Sternwarte 16, 14482 Potsdam, Germany\label{aff8}
\and
Department of Physics, Universit\'{e} de Montr\'{e}al, 2900 Edouard Montpetit Blvd, Montr\'{e}al, Qu\'{e}bec H3T 1J4, Canada\label{aff9}
\and
Departamento de F\'{i}sica Te\'{o}rica, At\'{o}mica y \'{O}ptica, Universidad de Valladolid, 47011 Valladolid, Spain\label{aff10}
\and
Instituto de Astrof\'isica e Ci\^encias do Espa\c{c}o, Faculdade de Ci\^encias, Universidade de Lisboa, Tapada da Ajuda, 1349-018 Lisboa, Portugal\label{aff11}
\and
INAF - Osservatorio Astronomico d'Abruzzo, Via Maggini, 64100, Teramo, Italy\label{aff12}
\and
Universit\"ats-Sternwarte M\"unchen, Fakult\"at f\"ur Physik, Ludwig-Maximilians-Universit\"at M\"unchen, Scheinerstrasse 1, 81679 M\"unchen, Germany\label{aff13}
\and
School of Physics and Astronomy, University of Nottingham, University Park, Nottingham NG7 2RD, UK\label{aff14}
\and
Centre de Recherche Astrophysique de Lyon, UMR5574, CNRS, Universit\'e Claude Bernard Lyon 1, ENS de Lyon, 69230, Saint-Genis-Laval, France\label{aff15}
\and
INAF-Osservatorio Astrofisico di Arcetri, Largo E. Fermi 5, 50125, Firenze, Italy\label{aff16}
\and
Observatoire Astronomique de Strasbourg (ObAS), Universit\'e de Strasbourg - CNRS, UMR 7550, Strasbourg, France\label{aff17}
\and
Kapteyn Astronomical Institute, University of Groningen, PO Box 800, 9700 AV Groningen, The Netherlands\label{aff18}
\and
UK Astronomy Technology Centre, Royal Observatory, Blackford Hill, Edinburgh EH9 3HJ, UK\label{aff19}
\and
Department of Astronomy, University of Geneva, ch. d'Ecogia 16, 1290 Versoix, Switzerland\label{aff20}
\and
INAF-Osservatorio Astronomico di Capodimonte, Via Moiariello 16, 80131 Napoli, Italy\label{aff21}
\and
Institut universitaire de France (IUF), 1 rue Descartes, 75231 PARIS CEDEX 05, France\label{aff22}
\and
Laboratoire d'Astrophysique de Bordeaux, CNRS and Universit\'e de Bordeaux, All\'ee Geoffroy St. Hilaire, 33165 Pessac, France\label{aff23}
\and
NRC Herzberg, 5071 West Saanich Rd, Victoria, BC V9E 2E7, Canada\label{aff24}
\and
Max Planck Institute for Extraterrestrial Physics, Giessenbachstr. 1, 85748 Garching, Germany\label{aff25}
\and
European Space Agency/ESTEC, Keplerlaan 1, 2201 AZ Noordwijk, The Netherlands\label{aff26}
\and
Max-Planck-Institut f\"ur Astronomie, K\"onigstuhl 17, 69117 Heidelberg, Germany\label{aff27}
\and
Johns Hopkins University 3400 North Charles Street Baltimore, MD 21218, USA\label{aff28}
\and
Sterrenkundig Observatorium, Universiteit Gent, Krijgslaan 281 S9, 9000 Gent, Belgium\label{aff29}
\and
INAF-Osservatorio Astronomico di Trieste, Via G. B. Tiepolo 11, 34143 Trieste, Italy\label{aff30}
\and
Jodrell Bank Centre for Astrophysics, Department of Physics and Astronomy, University of Manchester, Oxford Road, Manchester M13 9PL, UK\label{aff31}
\and
Institut d'Astrophysique de Paris, UMR 7095, CNRS, and Sorbonne Universit\'e, 98 bis boulevard Arago, 75014 Paris, France\label{aff32}
\and
Universit\'e de Strasbourg, CNRS, Observatoire astronomique de Strasbourg, UMR 7550, 67000 Strasbourg, France\label{aff33}
\and
Departamento de Astrof\'isica, Universidad de La Laguna, 38206, La Laguna, Tenerife, Spain\label{aff34}
\and
Instituto de Astrof\'isica de Canarias, Calle V\'ia L\'actea s/n, 38204, San Crist\'obal de La Laguna, Tenerife, Spain\label{aff35}
\and
Aix-Marseille Universit\'e, CNRS/IN2P3, CPPM, Marseille, France\label{aff36}
\and
Institute of Physics, Laboratory of Astrophysics, Ecole Polytechnique F\'ed\'erale de Lausanne (EPFL), Observatoire de Sauverny, 1290 Versoix, Switzerland\label{aff37}
\and
Universit\'e Paris Cit\'e, CNRS, Astroparticule et Cosmologie, 75013 Paris, France\label{aff38}
\and
Laboratoire de Physique de l'\'Ecole Normale Sup\'erieure, ENS, Universit\'e PSL, CNRS, Sorbonne Universit\'e, 75005 Paris, France\label{aff39}
\and
STAR Institute, Quartier Agora - All\'ee du six Ao\^ut, 19c B-4000 Li\`ege, Belgium\label{aff40}
\and
Space physics and astronomy research unit, University of Oulu, Pentti Kaiteran katu 1, FI-90014 Oulu, Finland\label{aff41}
\and
INAF-Osservatorio Astronomico di Roma, Via Frascati 33, 00078 Monteporzio Catone, Italy\label{aff42}
\and
Dipartimento di Fisica e Astronomia, Universit\`{a} di Firenze, via G. Sansone 1, 50019 Sesto Fiorentino, Firenze, Italy\label{aff43}
\and
University of Trento, Via Sommarive 14, I-38123 Trento, Italy\label{aff44}
\and
Institute of Astronomy, University of Cambridge, Madingley Road, Cambridge CB3 0HA, UK\label{aff45}
\and
ESAC/ESA, Camino Bajo del Castillo, s/n., Urb. Villafranca del Castillo, 28692 Villanueva de la Ca\~nada, Madrid, Spain\label{aff46}
\and
School of Mathematics and Physics, University of Surrey, Guildford, Surrey, GU2 7XH, UK\label{aff47}
\and
INAF-Osservatorio Astronomico di Brera, Via Brera 28, 20122 Milano, Italy\label{aff48}
\and
Dipartimento di Fisica e Astronomia, Universit\`a di Bologna, Via Gobetti 93/2, 40129 Bologna, Italy\label{aff49}
\and
INFN-Sezione di Bologna, Viale Berti Pichat 6/2, 40127 Bologna, Italy\label{aff50}
\and
INAF-Osservatorio Astronomico di Padova, Via dell'Osservatorio 5, 35122 Padova, Italy\label{aff51}
\and
INAF-Osservatorio Astrofisico di Torino, Via Osservatorio 20, 10025 Pino Torinese (TO), Italy\label{aff52}
\and
Dipartimento di Fisica, Universit\`a di Genova, Via Dodecaneso 33, 16146, Genova, Italy\label{aff53}
\and
INFN-Sezione di Genova, Via Dodecaneso 33, 16146, Genova, Italy\label{aff54}
\and
Department of Physics "E. Pancini", University Federico II, Via Cinthia 6, 80126, Napoli, Italy\label{aff55}
\and
INFN section of Naples, Via Cinthia 6, 80126, Napoli, Italy\label{aff56}
\and
Instituto de Astrof\'isica e Ci\^encias do Espa\c{c}o, Universidade do Porto, CAUP, Rua das Estrelas, PT4150-762 Porto, Portugal\label{aff57}
\and
Faculdade de Ci\^encias da Universidade do Porto, Rua do Campo de Alegre, 4150-007 Porto, Portugal\label{aff58}
\and
Dipartimento di Fisica, Universit\`a degli Studi di Torino, Via P. Giuria 1, 10125 Torino, Italy\label{aff59}
\and
INFN-Sezione di Torino, Via P. Giuria 1, 10125 Torino, Italy\label{aff60}
\and
INAF-IASF Milano, Via Alfonso Corti 12, 20133 Milano, Italy\label{aff61}
\and
Centro de Investigaciones Energ\'eticas, Medioambientales y Tecnol\'ogicas (CIEMAT), Avenida Complutense 40, 28040 Madrid, Spain\label{aff62}
\and
Port d'Informaci\'{o} Cient\'{i}fica, Campus UAB, C. Albareda s/n, 08193 Bellaterra (Barcelona), Spain\label{aff63}
\and
Institute for Theoretical Particle Physics and Cosmology (TTK), RWTH Aachen University, 52056 Aachen, Germany\label{aff64}
\and
Institute of Space Sciences (ICE, CSIC), Campus UAB, Carrer de Can Magrans, s/n, 08193 Barcelona, Spain\label{aff65}
\and
Institut d'Estudis Espacials de Catalunya (IEEC),  Edifici RDIT, Campus UPC, 08860 Castelldefels, Barcelona, Spain\label{aff66}
\and
Dipartimento di Fisica e Astronomia "Augusto Righi" - Alma Mater Studiorum Universit\`a di Bologna, Viale Berti Pichat 6/2, 40127 Bologna, Italy\label{aff67}
\and
Institute for Astronomy, University of Edinburgh, Royal Observatory, Blackford Hill, Edinburgh EH9 3HJ, UK\label{aff68}
\and
European Space Agency/ESRIN, Largo Galileo Galilei 1, 00044 Frascati, Roma, Italy\label{aff69}
\and
Universit\'e Claude Bernard Lyon 1, CNRS/IN2P3, IP2I Lyon, UMR 5822, Villeurbanne, F-69100, France\label{aff70}
\and
UCB Lyon 1, CNRS/IN2P3, IUF, IP2I Lyon, 4 rue Enrico Fermi, 69622 Villeurbanne, France\label{aff71}
\and
Mullard Space Science Laboratory, University College London, Holmbury St Mary, Dorking, Surrey RH5 6NT, UK\label{aff72}
\and
Departamento de F\'isica, Faculdade de Ci\^encias, Universidade de Lisboa, Edif\'icio C8, Campo Grande, PT1749-016 Lisboa, Portugal\label{aff73}
\and
Instituto de Astrof\'isica e Ci\^encias do Espa\c{c}o, Faculdade de Ci\^encias, Universidade de Lisboa, Campo Grande, 1749-016 Lisboa, Portugal\label{aff74}
\and
INAF-Istituto di Astrofisica e Planetologia Spaziali, via del Fosso del Cavaliere, 100, 00100 Roma, Italy\label{aff75}
\and
INFN-Padova, Via Marzolo 8, 35131 Padova, Italy\label{aff76}
\and
School of Physics, HH Wills Physics Laboratory, University of Bristol, Tyndall Avenue, Bristol, BS8 1TL, UK\label{aff77}
\and
Istituto Nazionale di Fisica Nucleare, Sezione di Bologna, Via Irnerio 46, 40126 Bologna, Italy\label{aff78}
\and
FRACTAL S.L.N.E., calle Tulip\'an 2, Portal 13 1A, 28231, Las Rozas de Madrid, Spain\label{aff79}
\and
Dipartimento di Fisica "Aldo Pontremoli", Universit\`a degli Studi di Milano, Via Celoria 16, 20133 Milano, Italy\label{aff80}
\and
Institute of Theoretical Astrophysics, University of Oslo, P.O. Box 1029 Blindern, 0315 Oslo, Norway\label{aff81}
\and
Leiden Observatory, Leiden University, Einsteinweg 55, 2333 CC Leiden, The Netherlands\label{aff82}
\and
Jet Propulsion Laboratory, California Institute of Technology, 4800 Oak Grove Drive, Pasadena, CA, 91109, USA\label{aff83}
\and
Department of Physics, Lancaster University, Lancaster, LA1 4YB, UK\label{aff84}
\and
Felix Hormuth Engineering, Goethestr. 17, 69181 Leimen, Germany\label{aff85}
\and
Technical University of Denmark, Elektrovej 327, 2800 Kgs. Lyngby, Denmark\label{aff86}
\and
Cosmic Dawn Center (DAWN), Denmark\label{aff87}
\and
NASA Goddard Space Flight Center, Greenbelt, MD 20771, USA\label{aff88}
\and
Department of Physics and Helsinki Institute of Physics, Gustaf H\"allstr\"omin katu 2, 00014 University of Helsinki, Finland\label{aff89}
\and
Universit\'e de Gen\`eve, D\'epartement de Physique Th\'eorique and Centre for Astroparticle Physics, 24 quai Ernest-Ansermet, CH-1211 Gen\`eve 4, Switzerland\label{aff90}
\and
Department of Physics, P.O. Box 64, 00014 University of Helsinki, Finland\label{aff91}
\and
Helsinki Institute of Physics, Gustaf H{\"a}llstr{\"o}min katu 2, University of Helsinki, Helsinki, Finland\label{aff92}
\and
Department of Physics and Astronomy, University College London, Gower Street, London WC1E 6BT, UK\label{aff93}
\and
NOVA optical infrared instrumentation group at ASTRON, Oude Hoogeveensedijk 4, 7991PD, Dwingeloo, The Netherlands\label{aff94}
\and
INFN-Sezione di Milano, Via Celoria 16, 20133 Milano, Italy\label{aff95}
\and
Universit\"at Bonn, Argelander-Institut f\"ur Astronomie, Auf dem H\"ugel 71, 53121 Bonn, Germany\label{aff96}
\and
Dipartimento di Fisica e Astronomia "Augusto Righi" - Alma Mater Studiorum Universit\`a di Bologna, via Piero Gobetti 93/2, 40129 Bologna, Italy\label{aff97}
\and
Department of Physics, Centre for Extragalactic Astronomy, Durham University, South Road, DH1 3LE, UK\label{aff98}
\and
Department of Physics, Institute for Computational Cosmology, Durham University, South Road, DH1 3LE, UK\label{aff99}
\and
Universit\'e C\^{o}te d'Azur, Observatoire de la C\^{o}te d'Azur, CNRS, Laboratoire Lagrange, Bd de l'Observatoire, CS 34229, 06304 Nice cedex 4, France\label{aff100}
\and
University of Applied Sciences and Arts of Northwestern Switzerland, School of Engineering, 5210 Windisch, Switzerland\label{aff101}
\and
Institut d'Astrophysique de Paris, 98bis Boulevard Arago, 75014, Paris, France\label{aff102}
\and
Institut de F\'{i}sica d'Altes Energies (IFAE), The Barcelona Institute of Science and Technology, Campus UAB, 08193 Bellaterra (Barcelona), Spain\label{aff103}
\and
Department of Physics and Astronomy, University of Aarhus, Ny Munkegade 120, DK-8000 Aarhus C, Denmark\label{aff104}
\and
Waterloo Centre for Astrophysics, University of Waterloo, Waterloo, Ontario N2L 3G1, Canada\label{aff105}
\and
Department of Physics and Astronomy, University of Waterloo, Waterloo, Ontario N2L 3G1, Canada\label{aff106}
\and
Perimeter Institute for Theoretical Physics, Waterloo, Ontario N2L 2Y5, Canada\label{aff107}
\and
Space Science Data Center, Italian Space Agency, via del Politecnico snc, 00133 Roma, Italy\label{aff108}
\and
Centre National d'Etudes Spatiales -- Centre spatial de Toulouse, 18 avenue Edouard Belin, 31401 Toulouse Cedex 9, France\label{aff109}
\and
Institute of Space Science, Str. Atomistilor, nr. 409 M\u{a}gurele, Ilfov, 077125, Romania\label{aff110}
\and
Dipartimento di Fisica e Astronomia "G. Galilei", Universit\`a di Padova, Via Marzolo 8, 35131 Padova, Italy\label{aff111}
\and
Departamento de F\'isica, FCFM, Universidad de Chile, Blanco Encalada 2008, Santiago, Chile\label{aff112}
\and
Satlantis, University Science Park, Sede Bld 48940, Leioa-Bilbao, Spain\label{aff113}
\and
Centre for Electronic Imaging, Open University, Walton Hall, Milton Keynes, MK7~6AA, UK\label{aff114}
\and
Infrared Processing and Analysis Center, California Institute of Technology, Pasadena, CA 91125, USA\label{aff115}
\and
Universidad Polit\'ecnica de Cartagena, Departamento de Electr\'onica y Tecnolog\'ia de Computadoras,  Plaza del Hospital 1, 30202 Cartagena, Spain\label{aff116}
\and
Institut de Recherche en Astrophysique et Plan\'etologie (IRAP), Universit\'e de Toulouse, CNRS, UPS, CNES, 14 Av. Edouard Belin, 31400 Toulouse, France\label{aff117}
\and
INFN-Bologna, Via Irnerio 46, 40126 Bologna, Italy\label{aff118}
\and
IFPU, Institute for Fundamental Physics of the Universe, via Beirut 2, 34151 Trieste, Italy\label{aff119}
\and
INAF, Istituto di Radioastronomia, Via Piero Gobetti 101, 40129 Bologna, Italy\label{aff120}
\and
Junia, EPA department, 41 Bd Vauban, 59800 Lille, France\label{aff121}
\and
Aurora Technology for European Space Agency (ESA), Camino bajo del Castillo, s/n, Urbanizacion Villafranca del Castillo, Villanueva de la Ca\~nada, 28692 Madrid, Spain\label{aff122}
\and
Department of Physics and Astronomy, University of British Columbia, Vancouver, BC V6T 1Z1, Canada\label{aff123}}    

%
%

\abstract{The \Euclid \ac{ERO} programme targeted the Perseus cluster of galaxies, gathering deep data in the central region of the cluster over $0.7\, {\rm deg^2}$, corresponding to approximately $0.25$ $r_{200}$. 
The data set reaches a point-source depth of $\IE = 28.0$ ($\YE,\,\JE,\,\HE\ = 25.3$) AB magnitudes at $5\,\sigma$ with a \ang{;;0.16} (\ang{;;0.48}) \ac{FWHM}, and a surface brightness limit of 30.1 (29.2)\,$\mathrm{mag}\,\mathrm{arcsec}^{-2}$. 
The exceptional depth and spatial resolution of this wide-field multi-band data enable the simultaneous detection and characterisation of both bright and low surface brightness galaxies, along with their globular cluster systems, from the optical to the \ac{NIR}. 
This study advances beyond previous analyses of the cluster and enables a range of scientific investigations summarised here. 
We derive the luminosity and stellar mass functions (LF and SMF) of the Perseus cluster in the \Euclid \IE band, thanks to supplementary $u$, $g$, $r$, $i$, $z$, and H$\alpha$ data from the \ac{CFHT}. 
We adopt a catalogue of 1100 dwarf galaxies, detailed in the corresponding \ac{ERO} paper, which includes their photometric and structural properties.
We identify all other sources in the \Euclid images and obtain accurate photometric measurements using \texttt{AutoProf} or \texttt{AstroPhot} for 138 bright cluster galaxies, and \texttt{SourceExtractor} for half a million compact sources. 
Cluster membership for the bright sample is determined by calculating photometric redshifts with \texttt{Phosphoros}. 
Our LF and SMF are the deepest recorded for the Perseus cluster, highlighting the groundbreaking capabilities of the \Euclid telescope. Both the LF and SMF fit a Schechter plus Gaussian model. 
The LF features a dip at $M(\IE) \simeq -19$ and a faint-end slope of $\alpha_{\rm S} \simeq -1.2$ to $-1.3$. 
The SMF displays a low-mass-end slope of $\alpha_{\rm S} \simeq -1.2$ to $-1.35$. 
These observed slopes are flatter than those predicted for dark matter halos in cosmological simulations, offering significant insights for models of galaxy formation and evolution.
}
%
%
    \keywords{Galaxies: clusters: individual: Perseus -- Galaxies: luminosity function -- mass function -- Galaxies: fundamental parameters}
%
%
\titlerunning{\Euclid: ERO -- Overview of the Perseus galaxy cluster and the luminosity \& stellar mass functions}
\authorrunning{Cuillandre et al.}
   
\maketitle
%
%
%
%

\begin{figure*}[htbp!]
\centering
\includegraphics[width=0.99\textwidth]{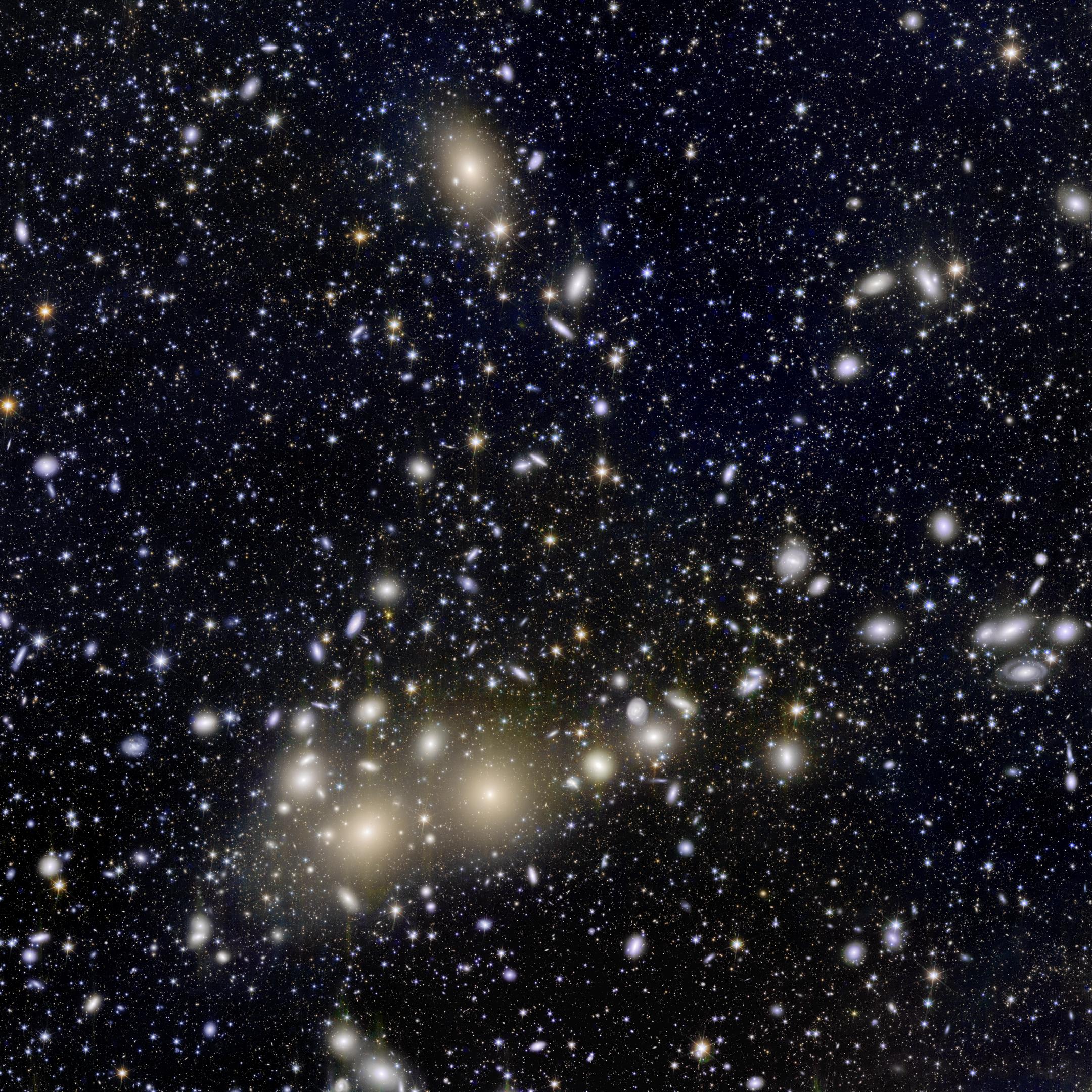}
\caption{Colour image of the Perseus cluster released worldwide by the \ac{ESA} in November 2023, created by combining \ac{VIS} and \ac{NISP} \Euclid images using the \IE\/ band in the blue, \YE in the green, and \HE in the red. The field of view of this image is $0.5\,\mathrm{deg}^2$, with the $x$- and $y$-axes aligned with the \Euclid camera native pixel geometry (see \cref{fig:3CompletenessRegions} for an equatorial projection). Credit: ESA/Euclid/Euclid Consortium/NASA, image processing by J.-C. Cuillandre (CEA Paris-Saclay), G. Anselmi.}
\label{fig:Perseus}
\end{figure*}

\section{\label{sc:Intro}Introduction}

The \ac{LF} and \ac{SMF} have often been used in the literature to constrain cosmological simulations and semi-analytic models of galaxy evolution \citep[e.g.,][]{1993MNRAS.264..201K,2000MNRAS.311..576K,2000MNRAS.319..168C,2003ApJ...599...38B,2015MNRAS.446..521S,2018MNRAS.473.4077P}. The number of objects detected in blind observations in given intervals of luminosity or mass can be easily compared to those predicted by simulations. When measured in the visible and \ac{NIR} bands they trace the statistical weight of galaxies of different luminosities and stellar masses to the total stellar emission of the Universe. Moreover, they can be measured in different intervals of redshift to study the evolution of galaxies with time \citep[e.g.,][]{1988MNRAS.232..431E,1992ApJ...390..338L,1995ApJ...455..108L,2001MNRAS.326..255C,2003ApJ...592..819B,2003ApJS..149..289B,2007ApJ...665..265F,2010A&A...523A..13P}.

The comparison of observed \ac{LF}s in the visible and \ac{NIR} bands with the predictions of simulations has been crucial for understanding the importance of feedback in shaping galaxy evolution. The observed flat slope of the distribution compared to the steep rise at the faint end predicted by the first generation of cosmological simulations (the `missing satellite problem') has been solved by including the contribution of feedback from supernovae. In low-mass systems feedback is able to sweep away a considerable fraction of the cold \ac{ISM} loosely bound to shallow gravitational potential wells, thus significantly reducing the star-formation activity in the disc. Feedback from an \ac{AGN} has also been claimed to explain the observed abrupt decrease of massive galaxies at the bright end of the \ac{LF} and \ac{SMF} \citep[e.g.,][]{1991ApJ...379...52W,1993MNRAS.264..201K,1999MNRAS.310.1087S,2003ApJ...599...38B,2015ARA&A..53...51S}, and in general the feedback processes and the interplay with the environment give rise to the different shape of the SMF compared to the halo mass function \citep{2006MNRAS.368....2D}.

The \ac{LF} and \ac{SMF} are powerful statistical tools for characterising the properties of galaxies in different environments, from rich clusters to groups, filaments, and voids. Several studies seem to indicate that the properties of the \ac{LF}s of galaxies in high-density regions are different than those derived in the field, where the slope at the faint end is generally flatter and the characteristic luminosity lower than in rich clusters (\citealp{1985AJ.....90.1759S,2001ApJ...557..117B,2003MNRAS.342..725D,2004ApJ...610..745L,2005ApJ...633..122H,2005A&A...433..415P,2006PASP..118..517B,2016ApJ...824...10F}). This systematic difference in the statistical distribution of galaxies has been explained as related to the different evolutionary paths that they undergo in rich environments where they suffer a large variety of perturbations \citep[e.g][]{2006PASP..118..517B,2014A&ARv..22...74B,2022A&ARv..30....3B}. The brightest galaxies (centrals) are formed by multiple merging events in the most massive halos \citep{1977ApJ...217L.125O,2003MNRAS.342..725D,2007MNRAS.375....2D,2011MNRAS.416.1680C,2014A&A...570A..69B}, while dwarf systems can be formed in rich environments by galaxy harassment \citep{1998ApJ...495..139M,2005MNRAS.363..509M,2006A&A...445...29P,2007ApJ...671.1471B,2011MNRAS.414.2771D} or by the simple fading of the star-formation activity of galaxies undergoing a ram-pressure-stripping event \citep[e.g.,][]{2008A&A...489.1015B,2008ApJ...674..742B,2014A&ARv..22...74B}. 
The shape of the \ac{LF} also changes as a function of wavelength, with redder galaxies dominant in high-density regions and star-forming galaxies in the field \citep[e.g.,][]{2005ApJ...629..143B}, as expected for the well-known morphological segregation effect \citep[e.g.,][]{1980ApJ...236..351D,1993ApJ...407..489W, 1997ApJ...490..577D}. 

There are, however, several unclear questions that still need to be answered. The first of them concerns the contribution of \ac{LSB} galaxies to the observed \ac{LF}, in particular at the faint end where these systems can be dominant. 
Using a complete set of data extracted from the \ac{SDSS}, \citet{2005ApJ...629..143B} have shown that the faint end slope of the LF measured in the visible bands and parameterised with a \citet{1976ApJ...203..297S} function can drastically change from $\alpha \simeq -1$ to $\alpha \simeq -1.5$ (where $\alpha$ is the slope of the power-law distribution) when \ac{LSB} galaxies are correctly accounted for. 
This steeper slope is still far from the one predicted for the dark matter halo mass function by cosmological simulations \citep[$\alpha \simeq -2$, e.g.,][]{2015ARA&A..53...51S}. 
It is rather similar to the one often observed in rich clusters where the \ac{LF}s are measured using deep and very sensitive imaging data gathered using wide-field cameras coupled with 4--8\,m class telescopes \citep[e.g.,][]{1985AJ.....90.1759S,2006A&A...445...29P,2016ApJ...824...10F}.
Presently, studies of the faint end of the \ac{LF} and \ac{SMF} can only be conducted on nearby clusters with deep imaging data able to detect \ac{LSB} systems \citep[e.g.,][]{2015ApJ...804L..26V,2015ApJ...807L...2K,2015ApJ...809L..21M,2017A&A...608A.142V,2017A&A...607A..79V,2020ApJ...899...69L,2024ApJS..271...52Z} and in field/group environments \citealt{Marleau2021}). 
Further study, including systems of varying halo mass would allow us to trace any possible dependence of the \ac{LF} on the mass of the host dark matter halo and, at the same time, minimising the impact of sampling variance on our results.

The \Euclid space mission \citep{EuclidSkyOverview} has been designed to map with an extraordinary sensitivity and image quality most of the extragalactic sky [\ac{EWS}, 14\,000\,deg$^2$] in the visible and \ac{NIR} wavelength range. 
In particular, the \ac{VIS} \citep{EuclidSkyVIS} and \ac{NISP} \citep{EuclidSkyNISP} instruments are the first wide-field cameras sensitive to surface brightness levels of 29.8 and 28.4$\,\mathrm{mag}\,\mathrm{arcsec}^{-2}$ \citep{Scaramella-EP1} in the visible and \ac{NIR} bands, respectively, values never achieved before over such a wide area.
The  \ac{EWS} will provide a unique set of photometric data to explore the \ac{LSB} Universe and thus quantify on solid statistical arguments the contribution of \ac{LSB} and \ac{UDGs}, and in general dwarf systems, at the faint end of the \ac{LF}. 
The four \Euclid photometric bands used during the observations (\IE, \YE, \JE, and \HE) are sensitive to the bulk of the stellar emission and are thus optimal to infer the \ac{SMF} with extreme accuracy. 
Covering a wide fraction of the sky, the \ac{EWS}  will be perfectly suited to derive these statistical functions for galaxies in different environments, from rich clusters to voids, and at different redshifts, thus providing a unique reference for comparison with the predictions of cosmological simulations on the mass assembly process in the Universe. 
The improvement with respect to \ac{SDSS} will be substantial, while the synergy with the \ac{UNIONS} in the northern hemisphere and \ac{LSST} in the south will be extremely useful for deriving photometric redshifts and physical properties of galaxies, hence enabling the study of the wavelength dependence of these statistical functions, fundamental for reconstructing the stellar evolution across time \citep{2022zndo...5836022G}. 

This paper presents the results obtained by the analysis of the data gathered during the \Euclid Early Release Observations \citep[ERO,][]{EROcite} programme for the Perseus cluster of galaxies.
This unique data set is used to derive the \ac{LF} and \ac{SMF} of galaxies located within $R\simeq0.25\, r_{200}$ of the cluster centre. Here, $r_{200}$ is the radius within which the cluster density is 200 times the Universe's critical density, a proxy for the virial radius. 
Being a nearby rich cluster of galaxies dominated by \ac{LSB} early-type objects, Perseus is an excellent target for testing the unique capabilities of the telescope in the study of the \ac{LSB} Universe. 
Located at a relatively low Galactic latitude ($b\sim -13^\circ$), Perseus is also an optimal target to quantify the possible contamination of Galactic cirrus emission in these bands, providing us with a unique reference for all future \Euclid extragalactic studies. 
The paper is structured as follows.  We describe in \cref{sc:Perseus} the main properties of the Perseus cluster, in \cref{sc:Data} the data used in the analysis, and in \cref{sc:Identification} the methodology used to identify all cluster members. 
The \ac{LF} and the \ac{SMF} are derived in \cref{sc:LF,sc:SMF}. The results are discussed in \cref{sc:Discussion} and the conclusions are given in \cref{sc:Conclusions}. 
In the following analysis we adopt a standard flat $\Lambda$CDM cosmology with $\Omega_{\rm m}=0.319$ and $H_0=67$\,km\,s$^{‐1}$Mpc$^{‐1}$ \citep{2020A&A...641A...6P}, and all magnitudes are given using the AB magnitude system.

\section{\label{sc:Perseus}The Perseus cluster}

The Perseus cluster (Abell\,426) is one of the most intensively studied high-density regions in the nearby Universe.
Located at only $72\,\mathrm{Mpc}$ \citep[$V_\mathrm{c} = 5258\,\mathrm{km}\,\mathrm{s}^{-1}$, $\sigma_\mathrm{c} = 1040\,\mathrm{km}\,\mathrm{s}^{-1}$;][]{2020MNRAS.494.1681A} at the edge of the Taurus void \citep{1985AJ.....90.1413B}, 
it belongs to the Perseus-Pisces supercluster, a large-scale structure in the southern sky extending more 
than $50\,\mathrm{Mpc}$ in the direction perpendicular to the line of sight \citep{1983A&A...121....5C, 1993AJ....105.1251W}.
Perseus is classified as a Bautz--Morgan type II-III cluster of richness class 2 \citep{1970ApJ...162L.149B,1999ApJS..125...35S}, 
and hosts one of the most spectacular known core-cooling flows tightly connected with the central type-D giant elliptical galaxy galaxy NGC\,1275 and its AGN activity \citep{2001AJ....122.2281C,2003MNRAS.344L..48F,2006A&A...454..437S}. 
Perseus is also the brightest known X-ray cluster of galaxies \citep{1990MNRAS.245..559E}, the target of several X-ray observations given its peculiar structure,
with large-scale motions in the \ac{IGM} and complex metal enrichment history \citep[e.g.,][]{2003ApJ...590..225C,2007MNRAS.381.1381S,2012ApJ...757..182S, 2019MNRAS.483.1701S,2013Natur.502..656W,2014PhRvL.113y1301B, 2017ApJ...837L..15A, 2017ApJ...849...54L,2020A&A...633A..42S}. X-ray observations also
indicate that the cluster might not be fully relaxed \citep{1992ApJ...397..430U,2003ApJ...590..225C,2019MNRAS.483.1744I}.
Additional peculiarities have been observed in the radio domain,
with the presence of a mini-halo and cavities associated with radio-mode feedback from NGC\,1275 \citep[e.g.,][]{1983SvAL....9..305S,1993MNRAS.264L..25B,2017MNRAS.469.3872G,2020MNRAS.499.5791G}.

The cluster has been the target of several spectroscopic surveys aimed at identifying its galaxy members, including dwarf \ac{LSB} systems \citep[e.g.,][]{1971ApJ...168..321C,1999A&AS..139..141B,2008MNRAS.383..247P,2017MNRAS.470.1512W,2019ApJS..245...10W,2020MNRAS.494.1681A,2020A&A...640A..30M}. The spectroscopic data have also been used to derive the mean properties of the cluster, such as its radius and mass \citep[$r_{200} = 2.2\,\mathrm{Mpc}$ and $M_{200} = 1.2 \times 10^{15}\,M_{\odot}$,][]{2020MNRAS.494.1681A}. These values can be compared to the estimates of $r_{200} = 1.79\,\mathrm{Mpc}$ and $M_{200} = 6.65 \times 10^{14}\,M_{\odot}$  derived from X-ray observations
by \citet{2012ApJ...757..182S}.
The cluster population is dominated by early-type systems with few spiral galaxies \citep{1983AJ.....88..697K,1994A&A...284..801A,1986AJ.....92..250G}, but it
hosts a few star-forming galaxies with radio continuum morphologies witnessing an ongoing transformation \citep[George et al. 2024 in prep.]{2022A&A...658A..44R}. 
The region analysed in this work covers $0.7\,{\rm deg}^2$ on the sky ($1\,{\rm Mpc}^2$ at the Perseus distance), located within the inner $\simeq  0.25\,r_{200}$, where the density of galaxies is very high (around $3000\,\mathrm{galaxies}\,\mathrm{deg}^{-2}$ at the distance of the cluster). Consistently with the other papers based on \ac{ERO} data on the cluster, we assume a distance of $72\,\mathrm{Mpc}$ ($m-M = 34.287$, $z = 0.0167$). 


\begin{figure*}[htbp!]
\centering
\includegraphics[width=0.7\textwidth]{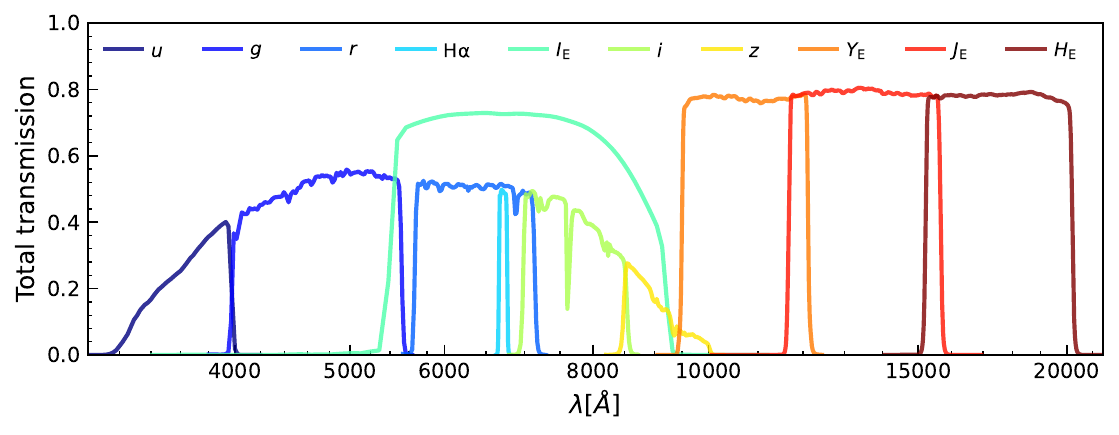}
\caption{Transmission of the \Euclid and \ac{CFHT} filters used in the present work.} 
\label{fig:filters}%
\end{figure*}

\section{\label{sc:Data}Observations and data processing}

The core of our data set are the new deep \Euclid observations of the Perseus cluster in the very broad filter \IE\,(equivalent to $r+i+z$) in the optical from the \ac{VIS} instrument \citep{2016SPIE.9904E..0QC,EuclidSkyVIS} and the three broad filters \YE, \JE, and \HE in the \ac{NIR} from the \ac{NISP} instrument \citep{2016SPIE.9904E..0TM,EuclidSkyNISP}. 

To fully harness the scientific potential of the \Euclid data, integrating complementary optical broad-band photometry is essential. This section briefly reviews the observations obtained with \Euclid and their performance, alongside deep complementary MegaCam data acquired at the \ac{CFHT} prior to the \Euclid launch.

The comprehensive range of science cases explored by the \ac{ERO} Perseus science team using this data set is detailed in \cref{appendix:ScienceOverview}. Each scientific programme demands high-quality data that preserve both resolution and photometric integrity for various astronomical objects, from compact sources such as stars, globular clusters, and field galaxies, to extended sources including large galaxies.
Details of the specific efforts undertaken to process the \ac{ERO} data set are documented in \cite{Cuillandre2024a}, while this report focuses solely on the performance achieved for the Perseus field.

Accompanying this report on the \ac{LF} and \ac{SMF} of the Perseus cluster, are two papers focusing on how these data have transformed our understanding of \ac{ICL} and \acp{ICGC} and cluster dwarf galaxies \citep{EROPerseusICL,EROPerseusDGs}.

\subsection{\label{sc:EuclidData}Euclid VIS and NISP data set}

The Perseus cluster data were collected during the \Euclid performance-verification phase in September 2023 \citep{Cuillandre2024a}. All \Euclid science observations adhere to a predetermined reference observing sequence \citep[ROS,][]{Scaramella-EP1}, which consists of four dithered exposures lasting 566\,s each in the \IE filter, and four dithered exposures of 87.2\,s each in the \YE, \JE, and \HE filters. Their passbands are shown in \cref{fig:filters} (\citealp{Schirmer-EP18,2016SPIE.9904E..0QC,Scaramella-EP1}).

Four ROSs were obtained in total on the Perseus cluster, two ROSs on 9 September, with a \ang{;3;} offset between the two ROSs, along the $x,y$ common axis of both instruments. 
Due to an inversion of the dither axis, this set was duplicated on 16 September to mitigate signal-to-noise ratio variations across the detector mosaic gaps. 
In total, with these four ROSs, the integration time over the common field of view of $0.7\,\mathrm{deg}^2$ between the two instruments is 7456.0\,s in the \IE filter and 1392.2\,s in the \YE, \JE, and \HE filters (see \cref{table:Euclidspecs}). 
This combination achieves a depth that is 0.75\,mag deeper than the \ac{EWS} for the compact sources, which relies on a single ROS \citep{Scaramella-EP1}. The data are of exceptional quality, benefiting from being gathered in a nominal spacecraft configuration.

The \ac{ERO} data are meticulously detrended, calibrated, resampled, and stacked, as detailed in \cite{Cuillandre2024a}. The relative internal photometric calibration accuracy is evaluated at 5\,\% -- this represents the internal scatter for stars matched with \textit{Gaia}-DR3 \citep{2023A&A...674A...1G} -- while the absolute photometric calibration (zero points) achieves precision at the percent level. Astrometric precision using \textit{Gaia}-DR3 as a reference reaches $8\,\mathrm{mas}$ for \ac{VIS} and $15\,\mathrm{mas}$ for \ac{NISP}, both representing less than a tenth of a pixel for each instrument across the entire field of view.

The \ac{ERO} pipeline \citep{Cuillandre2024a} produces two distinct types of stacks for each \Euclid band: an \ac{LSB} stack designed to preserve all extended emission; and a stack optimised for compact sources science, which suppresses all diffuse emission beyond approximately $\ang{;;6}$. This latter stack is referred to as the `compact-sources stack'. 

The pixel sizes for the \ac{VIS} and \ac{NISP} data are \ang{;;0.1} and \ang{;;0.3}, respectively, while the \ac{FWHM} in the \IE, \YE, \JE, and \HE science stacks measures \ang{;;0.16}, \ang{;;0.48}, \ang{;;0.49}, and \ang{;;0.50}, respectively. At the distance of the Perseus cluster, these values correspond to approximately $56\, \mathrm{pc}$ for \ac{VIS} and $171\, \mathrm{pc}$ for \ac{NISP}.

\begin{figure}[htbp!]
\centering
\includegraphics[width=0.49\textwidth]{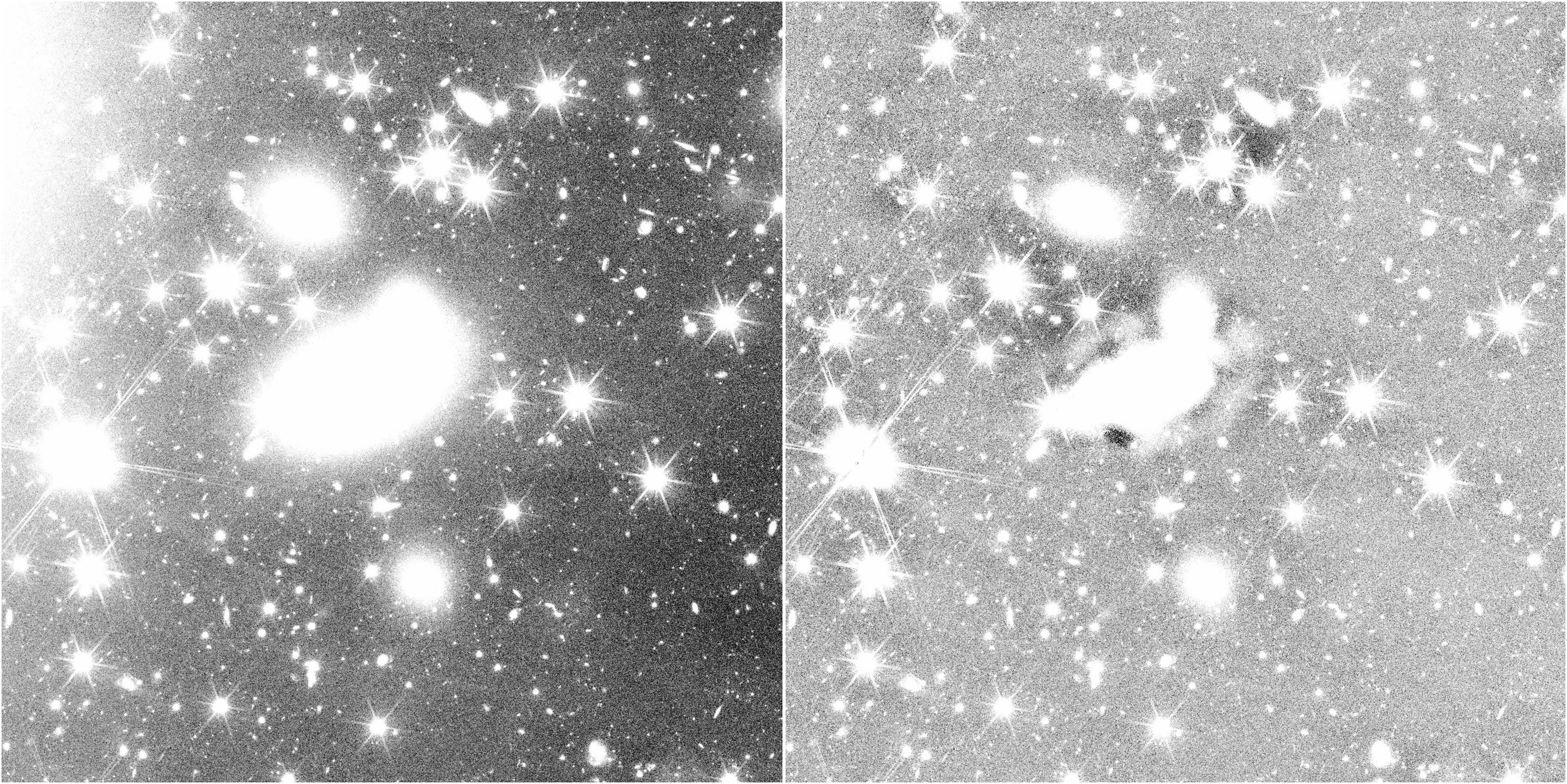}
\caption{{\it Left}: \ac{ERO} pipeline extended-emission stack preserving all scales across the image, optimised for the photometry of extended objects such as these Perseus cluster galaxies. {\it Right}: \ac{ERO} pipeline compact-sources stack where the internal background subtraction suppresses much of the signal from the large galaxies. The field size is \ang{;3;}\,by\,\ang{;3;}.}
\label{fig:LSBvsFlattened}
\end{figure}

\Cref{fig:Perseus} presents a \ac{RGB} colour image of the Perseus cluster, where extended diffuse emission is easily seen. 
The image background of each stack is dominated by zodiacal light \citep{Scaramella-EP1}, measured here at $22.3\,\mathrm{mag}\,\mathrm{arcsec}^{-2}$ in \IE, and 22.1, 22.3, and 22.5 \,mag\,arcsec$^{-2}$ in \YE, \JE, and \HE, respectively.

\begin{table*}[htbp!]
\caption{Properties of the \Euclid data set. The limiting surface brightness metric is the maximum depth reached for radially integrated profiles on Perseus cluster galaxies. The sixteen exposures originate from the execution of four \Euclid ROSs of four exposures each per filter. A point source stands for stars while a compact source stands for distant galaxies up to a few arcseconds in size at most.}
\begin{center}
\label{table:Euclidspecs}
\smallskip
\begin{tabular}{lcccc}
\hline\hline
\noalign{\vskip 1pt}
Filter & \IE & \YE & \JE & \HE \\
\hline
\noalign{\vskip 2pt}
Integration time per exposure [s] & 566.0 & 87.2 & 87.2 & 87.2\\
Total number of exposures & 16 & 16 & 16 & 16 \\
Mean image quality [arcsec] & 0.16 & 0.48 & 0.49 & 0.50\\
Sky background level [$\mathrm{mag}\,\mathrm{arcsec}^{-2}$] & 22.3 & 22.1 & 22.3 & 22.5\\
Point source depth $5\,\sigma$ PSF magnitude & 28.0 & 25.2 & 25.4 & 25.3\\
Compact source depth $10\,\sigma$ [Kron magnitude] & 26.1 & 23.8 & 24.0 & 24.1\\
Achieved limiting surface brightness [$\mathrm{mag}\,\mathrm{arcsec}^{-2}$]  & 30.1 & 29.1 & 29.2 & 29.2\\
\hline
\end{tabular}
\end{center}
\end{table*}

Section~9  of \citet{Cuillandre2024a} reports that the sensitivity to \ac{LSB} features of both instruments is exceptional thanks to \Euclid's unique optical design, and  based on a depth metric assuming the absence of contaminants in the sky background, such as high stellar density or Galactic cirrus, the following depths are expected to be reached: $\mu_{I_{\rm E}} = 30.5\sbmag$ and $\mu_{Y_{\rm E}, J_{\rm E},H_{\rm E}} = 28.7,\, 28.9,\, 28.9\sbmag$ $1\,\sigma$ limiting surface brightness at the $\ang{;;10}\times\ang{;;10}$ scale.

\begin{figure}[htbp!]
\centering
\includegraphics[width=0.49\textwidth]{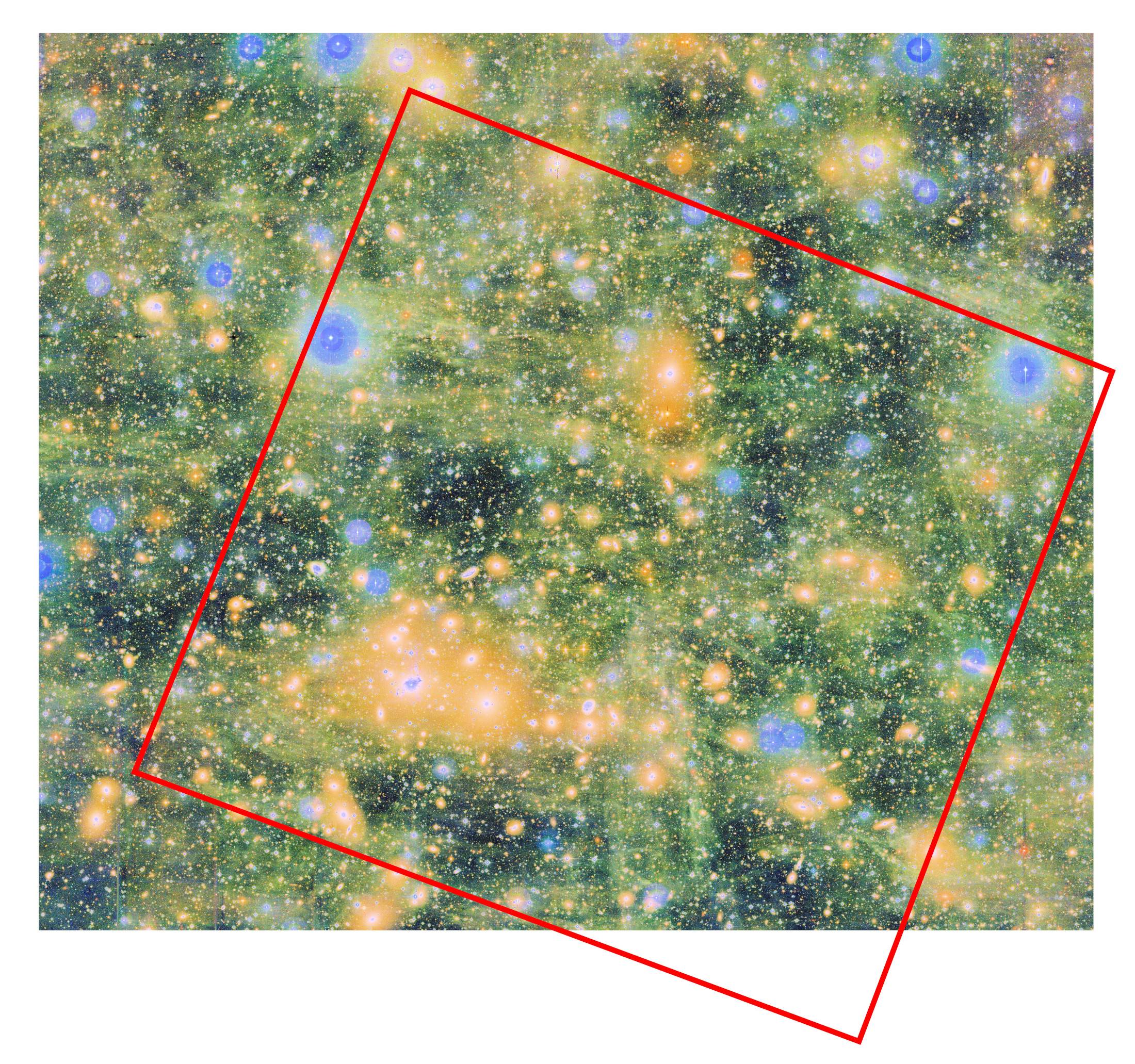}
\caption{The \Euclid Perseus area, marked in red and covering $0.7\,\mathrm{deg}^2$, is shown against the broader coverage of the \ac{CFHT}-MegaCam observations, which extend over $1.2\,\mathrm{deg}^2$. This high-contrast colour image is a composite \ac{RGB} picture created from the MegaCam $r$, $g$, and $u$ bands. The Galactic cirrus, which peaks in brightness in the $g$ band, highlights the extent of contamination in this area, with an average $E(B-V)$ of 0.17. While significantly fainter in the $r$ band, the Galactic cirrus still impacts the \ac{VIS} $\IE$ band, thereby limiting the low-surface brightness performance across the \Euclid field of view.} 
\label{fig:Perseus-ISM-MegaCam}%
\end{figure}

As detailed in \cref{sc:Identification}, we extensively employ the tools \texttt{AutoProf} and \texttt{AstroPhot} \citep{ConnorAP,2023MNRAS.525.6377S} to derive radially integrated profiles for all galaxies in the Perseus cluster. The Perseus field, filled with objects of various kinds, exhibits a notably patchy background due to Galactic cirrus, particularly prominent in the \IE band. This patchiness, combined with a relatively high stellar density, significantly challenges our ability to detect ultra-faint contrasts, even when integrating over thousands of square arcseconds in the outer annulus.

Based on an \texttt{AutoProf} analysis of isolated elliptical galaxies in the Perseus cluster (such as NGC\,1281), which possess certified faint extended stellar halos blending into the sky, we evaluate the \ac{LSB} performance relevant to our scientific objectives over several arcminutes. For the best cases -- isolated galaxies with minimal Galactic interference -- the Perseus data set achieves an \ac{LSB} performance of $\mu_{I_{\rm E}}  = 30.1\sbmag$ and $\mu_{{Y_{\rm E}} ,J_{\rm E},H_{\rm E}} = 29.1,\, 29.2,\, 29.2\,\sbmag$. Compared to the standard \ac{LSB} metric, \ac{VIS} slightly underperforms, largely due to the presence of Galactic cirrus that are brighter and more structured in the \IE band, with surface brightness levels ranging from 25 (maximum) to 26 (median) \sbmag.

\begin{figure}[htbp!]
\centering
\includegraphics[width=0.49\textwidth]{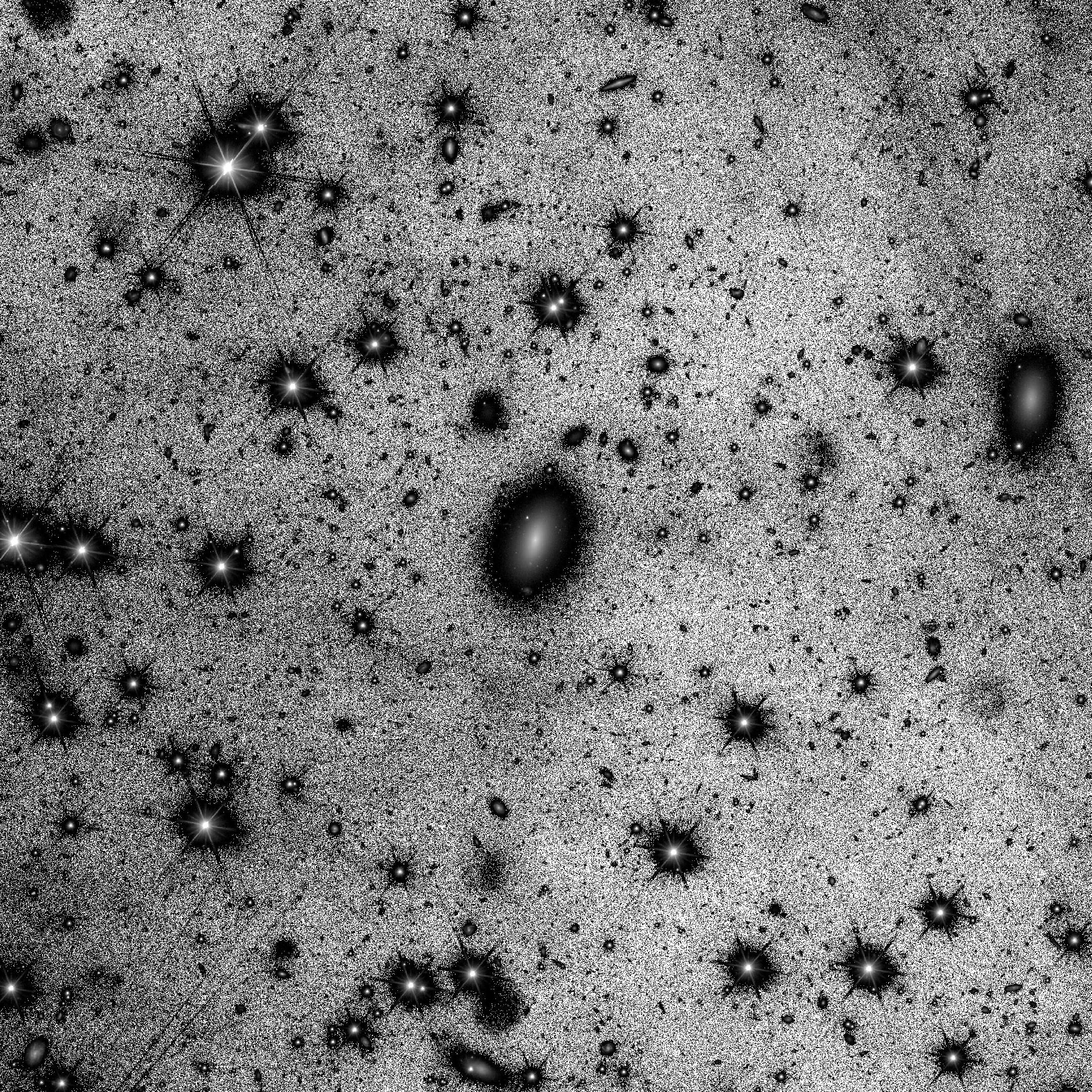}
\caption{Galactic cirrus contamination in the Perseus field.  This illustration, particularly around UGC\,2621 (located at the center of the image), demonstrates the non-uniformity of the sky background in the \IE\,filter. Here, the cirrus shines at surface brightness levels from 26.3\,mag\,arcsec$^{-2}$ (maximum) to 27.5\,mag\,arcsec$^{-2}$ (median), affecting the \ac{LSB} detection performance. This contamination limits our ability to detect other extended emission, such as the outskirts of diffuse stellar halos. The field of view here is \ang{;5;} by \ang{;5;}.}
\label{fig:SmallElliptical-VIS-Cirrus}%
\end{figure}

These \Euclid data are ideally suited for deriving the \ac{LF} and \ac{SMF} of the cluster, especially since their faint ends are dominated by \ac{LSB} systems \citep[e.g.,][]{1985AJ.....90.1759S}. The depths for compact sources, as determined from our multi-band photometric catalogue, are detailed in \cref{sc:Catalogues}.

\subsection{\label{sc:CFHTData}CFHT-MegaCam data set}

In the context of the Euclid Surveys \citep{Scaramella-EP1}, complementary ground-based optical data are crucial for enhancing the \Euclid data set, particularly for deriving colours and computing photometric redshifts. Observations were conducted using MegaCam at the \ac{CFHT} atop Maunakea, employing the $u$, $g$, $r$, $i$, and $z$ filters, along with the H$\alpha$ `off' filter (CFHT ID~9604). This filter is centred on $\lambda_{\rm c}=6719\,$\AA\ with a width of $\Delta \lambda=109\,$\AA, corresponding to a heliocentric velocity range of $4660\lesssim v_{\mathrm{hel}}/\mathrm{km}\,\mathrm{s}^{-1} \la 9600$, making it well-suited to cover the velocity range of the Perseus cluster.

The observations took place during January and February 2021, November 2021, and September 2022, under dark skies and photometric conditions at an airmass of 1.3. \Cref{table:CFHTspecs} details the main properties of this data set.

\begin{table*}[htbp!]
\caption{Properties of the \ac{CFHT}-MegaCam data set. The limiting surface brightness metric is the maximum depth reached for radially integrated profiles on Perseus cluster galaxies using \texttt{AutoProf}. Both compact source performance (Kron magnitude) and diffuse emission performance (\ac{LSB}) match well with the \Euclid \ac{NIR} observations, facilitating the production of a rich multi-band catalogue.}
\begin{center}
\label{table:CFHTspecs}
\smallskip
\begin{tabular}{lcccccc}
\hline\hline
\noalign{\vskip 1pt}
Filter & $u$ & $g$ & $r$ & $i$ & $z$ & ${\rm H}\alpha$\\
\hline
\noalign{\vskip 2pt}
Integration time per exposure [s] & 240 & 120 & 120 & 240 & 240 & 360\\
Total number of exposures & 7 & 14 & 51 & 10 & 6 & 7 \\
Mean image quality [arcsec] & 1.46 & 1.23 & 0.79 & 0.56 & 0.60 & 0.49\\
Sky background level [$\mathrm{mag}\, \mathrm{arcsec}^{-2}$] & 23.07 & 22.3 & 21.3 & 20.5 & 19.2 & 21.2\\
Compact source depth $10\,\sigma$ [Kron magnitude] & 24.3 & 25.1 & 25.1 & 24.0 & 22.3 & 22.9\\
Limiting surface brightness [$\mathrm{mag}\, \mathrm{arcsec}^{-2}$] & 29.5 & 29.9 & 29.8 & 29.5 & 28.0 & n/a\\
\hline
\end{tabular}
\end{center}
\end{table*}

\begin{figure}[htbp!]
\centering
\includegraphics[width=0.49\textwidth]{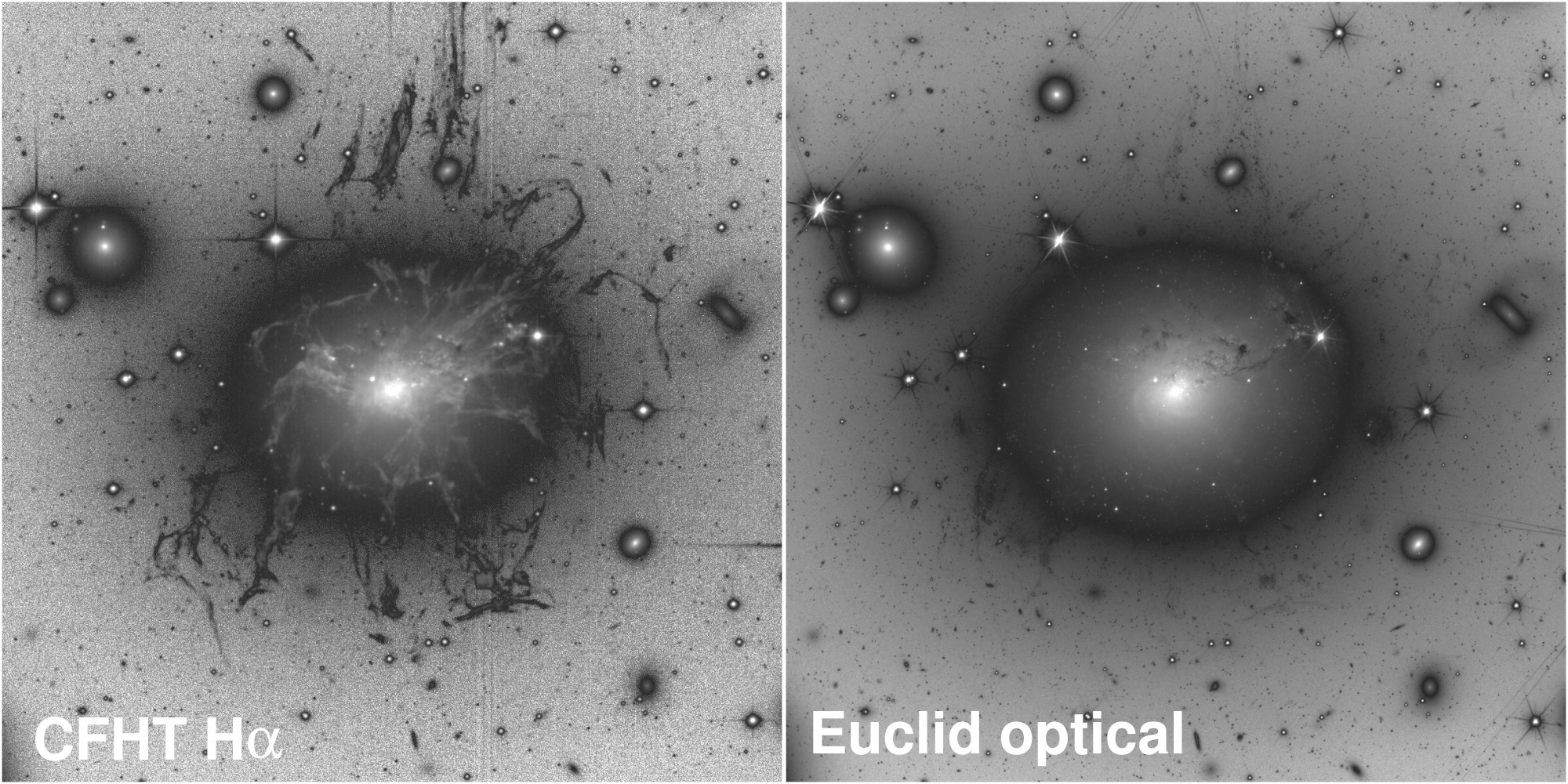}
\caption{NGC\,1275 in the \ac{CFHT}-MegaCam H$\alpha$ image in comparison to the \Euclid \IE image, demonstrating that our adopted narrow-band filter captures the H$\alpha$ emission line at the velocity range of the Perseus cluster, as demonstrated here with the showcase of NGC\,1275's spectacular filaments of ionised gas. The field of view per panel is \ang{;4;} by \ang{;4;}.}
\label{fig:NGC1275-CFHT-Euclid}%
\end{figure}

\begin{figure}[htbp!]
\centering
\includegraphics[width=0.49\textwidth]{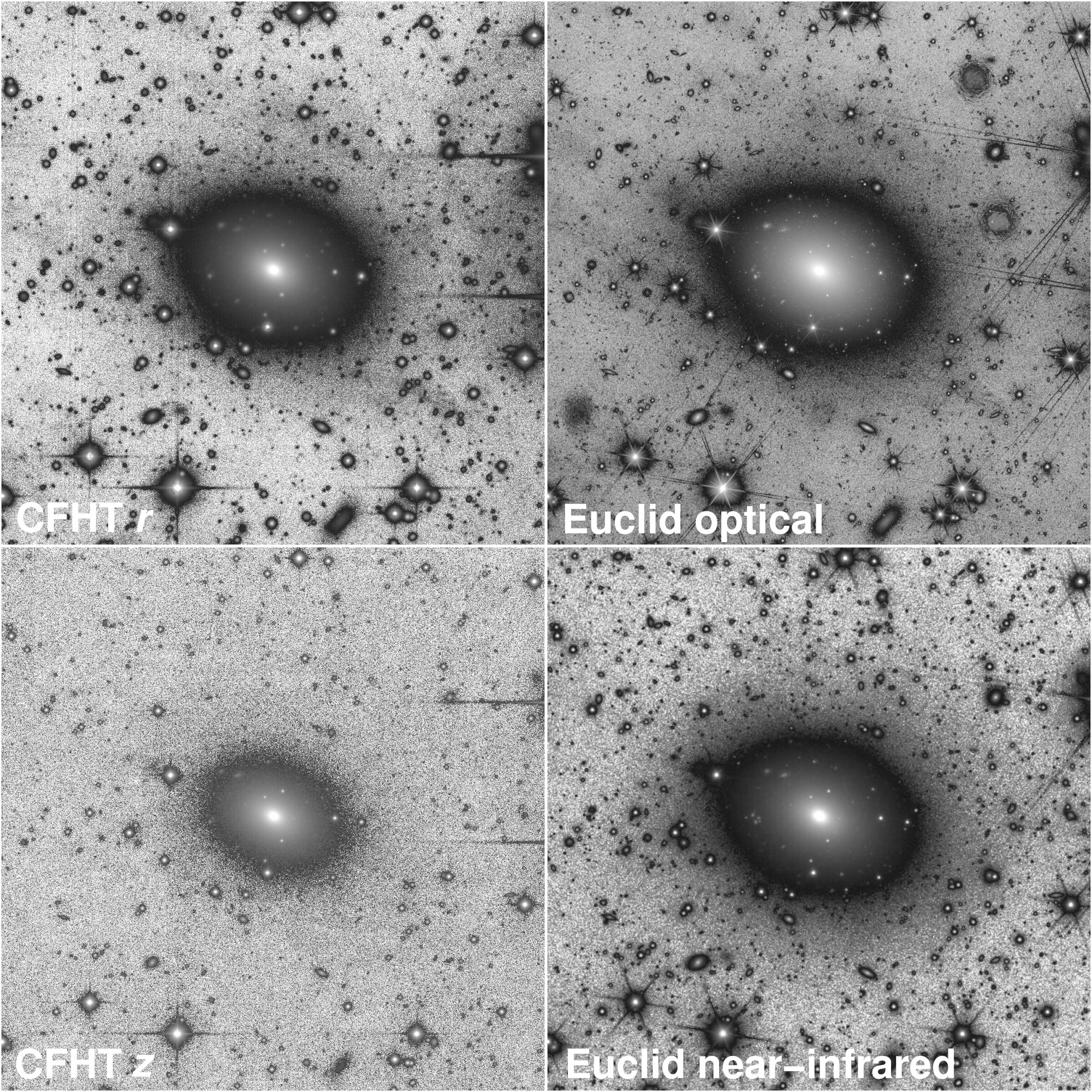}
\caption{Illustration of the comparative depth of the \ac{CFHT} and \Euclid Perseus cluster observations (the featured galaxy is PGC\,12254).  This indicates that our ground-based optical data ($r$-band here) match the \Euclid data well, both for compact sources and extended emission. The detection becomes challenging from the ground towards the \ac{NIR} where the Earth sky brightness increases rapidly (\ac{CFHT} $z$-band observations versus \Euclid \HE-band here). The field of view per panel is \ang{;4;} by \ang{;4;}.}
\label{fig:PGC12254-CFHT-Euclid}%
\end{figure}

A very large dithering approach was employed for these observations, moving objects by up to \ang{;10;} along both camera axes. This scale surpasses the size of all astronomical objects in the field, including the large elliptical galaxies that necessitated this dithering strategy. Such extensive dithering facilitates a clean median model of the background, although it results in a loss of field of view in the final stacked images. Only the central region, where the signal-to-noise ratio is relatively uniform, is retained.

The native camera field of view is $1.1\,\mathrm{deg}^{2}$, with a pixel resolution of 0.187\,arcsec\,pixel$^{-1}$. The final MegaCam stacks cover $1.2\,\mathrm{deg}^{2}$ and overlap with $95\%$ of the \Euclid observation (\cref{fig:Perseus-ISM-MegaCam}).

All images were detrended and calibrated using the \texttt{Elixir} pipeline \citep{Magnier2004}, and further processed with the \texttt{Elixir}-LSB pipeline, specifically designed for the Next Generation Virgo Survey \citep[NGVS,][]{2012ApJS..200....4F}, to detect extended \ac{LSB} features. This pipeline was also applied to the narrow-band imaging for the Virgo Environmental Survey Tracing Ionised Gas Emission \citep[VESTIGE,][]{2018A&A...614A..56B}. Thanks to MegaCam's well-calibrated and stable performance, no significant challenges were encountered in preparing this data set, which was captured under excellent sky conditions.

Exceptional image quality was achieved using the H$\alpha$ filter (\ac{FWHM}$ =\ang{;;0.49}$), comparable to the \Euclid \ac{NIR} bands, and in a subset of $r$-band images that matched the \ac{FWHM} criteria. The $r$-band image was employed to subtract a scaled continuum, producing a pure H$\alpha$ emission image used in this study to identify signs of star formation and assess the membership of some galaxies in the Perseus cluster.

\begin{figure*}
\centering
\includegraphics[width=0.99\textwidth]{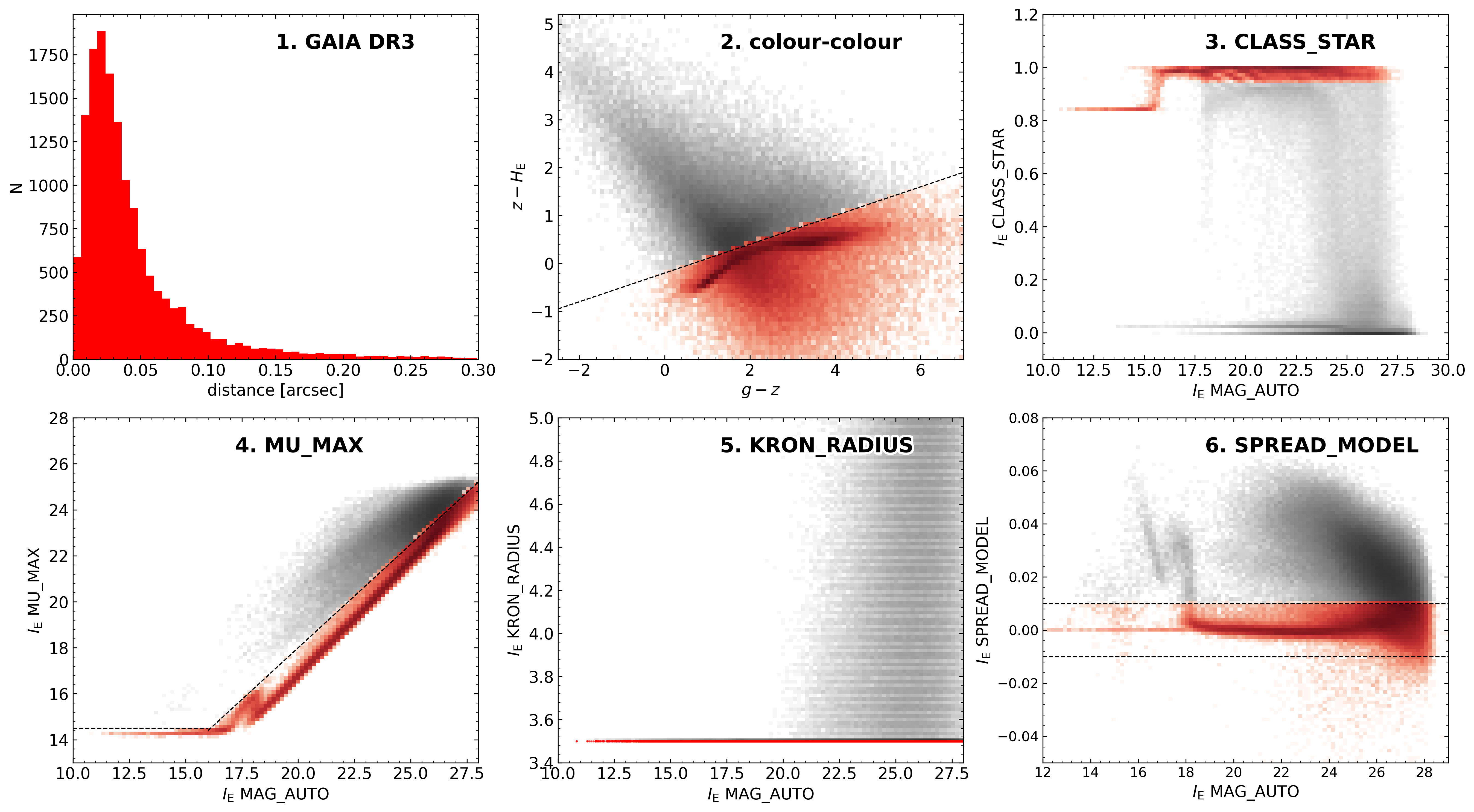}
\caption{The diagnostics used for star-galaxy separation: a source is classified as star if it fulfils at least four selection criteria. The panels illustrate: 1. the distance with the matched  {\it Gaia}-DR3 stars; 2. the colour-colour diagnostic; 3. the stellarity criterion provided by {\tt Sextractor}; 4. the relation between the magnitude in the Kron-aperture and the maximum surface brightness; 5. the size of the objects; 6. the {\tt SPREAD\_MODEL} sequence for point-like objects. The objects selected in each criterion are highlighted in red.}
\label{fig:stars}%
\end{figure*}

\subsection{\label{sc:Catalogues}Compact sources multi-band photometric catalogue}

The general effort for \ac{ERO} processing and calibration \citep{Cuillandre2024a} was focused on providing all science teams with astrometrically and photometrically calibrated image stacks across all four \Euclid bands, accompanied by comprehensive catalogues produced using the tool \texttt{SourceExtractor} \citep{1996A&AS..117..393B}. These versatile, science-ready catalogues are tailored exclusively for compact sources, utilising only the compact-sources stack while excluding the diffuse-emission stacks designed for tools such as \texttt{AutoProf}/\texttt{AstroPhot} used in this study. This processing enabled us to determine the depth of each band that we report in this work.

The \ac{VIS} catalogue is stand-alone, reflecting its distinct depth and resolution compared to the \ac{NISP} bands. For the \YE, \JE, and \HE\/ bands, a $\chi^2$ image combining these three bands is initially created by the \ac{ERO} pipeline to optimise detection across all bands. A $\chi^2$ image is produced by \texttt{SWarp} \citep{Bertin2002}, which combines all available signals to create very deep images. These images match the position, scale, and input size of the \YE, \JE, and \HE\ bands images. \Ac{PSF} models are derived for each band using \texttt{PSFex} to facilitate \ac{PSF} photometry by \texttt{SourceExtractor}. The detection threshold for both \ac{VIS} and \ac{NISP} is set at $5$ pixels above $1.5\,\sigma$, resulting in a total of 546\,562 sources in the \ac{VIS} catalogue and 335\,340 for each of the \ac{NISP} bands. These catalogues are densely populated with physical parameters, featuring approximately 200 columns that leverage the latest advancements in \texttt{SourceExtractor}, including spheroid and disc models. The depths are reported in \cref{table:Euclidspecs}.

\begin{figure}[ht!]
\centering
\includegraphics[width=0.49\textwidth]{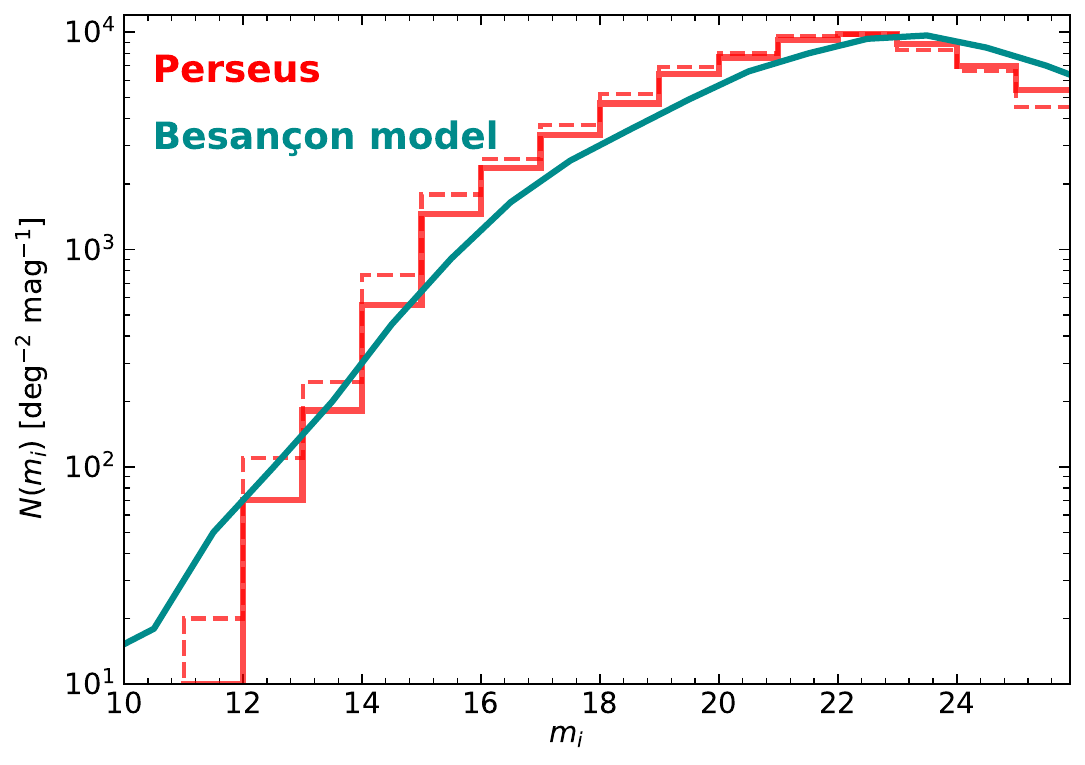}
\caption{Number counts in the {\tt MAG\_AUTO} $i_{\rm CFHT}$ band compared to the Besan\c{c}on model rendition in the same filter.  The reference counts have been extracted at the centre of the Perseus field in an area of 1\,deg$^2$ and compared to the number counts of stars selected with the multiple criteria method. The solid histogram refers to the data not corrected for \ac{MW} extinction, while the dashed histogram represents the corrected $i_{\rm CFHT}$.}
\label{fig:starcounts}%
\end{figure}
\begin{figure*}
\centering
\includegraphics[width=0.49\textwidth]{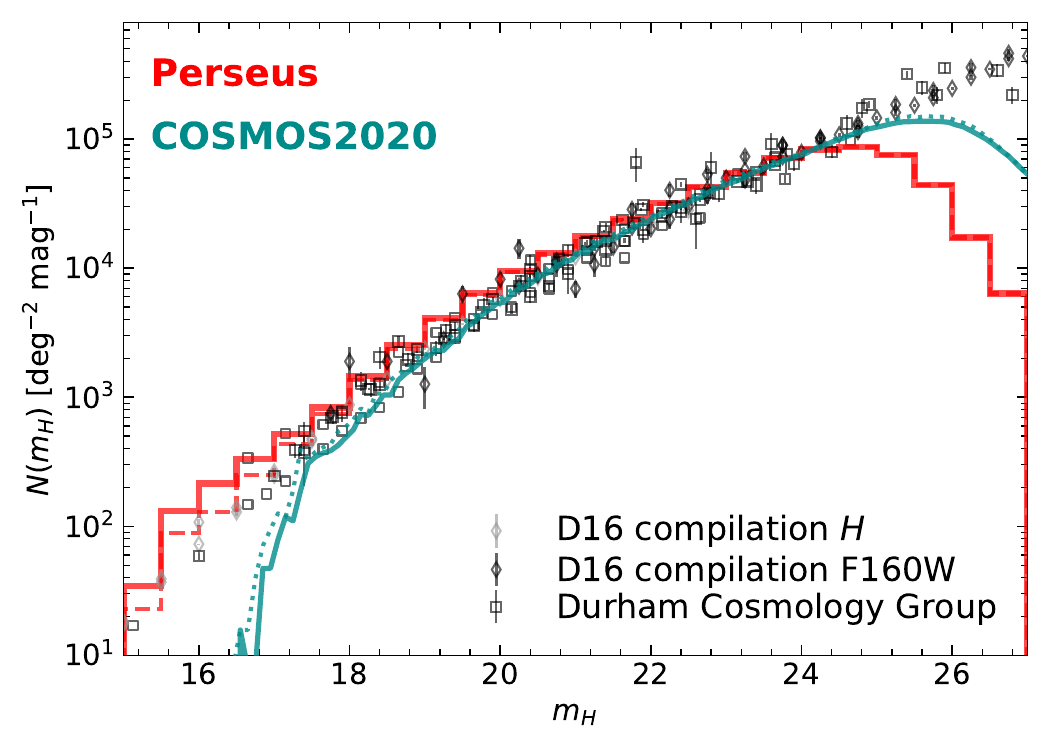}
\includegraphics[width=0.49\textwidth]{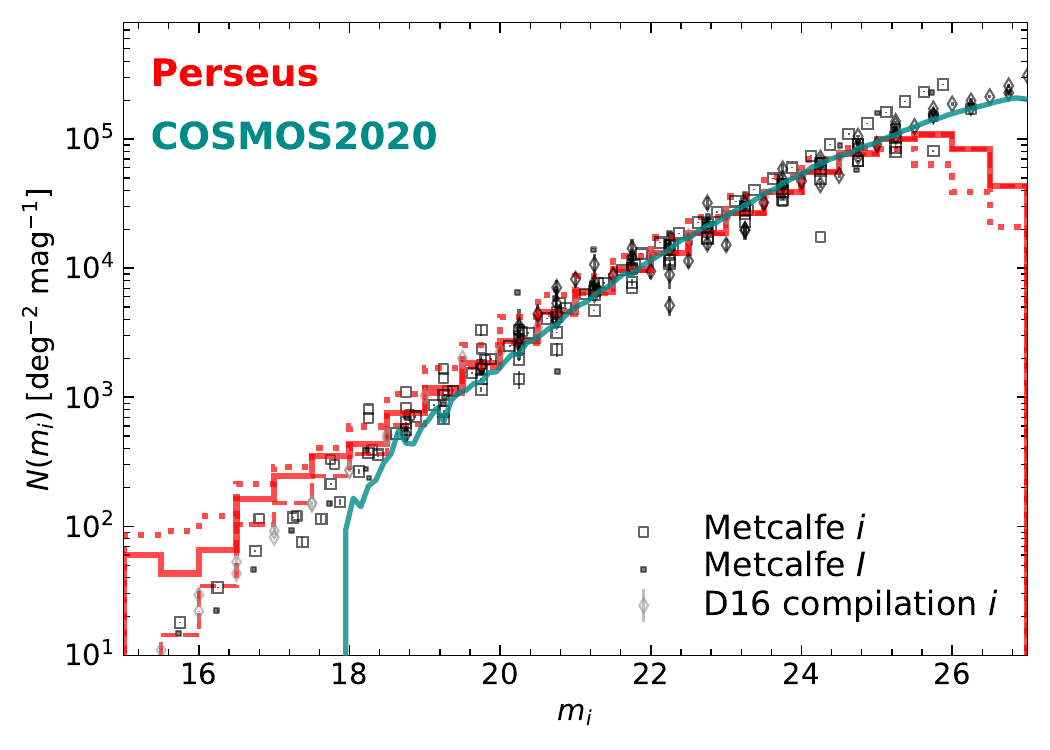}
\caption{Number counts of galaxies in the Perseus field compared to COSMOS2020 in \HE and $i_{\rm CFHT}$, both corrected for \ac{MW} extinction. 
Other data from the literature are taken from \cite{Driver:16}, while the Durham Cosmology Group's counts are from the compilation available at the dedicated web page (\url{http://star-www.dur.ac.uk/\textasciitilde nm/pubhtml/counts/counts.html}).
{\it Left}: \HE number counts. The COSMOS (from the Farmer catalogue described in 
\citealp{2022ApJS..258...11W}, after removing masked regions and objects classified as stars) counts are derived from UltraVISTA $H$-band magnitudes (dotted dark cyan line), also converted to \HE using the equation~D.22 in \citet[][solid dark cyan line]{Schirmer-EP29}.  Perseus galaxy number counts in \HE are shown as a solid red line, while the dashed red line represents the counts once  the members of the Perseus cluster are subtracted. 
{\it Right}: $i$-band number counts. The COSMOS2020 data are from the \ac{HSC} $i$-band, Perseus magnitudes are \IE (red solid line) and $i_{\rm CFHT}$ (red dotted line); the dashed red line represents the counts after the members of the Perseus cluster are subtracted.
}
\label{fig:galcounts}%
\end{figure*}

The need for photometric redshifts in the Perseus programme necessitated the creation of matching catalogues for all our \ac{CFHT}-MegaCam bands. Custom stacks were constructed from the individual images using \texttt{SWarp} \citep{Bertin2002} to resample individual frames just once -- from the native \ang{;;0.187} per pixel to the \ac{NISP} pixel scale of \ang{;;0.3}. This single resampling was mandated by the use of the \ac{NISP} $\chi^2$ image for detection. Consequently, the MegaCam catalogues contain the same number of entries as the \ac{NISP} catalogues and the fluxes are derived in the same apertures, to ensure the colours from the same physical regions are used when deriving photometric redshifts.

The multi-wavelength catalogue utilised in this analysis was constructed by matching the coordinates of \ac{NISP} and \ac{VIS} sources with a search radius of $1\arcsec$. This method effectively removes most spurious detections resulting from cosmic rays or border effects,  resulting in a catalogue of 263\,196 sources. 
The availability of colours and morphological parameters from \texttt{SourceExtractor}\footnote{See \url{https://SExtractor.readthedocs.io/en/latest/index.html} for the details on the used parameters.} allowed us to perform a thorough star-galaxy separation, combining different methods, with limits identified looking at positions of the matched SDSS and \textit{Gaia}stars and extrapolating to fainter magnitudes. We applied the following six criteria \citep[see e.g.,][for similar combinations of criteria]{2008A&A...482...81T,2023A&A...671A.146E}: 
\begin{enumerate}

\item match with \textit{Gaia} DR3 \citep{2016A&A...595A...1G, 2023A&A...674A...1G} stars:  candidate stars are identified within a search radius of $\ang{;;0.3}$; 

\item colour-colour $(g-z)$ versus $(z-\HE)$ diagram: with candidate stars characterised by $(z-\HE) < 0.3(g-z)-0.2$ \citep[similar to the well known $BzK$ criterion by][]{2004ApJ...617..746D}; 

\item stellarity classifier: high probability provided by {\tt CLASS\_STAR}, with ${\tt CLASS\_STAR} > 0.95$ in \IE if ${\tt MAG\_AUTO} > 16$ or ${\tt CLASS\_STAR} > 0.80$ if ${\tt MAG\_AUTO} \le 16$ to include saturated objects; 

\item compactness: the maximum surface brightness to be ${\tt MU\_MAX} < 14.5$ in \IE if ${\tt MAG\_AUTO} < 16$ (to include saturated objects) or ${\tt MU\_MAX} < 0.9\,{\tt MAG\_AUTO}$ if ${\tt MAG\_AUTO} > 16$; 

\item small size: ${\tt KRON\_RADIUS} = 3.5$\,px in \IE, corresponding to the minimum aperture adopted in \texttt{SourceExtractor}; 

\item comparison with local PSF: ${\tt SPREAD\_MODEL} < 0.01$ in \IE 
\citep[see also][]{EROGalGCs}, with {\tt SPREAD\_MODEL} being a \texttt{SourceExtractor} parameter that compares the objects with the local PSF model, and providing values close to zero for point sources, positive for extended sources, and negative for detections smaller than the PSF.

\end{enumerate}

The criteria and their selection regions are illustrated in \cref{fig:stars}, where in some of the panels the transition to bright saturated objects is producing discontinuities. 

We considered as stars the objects fulfilling at least four of the six above criteria.  
To validate the procedure, we verified that all the objects identified as stars in \ac{SDSS} are included in our sample. 
The final number of selected candidate stars is 49\,922 and their number counts are in fairly good agreement with the Besan\c{c}on model of stellar population synthesis of the Galaxy \citep{Robin+2003,Czekaj2012,Lagarde+2021} as shown in \cref{fig:starcounts}. We observe a mild overestimate of our star number counts at  magnitudes between $m_i = 15$ and $19$, that can be due to either misidentification of globular clusters or of compact galaxies. 
Very compact or nucleated galaxies could be classified as stars with the above criteria: we verified that among the sample of dwarf galaxies in \cite{EROPerseusDGs}, only $4$ out of $1100$ were incorrectly assigned to the sample of stars candidates. 
The above validations indicate that the combination of criteria for star-galaxy separation is effective out to faint magnitudes,  at the same time avoiding the failure of some criteria because of saturation issues.

The remaining 212\,975 objects are considered as galaxies in the following analysis. Their intrinsic fluxes are derived by correcting for the \ac{MW} extinction, which is significant in the Perseus field of view (see also \citealp{EROPerseusDGs}). 

We adopted the Planck 2013 \citep{2014A&A...571A..11P} dust opacity map,\footnote{\protect\url{{https://irsa.ipac.caltech.edu/data/Planck/release_1/all-sky-maps/}}} from which we extracted the values of the colour excess $E(B-V)$ to be associated to each galaxy in our catalogue (see \cref{fig:PerseusCatalogMap}). The colour excess is related to the magnitude absorbed at different wavelength through the \ac{MW} extinction curve $k(\lambda) = A(\lambda)/E(B-V) = R_V \; A(\lambda)/A_V$, where $A(\lambda)$ and $A_V$ are the magnitudes attenuated at the wavelength $\lambda$ and in the $V$ filter. We used the extinction curve by \cite{2023ApJ...950...86G}, assuming the extinction ratio $R_V=3.1$. 

Since the extinction depends on the wavelength and can vary substantially, especially in the UV, the exact derivation of the absorbed magnitude in broad-band photometry depends also on the shape of the \ac{SED} of the observed object  \citep[e.g.][]{2017A&A...598A..20G}. 
The correction process is implemented in a consistent way in \texttt{Phosphoros} (Paltani et al. in prep.), the photometric redshift code used in \cref{sc:ZP}, while we employed the SED of a $5700\,\mathrm{K}$ blackbody to derive an average correction in all the other cases. 
To this aim we derived a factor $c_x$ to derive the intrinsic magnitude in a filter $x$ as $m_{\rm int} = m_{\rm obs} - c_x E(B-V)$. This parameter has been derived including the above ingredients and the total transmission of the filters $R_x(\lambda)$ represented in \cref{fig:filters}, multiplied by $\lambda$ to take into account the photon transmission: 
\begin{equation}
c_x = 2.5\logten\frac{\int R_x(\lambda)\, \lambda\, F_{\rm BB5700}(\lambda)\, 10^{0.4\,k(\lambda)} \, \diff\lambda}{\int R_x(\lambda)\, \lambda\, F_{\rm BB5700}(\lambda)\, \diff\lambda}\, .
\label{eq:MWext}
\end{equation}
The values we obtained are $c_x = 4.633$, $3.552$, $2.516$, $1.919$, $1.487$, $2.122$, $1.066$, $0.726$, and $0.470$ in the $u$, $g$, $r$, $i$, $z$, $\IE$, $\YE$, $\JE$, and $\HE$ bands, respectively.
Given that the $E(B-V)$ values range from $0.12$ to $0.20$, the magnitudes absorbed in the \IE band range from $0.255$ to $0.424$. These corrections have been implemented for the objects not classified as stars. 

The galaxy number counts compared to COSMOS2020 \citep{2022ApJS..258...11W} and literature data are shown in \cref{fig:galcounts}. Overall we see a quite good agreement in the whole range of optical and \ac{NIR} magnitudes, especially when removing the members of the Perseus cluster: at bright magnitudes the COSMOS2020 catalogue is incomplete because the COSMOS field was chosen to explore the distant Universe, while the decrement at faint magnitudes is due to the incompleteness beyond the limiting magnitudes.

The photometric redshifts derived for the full sample of galaxies are described in \cref{sc:ZP}.


\section{\label{sc:Identification}Identification of cluster galaxies}

The major challenge in the determination of the \ac{LF} and \ac{SMF} in nearby clusters of galaxies is the accurate identification of all cluster members down to a given magnitude limit. The different criteria used to identify cluster members should also maximise the completeness and minimise the contamination of foreground and background sources. The analysis presented in this work is based on a selection performed on the \ac{VIS} data, which are of higher image quality in terms of sensitivity and angular resolution than the \ac{NISP} data. We also limit the identification of cluster members to resolved systems, thus to objects with an optical extension exceeding $56$\,pc (corresponding to the size of the FWHM in VIS, \ang{;;0.16}, see \cref{sc:EuclidData}). We will discuss in a following section how this assumption can affect the results.
For the identification of cluster members we adopt the following methodology used by \citet{2016ApJ...824...10F, 2020ApJ...890..128F}, and \citet{2016A&A...585A...2B}, which already proved to be efficient in the Virgo cluster. This identification is first based on morphological arguments, but then follows more stringent statistical and quantitative methods based on selected scaling relations to reject background systems and reduce incompleteness. Despite the greater distance to Perseus ($72\,\mathrm{Mpc}$ versus $16.5\,\mathrm{Mpc}$ for Virgo), this is possible thanks to the crisp image quality of the \Euclid data versus the ground-based imaging data of the NGVS in terms of angular resolution \citep[FWHM in the $i$ band $\lesssim \ang{;;0.6}$,][ versus \ang{;;0.16} in Perseus]{2012ApJS..200....4F} and sensitivity to \ac{LSB} features ($\mu_g = 29.0\,\mathrm{mag}\,\mathrm{arcsec}^{-2}$ versus $\mu_{I_{\rm E}} = 30.1\,\mathrm{mag}\,\mathrm{arcsec}^{-2}$ in Perseus).

\subsection{\label{sc:MainCatalog}Main catalogue}

The main sample of galaxies analysed in this work is the combination of dwarf Perseus members described in \citet{EROPerseusDGs} and the bright Perseus members (Mondelin et al. in prep.). 

The dwarf systems have been identified as cluster members after visual inspection of the VIS and VIS+NISP colour images performed by seven independent (human) classifiers. A detailed description of the criteria used to identify and select the dwarf galaxy candidates, as well as the comparison with other samples of dwarf systems in this cluster, can be found in that paper. We recall that cluster membership based on morphological classification has been successfully used in the past \citep[e.g.][VCC catalogue]{1985AJ.....90.1681B}, also in massive clusters such as Coma \citep{2008A&A...490..923M} and Perseus \citep{2019ApJS..245...10W}. A total of 1100 dwarf galaxies were counted. We then shift our focus to galaxies located at the periphery of the \ac{FOV} captured by the NISP and VIS instruments. The alignment between the two fields of view is not perfect, as depicted in \citet{Cuillandre2024a}. Regarding the \ac{LF} analysis, we concentrate on the shared \ac{FOV} between VIS and NISP. Thus 17 dwarf galaxies were excluded from the cluster source catalogue.

For the most luminous galaxies, a thorough selection process was carried out. Initially, we surveyed galaxies within the field with spectroscopic redshifts sourced from the literature, primarily from the \ac{SDSS} catalogue\footnote{\url{https://skyserver.sdss.org/dr18/}} but also from the NASA Extragalactic Database.\footnote{\url{ned.ipac.caltech.edu}} The selection criterion was a spectroscopic redshift ranging from 0.005 to 0.03. Following this, we examined galaxies with only photometric redshifts falling within the range of 0.005 to 0.03. Notably, we observed a small subset of \ac{SDSS} galaxies in the $z_{\rm spec}$ versus $z_{\rm phot}$ plane with $z_{\rm spec}$ below 0.03 but $z_{\rm phot}$ between 0.03 and 0.06, as depicted in \cref{phot_zSDSS}. Consequently, we expanded the selection criteria to include galaxies with $z_{\rm phot}$ up to 0.06. Subsequently, we visually confirmed galaxies with sizes exceeding several arcseconds and/or displaying visible substructure as potential cluster members. Finally, we conducted a thorough review of the entire VIS image, overlaying the source catalogue to ensure the inclusion of any bright galaxies. Additionally, we explored approximately 20 galaxies with sizes of a few arcseconds that could potentially be located in the outskirts of the cluster. 

The two final samples of dwarf and bright galaxies include 1083 and 137 galaxies, respectively, for a total of 1220 objects inside the common \ac{FOV} of \ac{VIS} and \ac{NISP}. This number drops to 1211 once the sample is cut to a completeness limit of $\IE \lesssim 23.6$ after correcting for the \ac{MW} extinction. The final sample of bright galaxies includes 105 galaxies with spectroscopic redshifts; out of the remaining 33 galaxies, 24 have photometric redshifts from SDSS, while 9 lack a redshift estimate. We made several tests to identify any possible missing cluster object, as well as to quantify the completeness as a function of the apparent magnitude and the possible contamination of background sources. Our derivation of the completeness function is described in \cref{appendix:Completeness}.

\begin{figure*}[htbp!]
\centering
\includegraphics[width=1.0\textwidth]{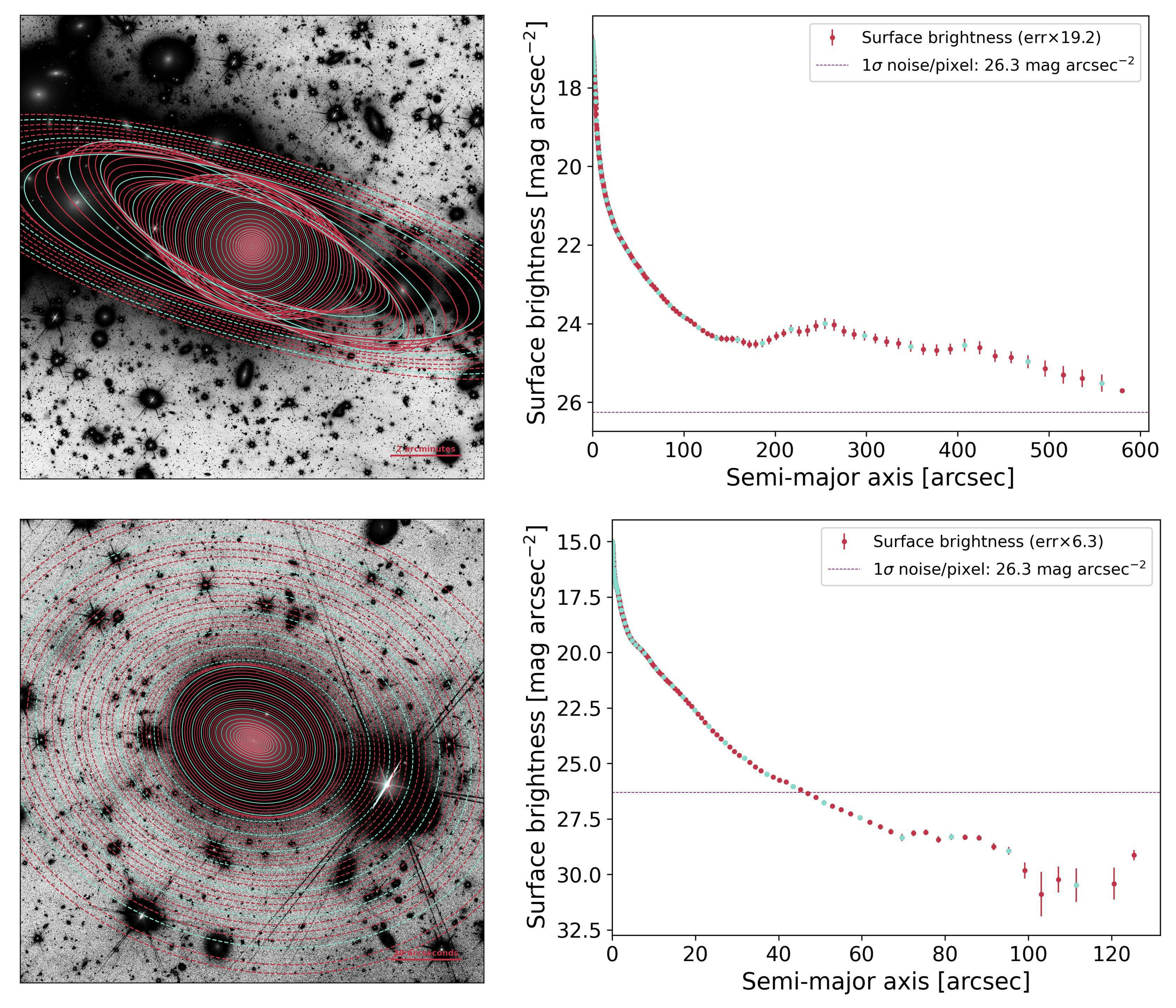}
\caption{Examples of \texttt{AutoProf} radial profile extraction on VIS data to derive photometric and morphological properties.
{\it Top left}: NGC\,1272 isophote fitting on an $8000\,\mathrm{pixel}\,{\times}\,8000\,\mathrm{pixel}$ pixel cutout ($\ang{;;800}\times\ang{;;800}$, red angular scale on the image: $\ang{;2;}$). 
{\it Top right}: Radial surface brightness profile for NGC\,1272, showing a dip at $r=\ang{;;180}$ due to overlap with NGC\,1275, marking the photometry's limit. In such complex cases, \texttt{AstroPhot} handles the photometric analysis.
{\it Bottom left}: NGC\,1281 isophote fitting on a $2000\,\mathrm{pixel}\,{\times}\,2000\,\mathrm{pixel}$ cutout ($\ang{;;200}\times\ang{;;200}$, red angular scale on the image: $\ang{;;30}$).
{\it Bottom right}: Radial surface brightness profile for NGC\,1281, reaching down to 30.1\,mag\,arcsec$^{-2}$ at $r=\ang{;;100}$ despite the proximity of a bright star. Cyan and red points on the radial profiles refer to the colours of isophotes in the associated {\tt AutoProf} visual, which help guide the eye when comparing plots.
}
\label{fig:AP}%
\end{figure*}

The size of the cutouts varied for all the dwarfs depending on their effective radius, as described in \citet{EROPerseusDGs}. For the bright galaxies, the cutouts also varied in size, ranging from $1000\times1000$, $2000\times2000$, $4000\times4000$, and $8000\times8000$ VIS pixels, depending on their angular scale on the sky, the tile size being optimised to facilitate accurate sky background estimation. Only the two largest galaxies in the cluster, NGC\,1272 and NGC\,1275, required a tile size of $8000\times8000$ VIS pixels, as illustrated in \cref{fig:AP}.

\begin{figure*}[htbp!]
\includegraphics[width=1.0\textwidth]{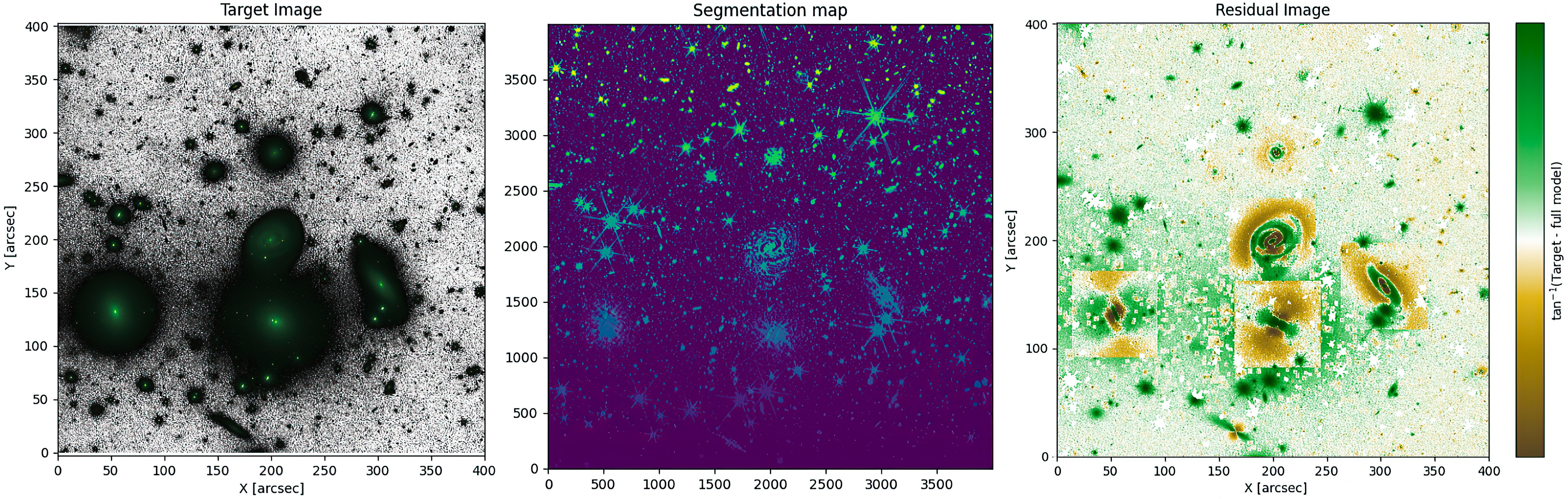}
\caption{Example of an \texttt{AstroPhot} fitting process on VIS data to derive photometric and morphological properties.
{\it Left}: Input image ($4000\,\mathrm{pixel}\,{\times}\,4000\,\mathrm{pixel}$ pixel cutout, $\ang{;;400}\times\ang{;;400}$) featuring multiple overlapping galaxies from the Perseus cluster, namely NGC\,1270, NGC\,1268, NGC\,1267, and CGCG540-089.
{\it Centre}: Segmentation map of the image by \texttt{SourceExtractor} (scale in pixels).
{\it Right}: Residual image (image minus models) for the initial fit, this gets substantially improved upon iterating.}
\label{fig:APh}%
\end{figure*}

For the dwarf galaxies, the galaxy modelling was initially performed using \texttt{AutoProf} \citep{ConnorAP} to obtain a first estimate of the structural parameters. These initial estimates were then used as input to both \texttt{Galfit} \citep{peng2010} and \texttt{AstroPhot} \citep{2023MNRAS.525.6377S}. The details of the dwarf galaxy modelling procedure and results are presented in \citet{EROPerseusDGs}.

For the bright galaxy sample, after initially selecting galaxies based on their apparent size, a second sorting was conducted based on the proximity of the galaxies to each other. This proximity significantly influenced the choice of photometric extraction tool. \texttt{AutoProf} was utilised for relatively isolated galaxies, where their outermost isophotes do not significantly overlap with those of neighbouring galaxies. Conversely, \texttt{AstroPhot} is preferred for galaxies that are in close proximity to each other.

Once the tiles were extracted, the photometry of isolated galaxies was determined using the \texttt{AutoProf} tool. To minimise contamination from \ac{MW} stars in the VIS photometry calculations, a star mask was created from each segmentation map generated by \texttt{SourceExtractor}. \texttt{AutoProf} then took this mask along with the galaxy image from the corresponding tile as inputs. \texttt{AutoProf} processed each galaxy to provide a detailed surface brightness profile, after fitting isophotes leading to an azimutally averaged profile averaging the pixels along the isophotes. Output products included characteristic fit parameters such as sky background estimation, position angle, and mean ellipticity, among others. \Cref{fig:AP} displays examples of profiles generated by \texttt{AutoProf}. Additionally, a S\'ersic model fit to the 1D profile was performed to derive the S\'ersic index. Other photometric parameters extracted include total magnitude and effective radius, providing a comprehensive picture of each galaxy.

The remaining galaxies found in a crowded environment, i.e., 34 of the 138 bright galaxies, were analysed using the \texttt{AstroPhot} tool, which provides a range of modelling options. For these galaxies, simple S\'ersic profiles were initially fitted. Tiles of 4000\,pixels\,$\times$\,4000\,pixels were preferred, to ensure ample sky background for accurate photometry estimation in these complex areas. To initialise \texttt{AstroPhot}, star masking was conducted using the same approach as with \texttt{AutoProf}, employing segmentation maps and parameters from \texttt{SourceExtractor}, such as ellipticity, position angle, and estimated flux for each galaxy. S\'ersic models were then initialised over $100\times100$ pixel windows, typically encompassing the galaxy's core. The process began with a first iteration using an iterative method to refine the model. This was followed by a second iteration that utilised the parameters from the first iteration but expanded the analysis to a $1000\times1000$ pixel window. Ultimately, a S\'ersic profile was fitted for each galaxy, detailing associated morphological and photometric parameters, as depicted in \cref{fig:APh}. The results of the photometric analysis for bright galaxies are described in the tables of the \cref{appendix:tables}.

\begin{figure*}
\centering
\includegraphics[width=0.85\textwidth]{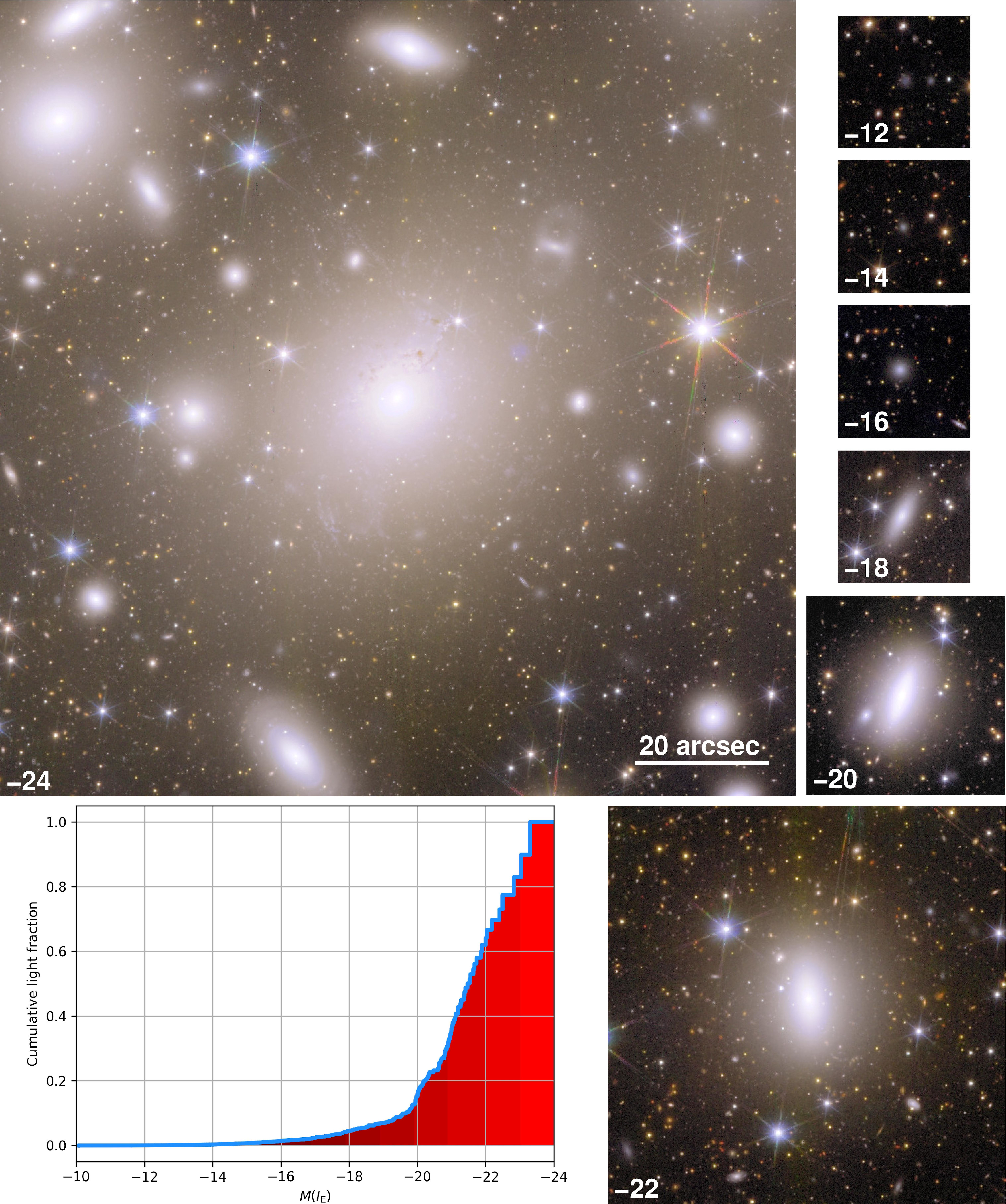}
\caption{\textit{Bottom left}: Cumulative light distribution (blue line) for the Perseus cluster catalogue of 1220 galaxies, which reaches down to $M(\IE)=-10$. The absolute magnitude of the galaxy at the centre of each image cutout is indicated in the \IE band. While the 1083 dwarf galaxies dominate by number, 90\,\% of the light in Perseus comes from the 83 galaxies brighter than $M(\IE)=-19.5$ (the remaining 48 galaxies plus all dwarf galaxies account for just 10\,\% of the light in Perseus). This is demonstrated with examples of relative brightness and physical size, all presented at the same pixel scale. The range spans from the giant elliptical NGC\,1275 (top left), which alone contributes 10\,\% of the light in Perseus, to a dwarf 12 magnitudes fainter (top right). The colour for each magnitude bin matches the colour used to represent galaxies on the map of \cref{fig:PerseusCatalogMap}. All dwarf galaxies are fainter than $M(\IE)=-18$, while all other galaxies are brighter than $M(\IE)=-16$.}
\label{fig:PerseusZoo}%
\end{figure*}

\begin{figure*}
\centering
\includegraphics[width=0.9\textwidth]{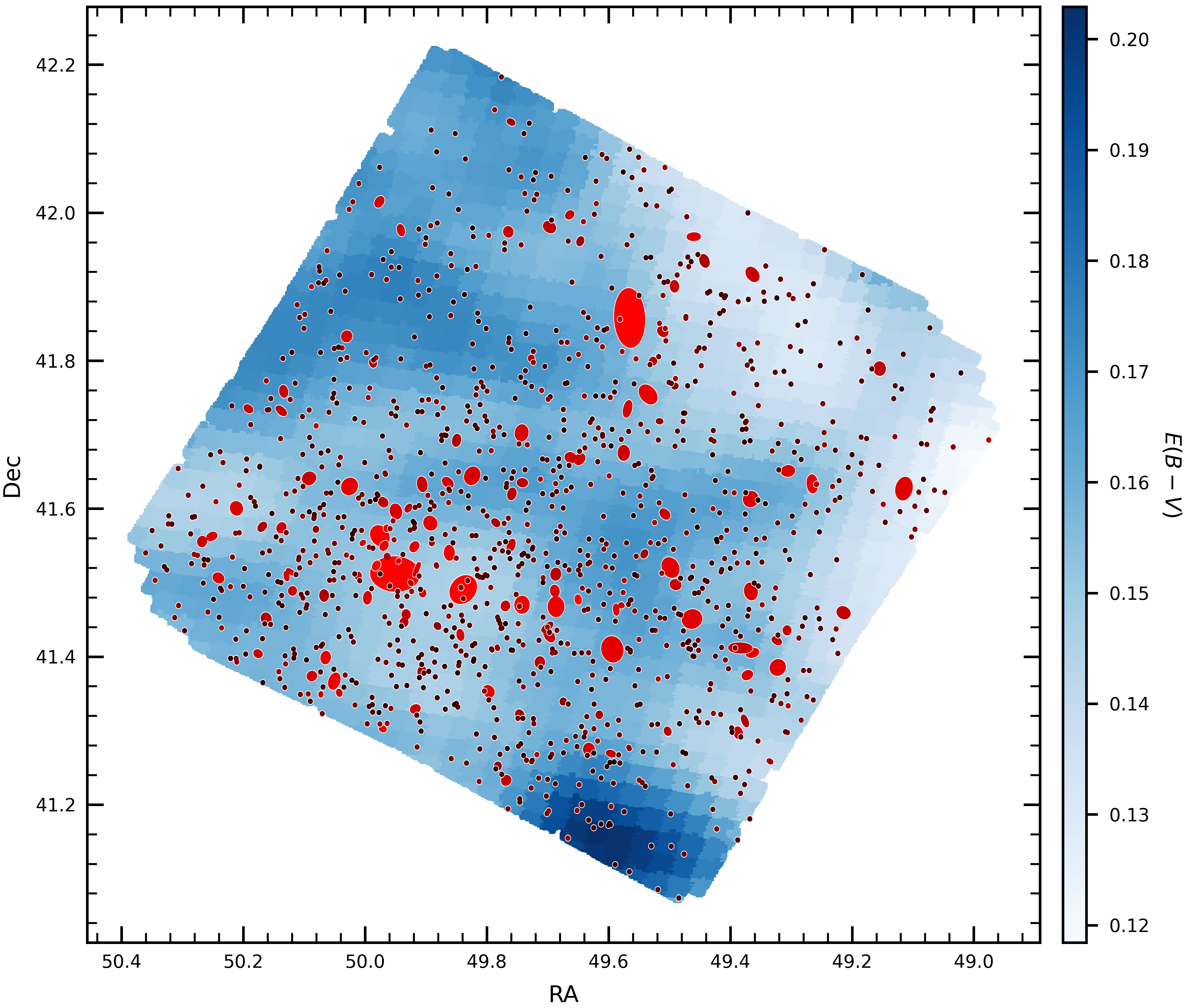}
\caption{The Perseus catalogue of 1220 galaxies plotted over the $0.7\, {\rm deg}^2$ sky coverage, alongside the \textit{Planck} dust map [average $E(B-V)= 0.156$]. The colour map for the galaxies is scaled from red to black according to the absolute magnitude ranging from $M(\IE)=-24$ to $-10$ (\cref{fig:PerseusZoo}). Measured ellipticity and angular size are depicted for the bright galaxies, while all 1083 dwarfs are represented as round dots, exaggerated in size relative to their physical dimensions. The three largest ellipses are NGC\,1275 and NGC\,1272 on the lower left, and NGC\,1265 at the top.}
\label{fig:PerseusCatalogMap}%
\end{figure*}

The two tools, \texttt{AutoProf} and \texttt{AstroPhot}, while demonstrating equivalent effectiveness for isolated galaxies, complement each other well within the highly diverse environment of the Perseus cluster. Together, they deliver quality photometry for all types of galaxy. \Cref{fig:PerseusZoo} illustrates the diversity of galaxy properties in the catalog, with a cumulative light distribution showing that 90\,\% of the light from the cluster originates from the top 7\,\% of the most brilliant galaxies. \cref{fig:PerseusCatalogMap} is a map displaying the distribution of light across the cluster, superimposed on the adopted \textit{Planck} dust map [average $E(B-V)= 0.156$].

\subsection{\label{sc:Photoz}Photometric redshifts}

\subsubsection{Photometric redshifts from SDSS}

\begin{figure*}
\centering
\subcaptionbox*{}[.48\linewidth]{%
\includegraphics[width=\linewidth]{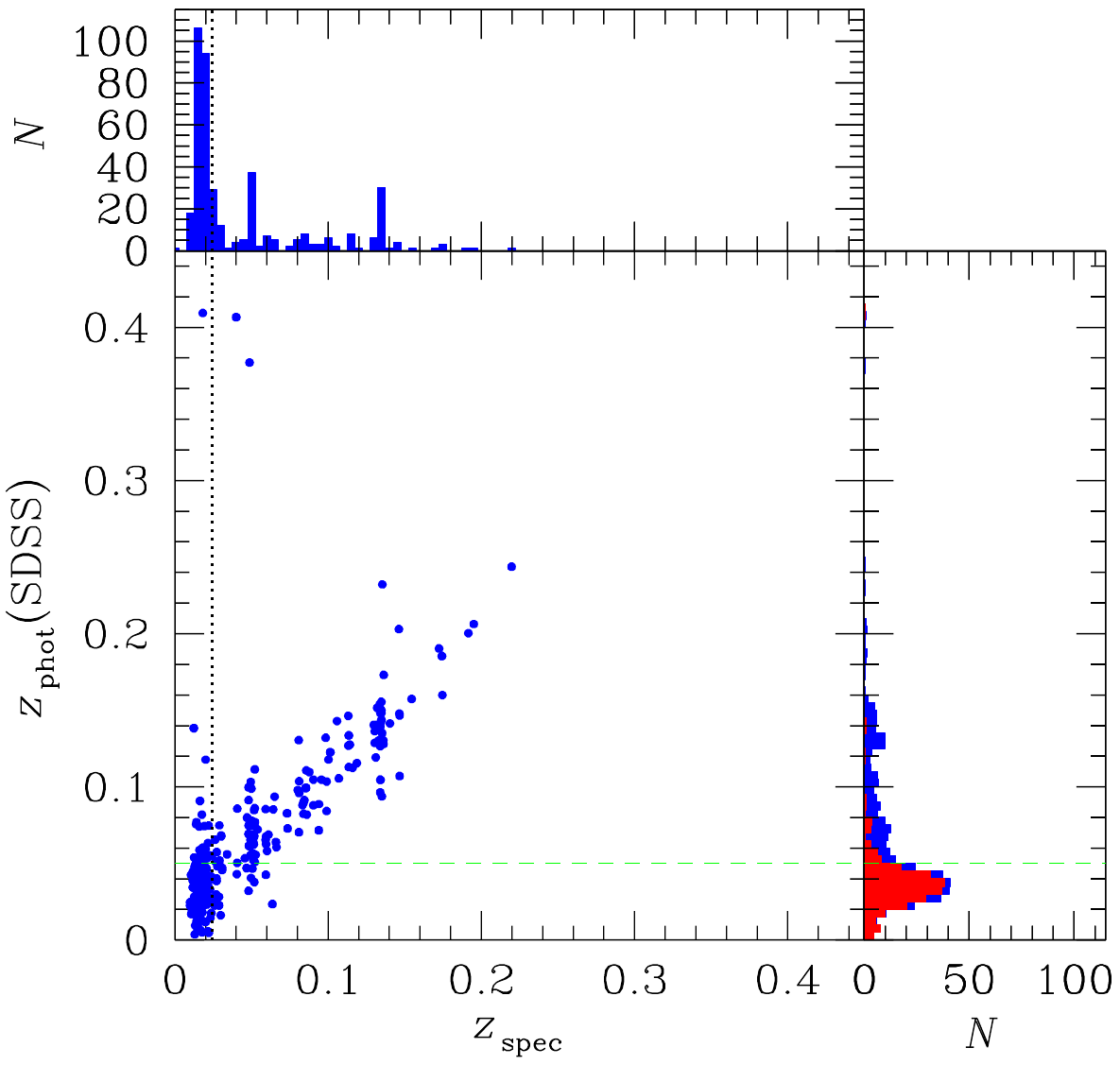}
}
\hfill
\subcaptionbox*{}[.48\linewidth]{%
\includegraphics[width=\linewidth]{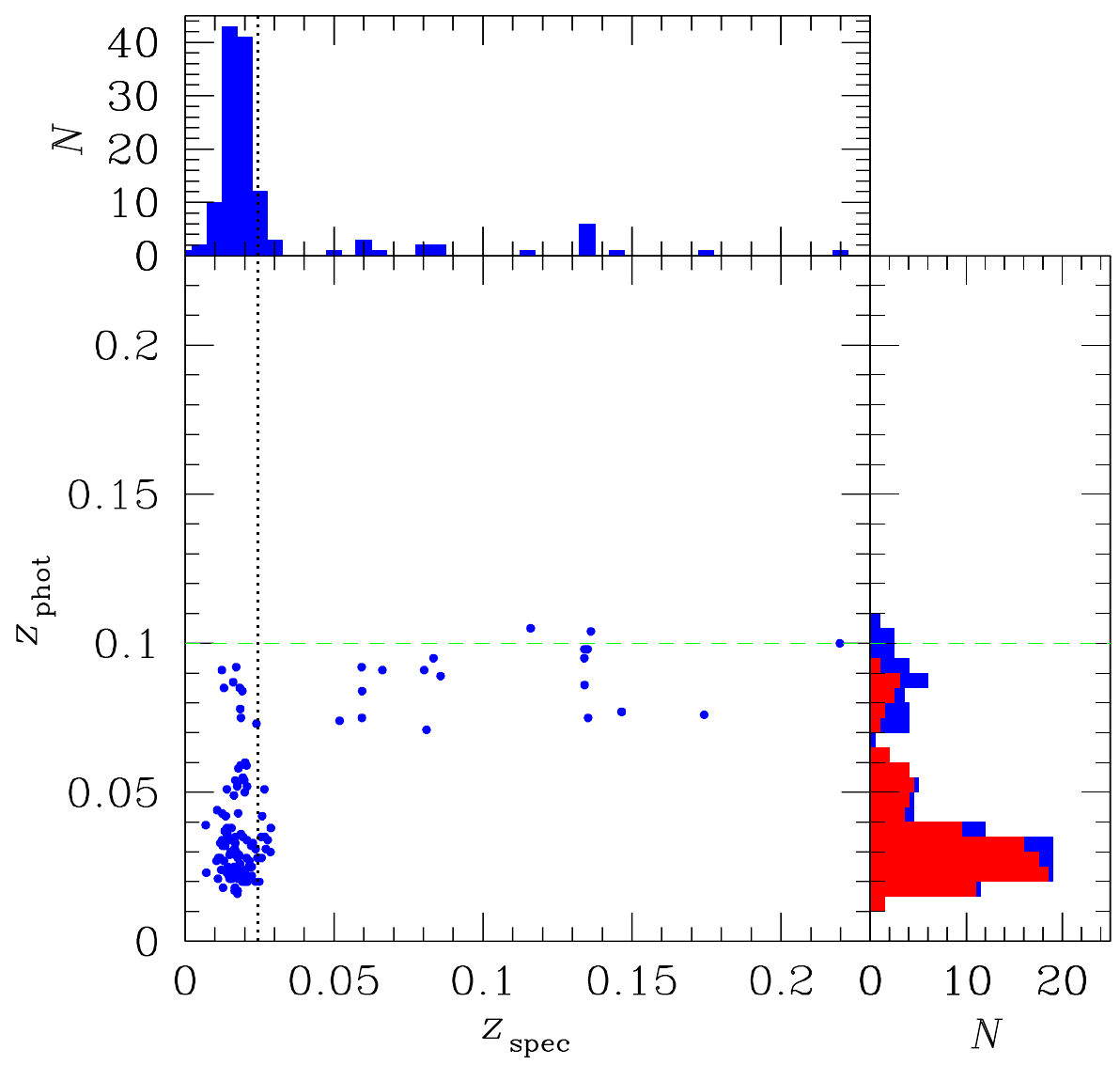}
}
\vspace{-0.4cm}
\caption{{\it Left}: Comparison between the \ac{SDSS} photometric redshift and the spectroscopic one for targets located within $R \leq r_{200}$ in the Perseus cluster with an $r$-band magnitude $r \lesssim 17.7$. 
{\it Right}: Comparison between the photometric redshifts derived with \texttt{Phosphoros} and the spectroscopic ones for targets located within the \Euclid field of the Perseus cluster ($R \leq 0.25 r_{200}$). 
The vertical black dotted line shows the adopted limit in spectroscopic redshift used for identifying Perseus members [$z_{{\rm spec}}\lesssim 0.02446$, corresponding to $V_\mathrm{c} + 2\,\sigma_\mathrm{c} = (5258 + 2\times1040)\, {\rm km\, s}^{-1}$]. The green dashed line shows the limit in photometric redshift to identify potential cluster candidates [$z_{{\rm phot}}({\rm SDSS})\lesssim 0.05$, left; $z_{\rm phot}\lesssim 0.1$, right]. The upper and right histograms show the spectroscopic and photometric redshift distribution, respectively, for all galaxies and for those objects classified as cluster members using their spectroscopic redshift ($z_{{\rm spec}}\lesssim 0.02446$, red).}
\label{phot_zSDSS}%
\label{zphotPhosphoros}%
\end{figure*}

\ac{SDSS} photometric redshifts are available for galaxies with $r$-band magnitudes brighter than $r\lesssim21$ \citep{2003AJ....125..580C,2007AN....328..852C}. Being trained on a large sample of galaxies mainly located in the local Universe, these photometric redshifts are optimised to identify local systems with excellent accuracy.\footnote{\url{https://www.sdss4.org/dr17/algorithms/photo-z/}} This is particularly true for galaxies with magnitudes $r\lesssim 17.7$, the redshift completeness limit of \ac{SDSS}. \Cref{phot_zSDSS} shows the relationship between the \ac{SDSS} photometric redshifts and the spectroscopic redshifts for all galaxies with available spectra within the Perseus cluster ($R<r_{200}$). It shows that a cut in $z_{{\rm phot}}({\rm SDSS})\lesssim 0.05$ secures an accurate identification of cluster members (87\,\%\ of those spectroscopically identified) with a minor contamination (11\,\%) when cluster members are defined as those with a redshift $z_{\mathrm{spec}}\lesssim 0.02446$ corresponding to $V_\mathrm{c}+2\,\sigma_\mathrm{c}$.\footnote{A cut in the spectroscopic redshift at $z_{\mathrm{spec}}\lesssim 0.02446$ is appropriate given that a group of galaxies has been observed in projection close to Perseus at a mean velocity of approximately $9000\,{\rm km\, s}^{-1}$ \citep{2000AJ....119.1638A}.} 

We can thus conclude that this selection on the \ac{SDSS} photometric redshift $z_{\mathrm{phot}}(\mathrm{SDSS})\lesssim 0.05$ for galaxies with $r\lesssim 17.7$ leads to a purity of 87\,\%\ and a completeness of 89\,\%. We can thus select all sources within the footprint of the \Euclid observations with $z_{{\rm phot}}({\rm SDSS})\lesssim 0.05$ and $r\lesssim 17.7$ not yet included in the main catalogue. After ignoring all stars and objects already included in the main catalogue, we identify three potential cluster member candidates with the SDSS photometric redshift cut. We visually inspected the deep \ac{VIS} images of these three objects: one is severely contaminated by the presence of a bright star close to its centre, which prevents the derivation of any structural and photometric parameters. The other two have morphological properties consistent with those observed in other cluster members. We thus included these two objects in the main catalogue and derived their structural parameters consistently with the other cluster members.

\subsubsection{\label{sc:ZP} Full-sample photometric redshifts}

\begin{figure*}
\centering
\includegraphics[width=0.8\textwidth]{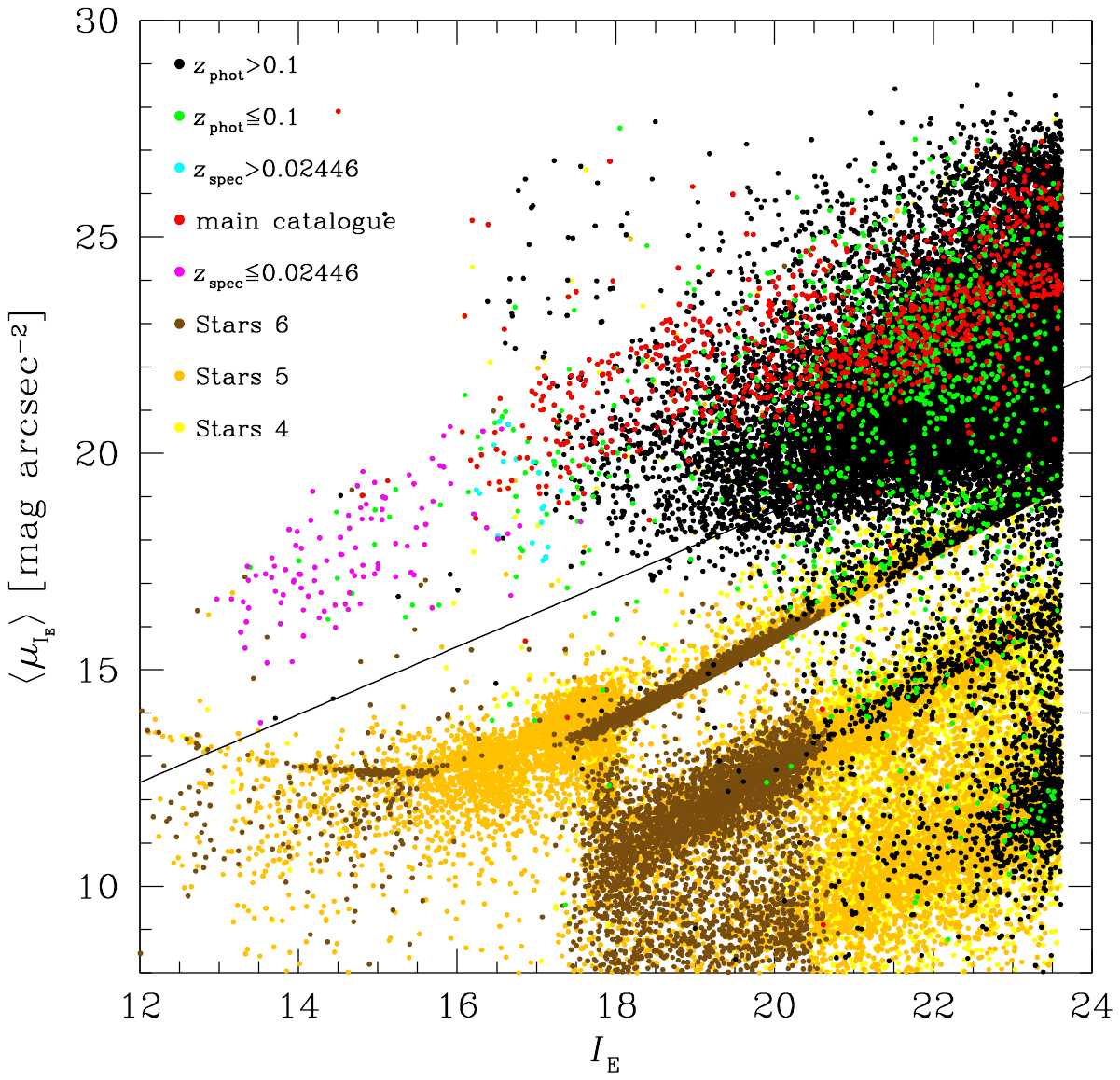}
\vspace{-4.cm}

\caption{Relation between the mean versus model effective surface brightness and the total magnitude for all sources in the full \texttt{SourceExtractor} catalogue with $\IE\lesssim 23.6$. Brown, orange, and yellow dots are for targets classified as stars consistently by six, five, and four independent criteria, respectively. Black and green dots are for galaxies with photometric redshifts from \texttt{Phosphoros} $z_{\rm phot}> 0.1$ and $\lesssim 0.1$,
respectively, cyan dots for spectroscopically confirmed background galaxies ($z_{{\rm spec}}>0.02446$), red dots for optically identified cluster members, and magenta dots for spectroscopically confirmed cluster members ($z_{{\rm spec}}\lesssim0.02446$). The black solid line ($\langle\mu_{I_{\rm E}}\rangle = 0.783\,\IE + 3\,{\rm mag\,arcsec}^{-2}$) delimits the region where most (17/1018, 1.7\,\%) of the spectroscopically and visually identified members are located.}
\label{fig:GUVICS_SB}%
\end{figure*}

The photometric parameters of all the emitting sources have been extracted using \texttt{SourceExtractor} on the compact-source stacks. 
This extraction pipeline has been run in dual mode on all the \ac{CFHT} and \Euclid \ac{NISP} images, while in an independent mode on the \Euclid \ac{VIS} image, producing a different number of detected sources, as mentioned in \cref{sc:Catalogues}.

The photometric redshifts of all sources identified as galaxies in \cref{sc:Catalogues} have been derived using the \ac{SED} fitting code \texttt{Phosphoros}\footnote{\url{https://phosphoros.readthedocs.io/}} (\citealp{2020A&A...644A..31E}), and are denoted as $z_{\rm phot}$ hereafter. We used the \ac{CFHT} broad bands and \ac{NISP} for which all the parameters have been derived within the same Kron aperture determined by \texttt{SourceExtractor} in the $\chi^2$ image (see \cref{sc:Catalogues}), while the \ac{VIS} catalogue has been derived separately, so as not to lose its superior resolution. For this reason the VIS photometry has not been used to derive photometric redshifts. 

Photometric redshifts have been derived by comparing the observed photometry values with that derived from the templates used in COSMOS \citep{2009ApJ...690.1236I}, suitable for high-redshift galaxies that need to be separated from the potential Perseus members. 
In template \ac{SED} fitting, a grid of model photometry is generated for plausible amounts of the internal dust attenuation and the redshift, with finer redshift steps (three equally-spaced binning schemes in three redshift ranges) at very low redshift.
However, photometric redshifts derived from template fitting techniques inherently face challenges at low redshifts. The key features used to estimate the distance of a galaxy from broad-band photometry are the slope of the \ac{SED} and spectral breaks, notably the $4000$\,\AA\ break. While this break can be bracketed by the $u$ and $g$ filters out to a redshift $z\simeq 0.35$, degeneracy in the properties regulating the slope of the blue part of the \ac{SED} result in poor constraints at low redshift with this combination of broad-band filters.

Promising improvements have been obtained for bright and resolved objects when using machine-learning techniques complementing the flux measurements with the images \citep{2024MNRAS.527..651T}. Nonetheless template-fitting methods are still required for faint galaxies for which no training sample is available. 

A compromise between the need to remove high-$z$ sources without rejecting potential members has been achieved with the use of a prior. 
The luminosity function prior, implemented in a fully Bayesian manner in \texttt{Phosphoros}, adjusts the likelihood when a degenerate probability distribution function is encountered based on the probability of a galaxy having a given luminosity at the photometric redshift. For our prior, we adopted the \ac{LF} in the $B$ band derived by \cite{2005ApJ...622..116G} at a redshift of approximately $1.2$ (chosen to avoid overly constraining the bright local galaxies).

The robustness of these photometric redshifts is tested here by comparing their values to those of the spectroscopic redshifts, where available, within the observed field (130 objects, see \cref{zphotPhosphoros}).\footnote{The dynamic range in redshift sampled in right panel of \cref{zphotPhosphoros} is limited to $z \leq 0.23$ where spectroscopic redshifts of galaxies are available in the literature.}
Despite the very limited number of objects (the \ac{ERO} Perseus field is the first deep field ever observed within the \Euclid photometric bands), \cref{zphotPhosphoros} shows that a selection of $z_{\rm phot}\lesssim 0.1$ is appropriate to include all galaxies identified as Perseus members, thanks to their spectroscopic redshift.
With this cut, the contamination of background galaxies is $27/130$ ($\simeq 21\,\%$). This number drops to 17/130 ($\simeq 13\,\%$) with a more stringent limit on spectroscopic redshift ($z_{\rm spec} \lesssim 0.03$) for cluster membership.

\subsection{\label{sc:Stellarmass}Stellar masses}

Similarly to photometric redshifts, stellar masses $\cal{M}$\footnote{We use that symbol in this paper to distinguish stellar masses from absolute magnitudes.} have been derived with the \ac{SED}-fitting technique, using the code \texttt{hyperz} \citep{2000A&A...363..476B,2010A&A...524A..76B}.
To derive the physical properties of galaxies fitting their optical and \ac{NIR} emission, the templates need to be derived from stellar population synthesis models \citep[e.g.,][]{1997A&A...326..950F,2003MNRAS.344.1000B,2005MNRAS.362..799M,2010ApJ...712..833C}. In the present work we have used the composite stellar populations derived from \cite{2003MNRAS.344.1000B} models revised in 2016,\footnote{\url{https://www.bruzual.org/bc03/Updated_version_2016/}} assuming the initial mass function of \cite{2003PASP..115..763C}. We adopted different star-formation histories (exponentially declining and delayed) with different timescales and two stellar metallicities (Solar and subsolar) and let the code chose the best-fit age, provided that it is smaller than the age of the Universe at the given redshift. For the members of the Perseus cluster we have fixed the redshift at $z=0.0167$ and determined the best-fit template.  

\begin{figure*}[htbp!]
\centering
\subcaptionbox*{}[.48\linewidth]{%
\includegraphics[width=\linewidth]{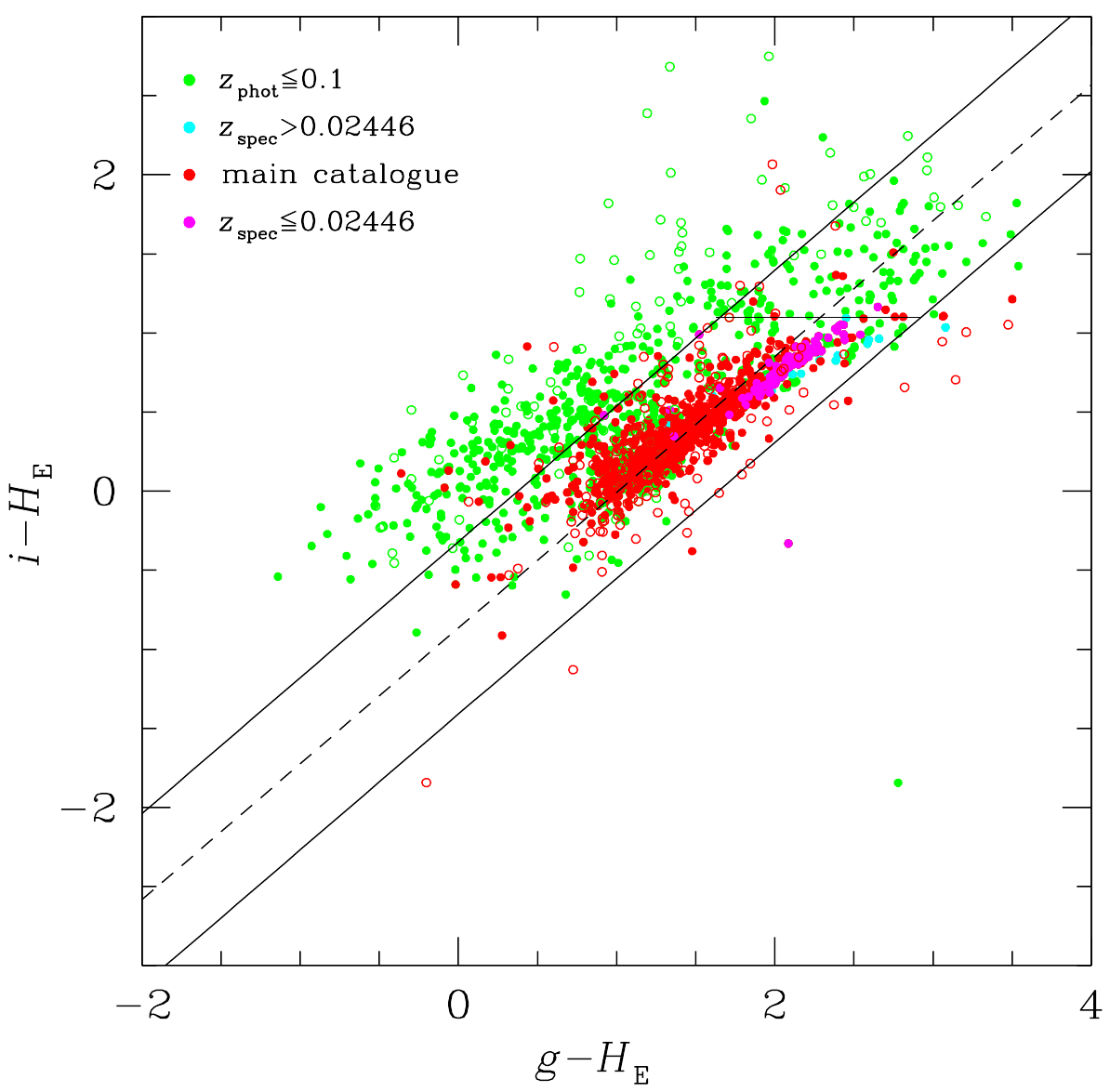}
}
\hfill
\subcaptionbox*{}[.48\linewidth]{%
\includegraphics[width=\linewidth]{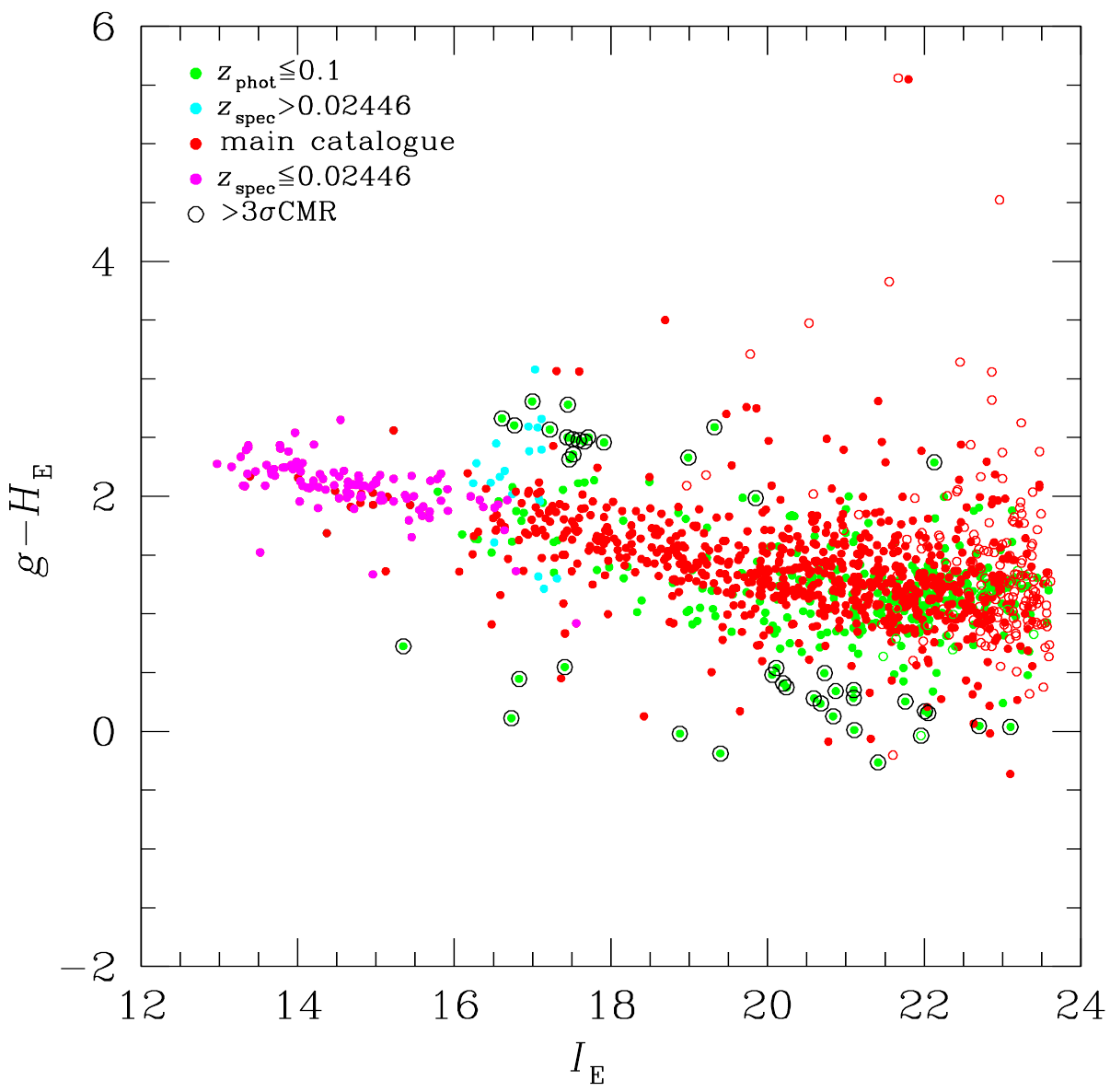}
}
\vspace{-0.4cm}
\caption{{\it Left:} $i-\HE$ versus $g-\HE$ colour-colour diagram for all sources in the full \texttt{SourceExtractor} catalogue with $\IE\lesssim 23.6$ identified as galaxies in \cref{fig:GUVICS_SB} with $z_{\rm phot} \lesssim 0.1$ (green dots), spectroscopically confirmed background galaxies ($z_{{\rm spec}}>0.02446$; cyan dots), and optically identified (red dots) or spectroscopically confirmed (magenta dots) cluster members ($z_{{\rm spec}}\lesssim0.02446$). Filled dots are for galaxies with uncertainty in the colour larger than 0.1\,mag. The black dashed line shows the bisector fit to the data for galaxies with a colour uncertainty $\sigma_{{\rm col}}\lesssim 0.1$, while the solid lines delimit the region within the 3\,$\sigma$ dispersion of the relation, where 99.7\,\%\ of the spectroscopically and visually identified members are located. The black solid horizontal line shows the upper limit in colour observed in the spectroscopically confirmed cluster members ($i-\HE < 1.1$).
{\it Right:} $i-\HE$ versus $\IE$ colour-magnitude relation for the same sample of galaxies but limited to those objects identified as potential cluster members in the $i-\HE$ versus $g-\HE$ colour-colour diagram (left panel). Galaxies are identified using the same symbols, with the exception of green dots within black circles, which are systems lying $>3\,\sigma$ from the mean colour-magnitude relation, where 99.7\,\%\ of the spectroscopically and visually identified cluster members are located.  
}
\label{fig:GUVICS_colori_2CFHTgH}%
\label{fig:GUVICS_coloriMAGgH}%
\end{figure*}

The stellar mass, defined as in equation~2 of \cite{2009MNRAS.394..774L}, is determined for each galaxy by the best-fit model and its normalisation. It is therefore important to have an estimate of the integrated colours for each galaxy, since they are related to the stellar population, and they can largely vary from the inner regions to the outskirts, as seen from the spatially resolved analysis (see \cref{sc:SpatiallyResolved}). Moreover, the total stellar mass can only be recovered from the total flux. 
The photometry derived from the compact-sources stacks, despite being robust for the estimate of colours, does not match the requirement of representing a good fraction of the stellar light, especially for large galaxies. 
To estimate the total stellar masses we used instead the photometry derived from the ring-filtered images, described in \cref{appendix:Completeness}, with the flux measured in larger apertures. The {\tt MAG\_AUTO} values in the different bands as measured in double image mode and using the ring-filtered \IE image provides a better representation of the total flux, but it can still lack part of the flux for the brightest galaxies and not be accurate in the presence of close bright objects. We therefore rescaled the stellar mass output from \texttt{hyperz} with the \IE magnitude derived by \texttt{AutoProf}, i.e., $\logten ({\cal M}/M_\odot) = \logten ({\cal M^{\rm bf}}/M_\odot) - 0.4\,(\IE^{\rm bf} -\IE^{\rm AutoProf})$, where 
${\cal M^{\rm bf}}$ is the stellar mass derived from the SED fitting applied to MAG\_AUTO magnitudes, $\IE^{\rm bf}$ is the magnitude derived from the best-fit template, and \IE is the total magnitude derived from the \texttt{AutoProf} measured profile. 

Deriving the completeness in stellar mass is not straightforward. For extragalactic surveys characterised by a flux limit, the smallest stellar mass used to derive the \ac{SMF} with negligible incompleteness is obtained by rescaling the stellar masses at the limiting magnitudes and checking their distributions in redshift bins \citep{2010A&A...523A..13P}. 
This approach is not feasible in this case; instead we propagated the information on the incompleteness function derived in \cref{appendix:Completeness}, taking into account that there is a scatter introduced by the different mass-to-light ratios. The median stellar mass at which we reach 90\,\% completeness ($M(\IE) \simeq -14$) is $\logten ({\cal M}/M_\odot) \simeq 7 $, and in \cref{sc:SMF} we will use this limit. 

A typical uncertainty in stellar mass estimate is of the order of $0.2$\,dex \citep{2007A&A...474..443P, 2009ApJ...696..348W, 2009ApJ...701.1765M, 2012MNRAS.422.3285P, Roediger2015,2015ApJ...808..101M}, depending on the assumptions on the \ac{SED} fitting method, the star-formation history, and the model of stellar population synthesis. Besides these uncertainties, the statistical $1\,\sigma$ error we derived from the \ac{SED} fitting on Perseus galaxies is of the order of $0.05$\,dex \citep[consistent with e.g. ][]{Roediger2015}. 
In \cref{sc:SMF} we will therefore adopt a binning larger than $0.25$\,dex to ensure our \ac{SMF} does not depend on the scatter in the stellar mass estimate.

\subsection{\label{sc:ScalingRelations}Refining cluster membership through scaling relations}

\subsubsection{Scaling relations for the full catalogue}

Following \citet{2016A&A...585A...2B} we use the output parameters of \texttt{SourceExtractor} to construct several scaling relations to identify all possible cluster members. Given that the completeness test shown in \cref{appendix:Completeness} has indicated that the main catalogue is approximately 50\,\% complete at $\IE \simeq 23.6$ (including a mean Galactic attenuation of $A_{I_{\rm E}}\simeq 0.33$, see \cref{{sc:Catalogues}}), we limit the analysis to all detected sources satisfying this magnitude limit. These are 33\,083 galaxies and 34\,649 stars. \Cref{fig:GUVICS_SB} shows the relation between the mean model effective brightness ({\tt MU\_MEAN\_MODEL}, representing the mean effective surface brightness above the background from the model-fitting photometry) and the proxy for the total magnitude ({\tt MAG\_AUTO} in the versus) both corrected for dust attenuation using the extinction law from \cite{2023ApJ...950...86G} and the \textit{Planck} 2013 dust map \citep{2014A&A...571A..11P}, as described in \citet{EROPerseusDGs}.

As already noticed in \citet{2016A&A...585A...2B}, this scaling relation well segregates stars from galaxies. 228/34\,649 (0.6\,\%) of the sources identified as stars using four, five, or six independent methods (as described in \cref{sc:Catalogues}) are located below the black solid line in \cref{fig:GUVICS_SB}, defined as $\ave{\mu_{I_{\rm E}}} = 0.783\,\IE + 3\,{\rm mag\,arcsec}^{-2}$, while 1001/1018 of the cluster members (spectroscopically confirmed or visually identified) are above it (98.3\% completeness). It is thus likely that the large majority of the sources with $z_{\rm phot}$ greater than $0.1$ (black dots) and smaller than or equal to $0.1$ (green dots) located below this line are misclassified stars that can be removed. We thus adopt this cut to exclude all possible stars from the following analysis.

Again following \citet{2016A&A...585A...2B} we construct other diagnostic diagrams to identify background galaxies from those located within the Perseus cluster. 
\Cref{fig:GUVICS_colori_2CFHTgH} shows the $i-\HE$ versus $g-\HE$ colour-colour diagram for the galaxies satisfying the previous criteria (i.e., with $z_{\rm phot}\lesssim 0.1$ and identified as galaxies in \cref{fig:GUVICS_SB}). The figure compares the distribution of these objects to that of the galaxies of the main catalogue and of spectroscopically confirmed background objects. 

The spectroscopically and visually identified cluster members are well located along a very tight relation, while the galaxies with a photometric redshift $z_{\rm phot} \lesssim 0.1$ follow the same relation but with a substantially larger scatter, with on average redder $i-\HE$ colours than cluster members. Their distribution within this diagram suggests that some of them are background galaxies, and can thus be removed from the following analysis. We do so by removing all objects located outside $\pm3\,\sigma$ of the best-fit relation. We also exclude all objects with $i -\HE > 1.1$, since all the massive and spectroscopically confirmed galaxies of the cluster have colours bluer than this value. This choice is justified by the fact that possible members are likely low-luminosity metal-poor objects, thus their colour is bluer than that of massive cluster members. The resulting sample of potential members missing from the main catalogue drops to 350 candidates.

Finally, as is common in the literature, we use a colour-magnitude relation \citep[e.g.,][for Perseus]{1992MNRAS.254..589B,2020MNRAS.494.1681A}, $g-\HE$ versus \IE\ in this case, to further reject background galaxies from the sample of potential members (\cref{fig:GUVICS_coloriMAGgH}). Most spectroscopically or visually identified cluster members are located along a tight colour-magnitude relation, as expected for a galaxy population highly dominated by quiescent early-type systems. Those located 3\,$\sigma$ above the relation (redder colours, 17 objects) are either galaxies with uncertain colours as given by \texttt{SourceExtractor}, which is optimised to derive photometric parameters for point sources, or background galaxies, as also suggested by the distribution of the spectroscopically confirmed background objects. Those located below the relation (25) are mainly background objects or star-forming systems, rare but still present in massive clusters of galaxies \citep[e.g.,][]{2010A&A...517A..73G,2014A&A...570A..69B}. Indeed, visual inspection of those already identified as cluster members show a spiral/irregular morphology typical of star-forming systems, some of which also host star-forming regions evident in the narrow-band H$\alpha$ image (see next section). After removing the objects located above the colour-magnitude relation, the sample of potential members not included in the main catalogue drops to 333 objects, 25 of which are located $>3\,\sigma$ below the relation traced by quiescent, early-type systems.

\subsubsection{H\texorpdfstring{$\alpha$}{Halpha} excess}

Galaxies located more than 3\,$\sigma$ below the colour-magnitude relation traced by quiescent early-type systems (25 objects) are either blue background galaxies or star-forming systems within the Perseus cluster. 
In this second case, their ongoing star-formation activity would produce ionising radiation detectable in the Balmer H$\alpha$ emission line at $6563\,$\AA. 
If present, this line emission can be detected in the narrow-band H$\alpha$ image of the cluster, which has been taken within a filter covering the velocity range $4660\,{\rm km\, s}^{-1} \lesssim v_{\mathrm{hel}} \lesssim 9600\,{\rm km\, s}^{-1}$ thus perfectly matching that of the cluster. 
We visually inspected the continuum-subtracted images of these objects, and found that only one of the 25 objects has detectable H$\alpha$ emission. 
This source, however, is discarded because it is associated with the perturbed gas of the galaxy UGC\,2665, which is already included in the main catalogue of spectroscopically confirmed objects.

\subsubsection{Visual inspection}

\begin{figure}
\centering
\includegraphics[width=0.49\textwidth]{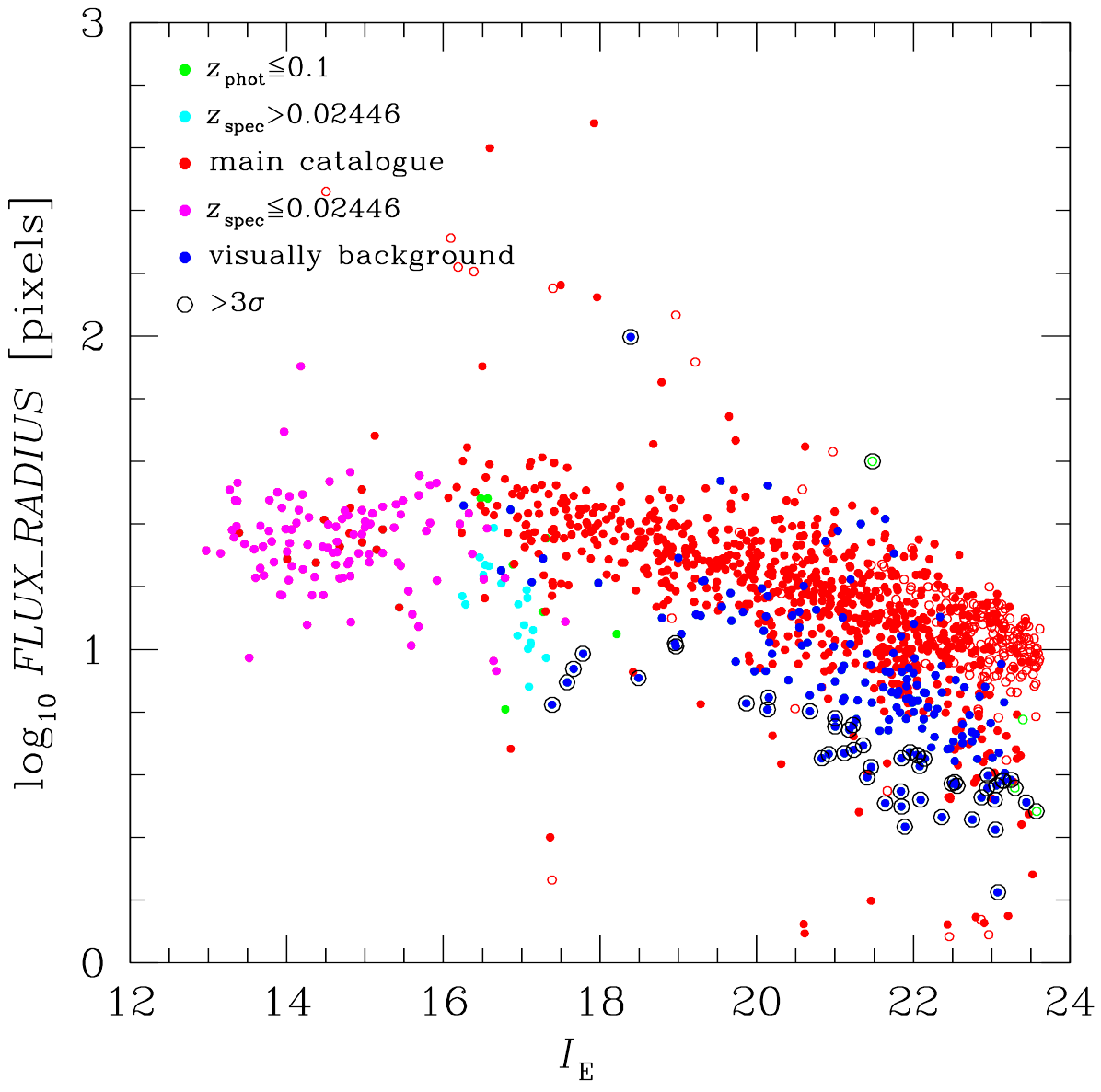}
\caption{Relation between the FLUX\_RADIUS and the \IE-band magnitude for all sources in the \texttt{SourceExtractor} catalogue with $\IE\lesssim 23.6$ identified as galaxies that satisfy the surface brightness-magnitude selection (\cref{fig:GUVICS_SB}), the colour-colour selection, and the colour-magnitude selection
(\cref{fig:GUVICS_coloriMAGgH}, left and right panel) for potential cluster members. Filled and empty dots are for galaxies with uncertainty in the $g-\HE$ colour smaller or larger than 0.1, respectively. Magenta dots are for spectroscopically confirmed cluster members ($z_{{\rm spec}}\lesssim 0.02446$), red dots for optically identified members, cyan dots for spectroscopically confirmed background galaxies ($z_{{\rm spec}}>0.02446$), and green dots potential members with $z_{\rm phot}\lesssim 0.1$ after visual inspection of their \ac{VIS} images. Blue dots are faint objects of small size considered as background systems.}
\label{fig:Radiusmag}%
\end{figure}

We then visually inspected all the remaining objects selected using the \ac{VIS} \IE-band images. Out of these, 85 are misclassified stars, seven are bright objects at the edge of the field where the image of the galaxy is truncated, 37 are problematic sources in low signal-to-noise ratio regions at the periphery of the frame where only a few exposures are available, or sources close to very bright stars where the photometry cannot be accurately measured. One source is another part of the fragmented ram-pressure-stripped galaxy UGC\,2665, four are different parts of two independent merging, star-forming systems. Since they do not have any associated emission in the H$\alpha$ continuum-subtracted image, these must be  background systems. All 129 of these sources can be removed from the following analysis. Of the remaining 180, seven are relatively bright galaxies that might have been missed during the first visual inspection of the \Euclid image for the definition of the bright galaxy sample mentioned in \cref{sc:Identification}. The remaining 173 are faint objects of small size when compared to those of the main cluster catalogue, as depicted in \cref{fig:Radiusmag}. If we remove those located at more than 3\,$\sigma$ from the scaling relation drawn by the cluster members in the main catalogue, only 123 galaxies remain as potential candidates.

\begin{figure*}
\centering
\includegraphics[width=0.8\textwidth]{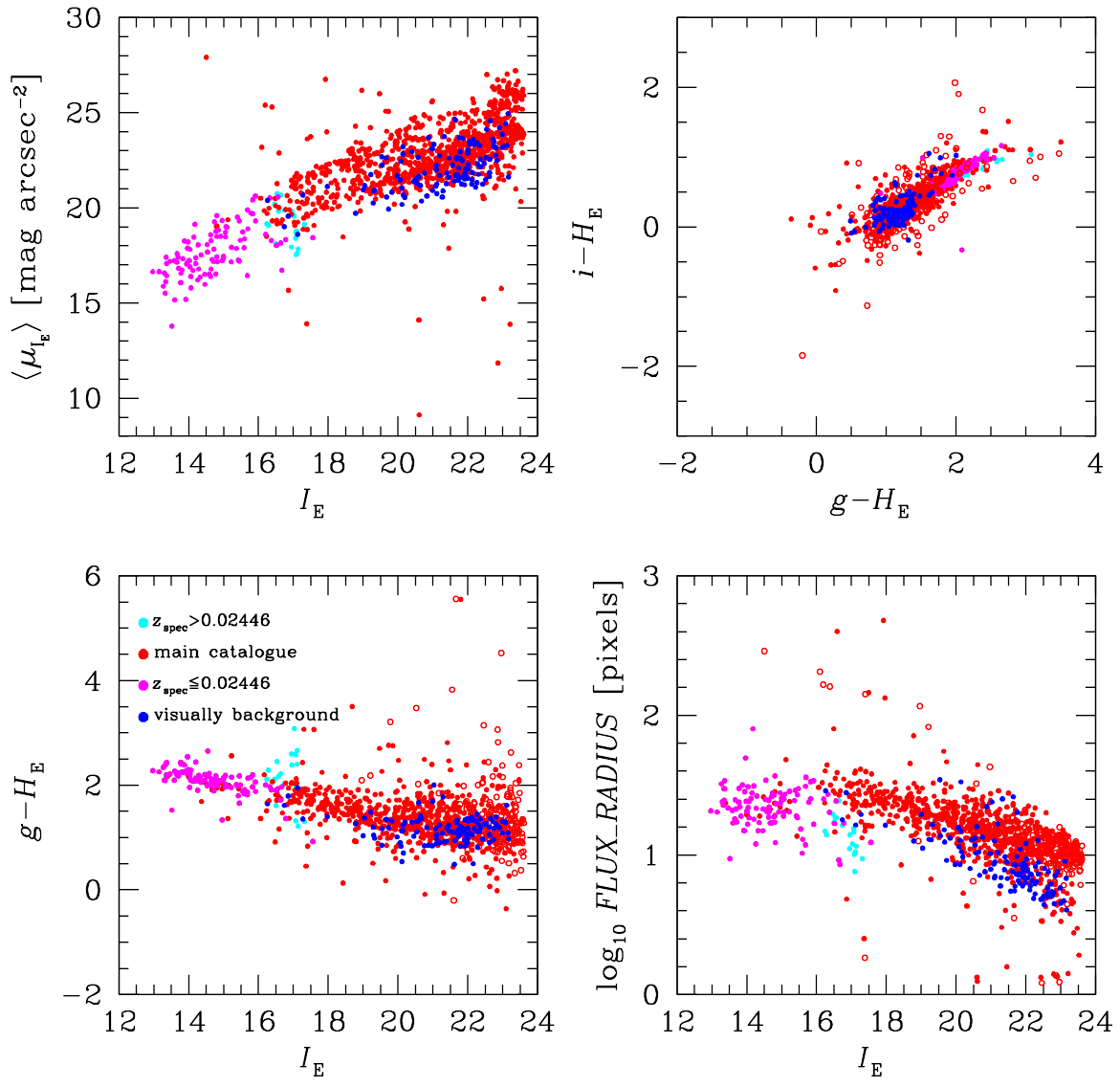}
\caption{Distribution of spectroscopically confirmed cluster members ($z_{{\rm spec}}\lesssim  0.02446$, magenta dots), optically identified cluster members (red dots), spectroscopically confirmed background objects ($z_{{\rm spec}}>0.02446$, cyan dots), and of the 123 possible members considered as likely background systems (blue dots) in the four scaling relations (surface brightness versus \IE, {\it upper left panel}; $i-\HE$ versus $g-\HE$, {\it upper right}; 
$g-\HE$ versus \IE, {\it lower left}; FLUX\_RADIUS versus \IE, {\it lower right}) used for their identification.}
\label{summary}%
\end{figure*}

\Cref{summary} shows how the possible cluster members identified using the surface brightness versus \IE, $i-\HE$ versus $g-\HE$, $g-\HE$ versus \IE, and FLUX\_RADIUS (the estimate of the half-flux radius as derived by \texttt{SourceExtractor}) versus \IE scaling relations are distributed within these relations in comparison to spectroscopically confirmed or optically identified members. 
They perfectly match the distribution of other cluster members in the photometric scaling relations ($i-\HE$ versus $g-\HE$, $g-\HE$ versus \IE) as expected, given that they have all been selected to have a photometric redshift below 0.1. 
However, their distribution along the two structural scaling relations (surface brightness versus \IE, FLUX\_RADIUS versus \IE) is systematically different from the other cluster members. 
On average, they have higher surface brightness and a smaller size than cluster dwarf galaxies, as is expected for background objects. 
Visual inspection shows that the majority of them are tiny objects, with no obvious resolved structures or the presence of any globular clusters. 
We recall that the purity of the photometric redshift selection adopted at the beginning of their identification, which reduced the original sample down to 1419 sources, is of order 21\,\%. 
Our criteria adopted so far have been tuned to reject background sources and we are left with these 123 potential candidates that could be true cluster members. 
We thus consider them in the following analysis as background systems, but discuss how their inclusion within the sample of Perseus members would affect the determination of the \ac{LF} and \ac{SMF}.

\subsubsection{Scaling relations for the cluster members}

\begin{figure*}
\centering
\includegraphics[width=0.80\textwidth]{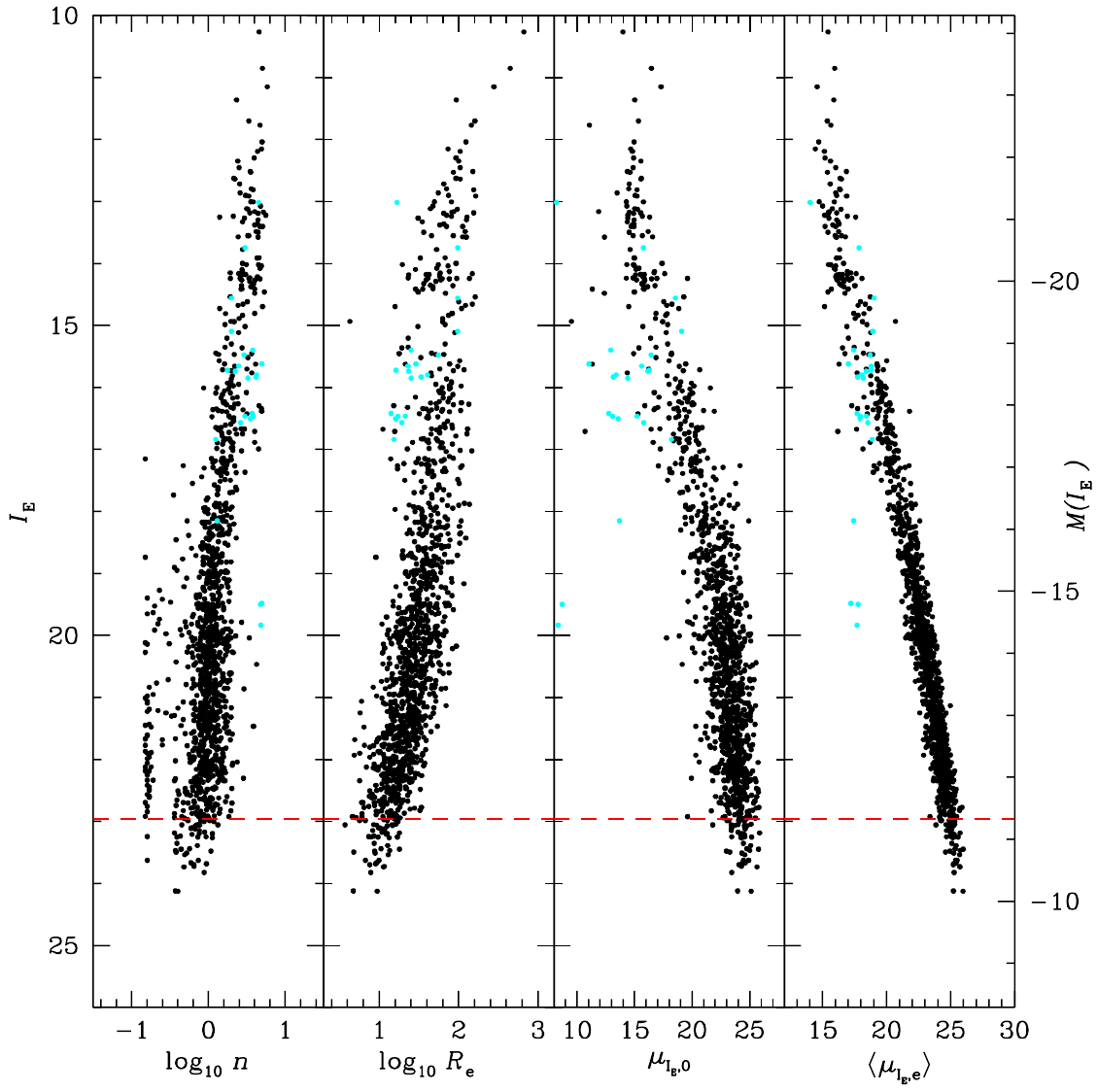}
\caption{Scaling relations between the S\'ersic index $n$, the effective radius $R_{\rm e}$, the central surface brightness $\mu_{I_{\rm E}, 0}$, and mean effective surface brightness within $R_{\rm e}$, $\langle\mu_{I_{\rm E} , {\rm e}}\rangle$, and the total magnitude measured using \texttt{AutoProf}/\texttt{AstroPhot}. Magnitudes are corrected for Galactic attenuation. Black dots are for cluster member galaxies, cyan dots for spectroscopically confirmed background objects.
The red dashed line shows the 50\,\%\ completeness of the sample, corresponding to $M(\IE) = -11.33$, assuming a mean Galactic attenuation $A_{I_{\rm E}} = 0.33$. The effective radius is in pixel units and surface brightness in mag\,arcsec$^{-2}$.
}
\label{Ferrarese2avg}%
\end{figure*}

To further test the purity of the selected sample of cluster members, we determine some scaling relations but using a more accurate measure of  different structural parameters derived for the main catalogue of galaxies by means of \texttt{AutoProf}, \texttt{AstroPhot}, and \texttt{Galfit} \citep[Mondelin et al. 2024, in prep.]{EROPerseusDGs}. Compared to those coming from \texttt{SourceExtractor}, where the different photometric parameters are derived automatically, here the ad-hoc extraction minimises the uncertainties, thus reducing any systematic scatter in the scaling relations. 
\Cref{Ferrarese2avg} shows some typical scaling relations between several structural parameters (S\'ersic index $n$, effective radius, central surface brightness, and mean effective surface brightness) as a function of the total magnitude. Surface brightness and magnitudes are corrected for Galactic dust attenuation as described in \citet{EROPerseusDGs}. Such scaling relations have been successfully used to identify cluster membership in Virgo by \citet{2020ApJ...890..128F}.

\Cref{Ferrarese2avg} shows that all galaxies identified as cluster members are well distributed along tight relations. The distribution of the few spectroscopically confirmed background galaxies (23 objects) is generally more scattered than that of cluster members, in particular in the $r_{\rm e}$ (effective radius) versus \IE and $\mu_0$ (central surface brightness) versus \IE (total magnitude) relations. Although spectroscopically confirmed background sources have only a limited range of magnitudes ($13\lesssim \IE \lesssim 20$), preventing the comparison on the whole dynamic range of the relations, \cref{Ferrarese2avg} consistently indicates that the galaxies included within the main catalogue are indeed all cluster members. This suggests that the possible contamination of background galaxies in the selected catalogue, if present, is minimal and thus will not strongly affect the estimation of the LF.

\subsection{The final catalogue}

To conclude, after the different selection criteria adopted to identify cluster members we end up with 1172 galaxies of $M(\IE)\lesssim -11$ observed magnitude, or equivalently $M(\IE)\lesssim -11.33$ when corrected for Galactic attenuation, above a completeness limit of approximately 50\,\%. Indeed, in this final step, 48 dwarf galaxies have corrected magnitudes $M(\IE)$ greater than $-11.33$. This is the sample of galaxies used to derive the \ac{LF}. To summarise, Perseus cluster members have been selected using the following criteria: 1) morphology (main catalogue, from \citealp{Marleau2024} and Mondelin et al., in prep.), or 2) SDSS photometric redshift $z_{{\rm SDSS}}\leq 0.05$. 
A strict selection on the full catalogue of SExtractor detections with magnitude brighter than this limit for sources satisfying the following criteria: 3) \texttt{Phosphoros} photometric redshift $z_{\rm phot} \leq 0.1$; 
4) $\ave{\mu_{I_{\rm E}}}  \leq 0.783\,\IE + 3\,{\rm mag\,arcsec}^{-2}$; 5) within $3\,\sigma$ of the $i-\HE$ vs. $g-\HE$, $g-\HE$ versus \IE, and
$\logten {\rm FLUX\_{RADIUS}}$ vs. \IE relation.
would add only 123 possible cluster candidates, a number roughly comparable to the possible contamination of background objects selected using these standard relations. We thus did not included them in the following analysis.
We recall that the completeness of the main catalogue described in \cref{sc:MainCatalog}, on which most of the following analysis is based, is $\sim$ 90\,\% (see \cref{appendix:Completeness}). The one of possible Perseus members derived from the full catalogue using photometric redshifts and scaling relations (\cref{sc:Photoz}) is of the order of $\sim$ 87\,\% , 89\,\%\ photometric redshifts, 98.3\,\%\ the mean effective surface brightness versus total mag relation (\cref{fig:GUVICS_SB}), 
99.7\%\ in the colour-colour, colour-magnitude, and FLUX\_RADIUS vs. magnitude relations (\cref{fig:GUVICS_coloriMAGgH,fig:Radiusmag}). 
We also stress that the current selection, which is principally based on morphology, does not allow to discriminate cluster members from foreground objects or from systems in the very close background from the cluster ($v_{{\rm hel}}\lesssim 12\,000\, {\rm km\, s}^{-1}$). 
The possible contamination of these objects, however, is very marginal and can be estimated by integrating the $i$-band \ac{LF} of the field derived at $z = 0.1$ by \citet{2003ApJ...592..819B} using SDSS data in the range $-25 \leq M_i\leq -11 $ and multiplying it by the sampled volume up to a distance of 170\,{\rm Mpc} ($\simeq 350\,{\rm Mpc}^3$, 15 objects in total).


\section{\label{sc:LF}The luminosity function}
\begin{figure*}
\centering
\includegraphics[width=0.8\textwidth]{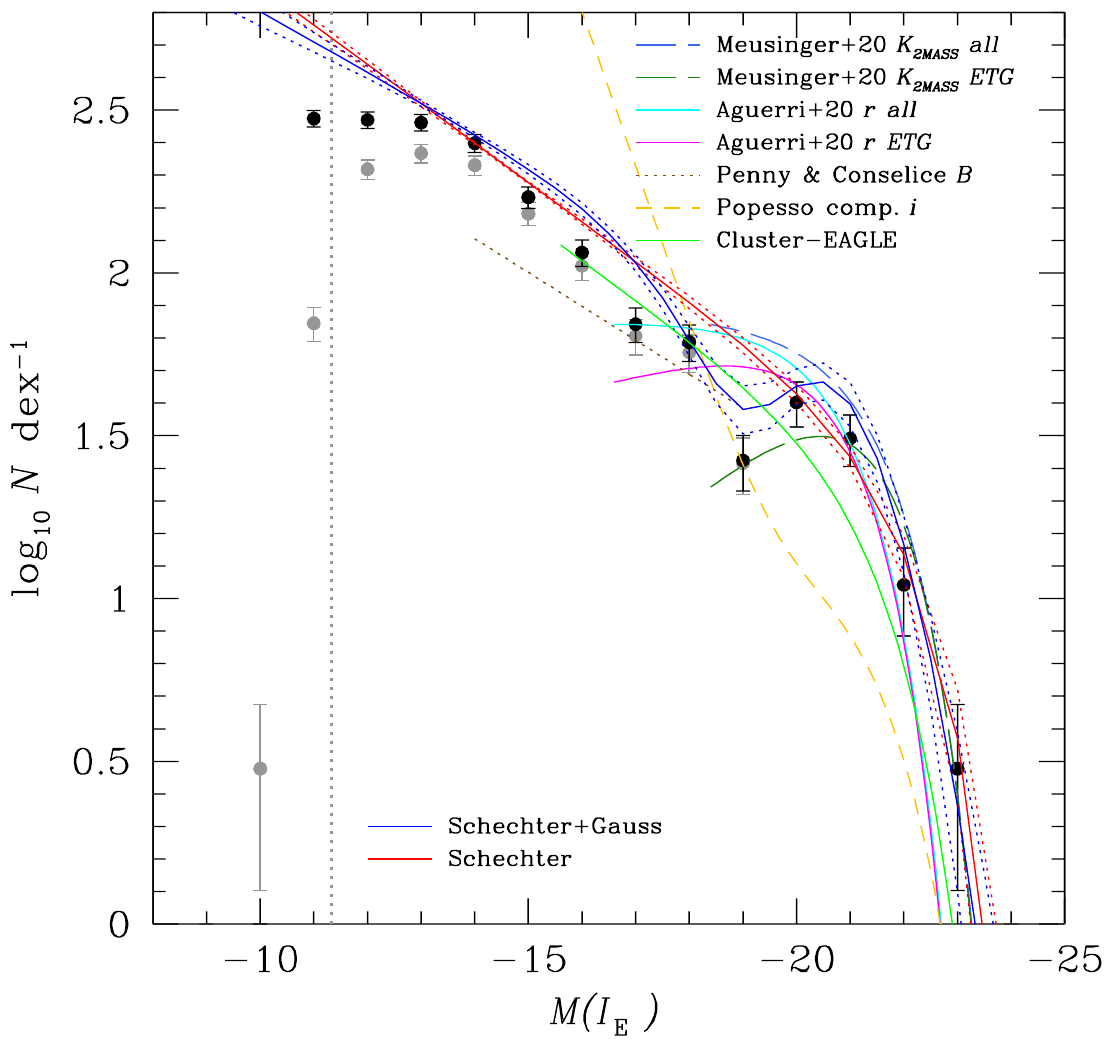}
\caption{Observed (grey dots) and corrected for completeness (black dots) \IE-band \ac{LF} derived excluding the three cD galaxies NGC\,1275, NGC\,1272, and NGC\,1265. The solid red and blue lines and the associated dotted lines show the best-fit and 1$\sigma$ confidence regions of the Schechter and Schechter $+$ Gaussian \ac{LF} parametrisations, respectively. The vertical grey dashed line shows the 50\,\%\ completeness limit. The two \ac{LF}s are compared to those available in the literature for the Perseus cluster in the $K$ band \citep{2020A&A...640A..30M}, in the $r$ band \citep{2020MNRAS.494.1681A}, and in the $B$ band \citep{2008MNRAS.383..247P}, to the mean $i$-band \ac{LF} of \ac{SDSS} local clusters derived by \citet{2005A&A...433..415P} within $0.5\,{\rm h^{-1}\,Mpc}$ from the cluster centre, and to the Cluster-EAGLE simulations \citep{2022MNRAS.515.2121N}. The \ac{LF}s of Perseus have been derived on different fractions of the cluster. They have been shifted to the \IE band using the mean colour of galaxies in the cluster, and in the $Y$-axis direction to visually match the data. The best fit parameters are given in \cref{table:TabLF}.}
\label{LF}%
\end{figure*}

For presentation purposes, we derive a binned \ac{LF} of the cluster by counting the number of objects per 1.0\,dex bin of \IE-band luminosity (corrected for Galactic extinction) and we correct it for completeness by dividing these numbers using the correction function shown in  \cref{fig:Completeness}. The two \ac{LF}s of the Perseus cluster, that have been measured within $R\lesssim 0.25\, r_{200}$, are shown in \cref{LF}. 

The \IE-band \ac{LF} shows three clear regimes: a Gaussian distribution at the bright end with a clear dip at $M(\IE)\simeq -19$; a steep rise out to $M(\IE)\simeq -14$; and a flattening down to the completeness limit of $M(\IE) \simeq -11$, with an abrupt decrease at fainter luminosity that is clearly due to incompleteness. 
The dip at $M(\IE)\simeq -19$ seems real, since the number of objects in that magnitude bin (19) would need to be approximately double to reach those measured in the adjacent bins, which is impossible given that the survey is 100\,\%\ complete at these luminosities. Indeed, the dip is 4.6\,$\sigma$ below the best-fit single Schechter function (see below), where $\sigma$ is the Poisson uncertainty (for a critical discussion on the possible relation between incompleteness and the presence of the dip see \citealp{2006MNRAS.372...60A}). The flattening observed at $M(\IE)\simeq -14 $ might be partly due to incompleteness; however, the analysis presented in the previous sections suggests that the observed trend is real. 
Indeed, to keep the same slope observed in the $ -19 \lesssim M(\IE)\lesssim -14$ range ($\alpha\simeq -1.5$, see below), the two following bins [$M(\IE)=-13$ and $-12$] should be missing about 400 objects, which is very unlikely. 
The statistical analysis presented in \cref{sc:Identification} identified 123 potential cluster candidates that the selection might have missed, which includes the fainter magnitude bin $M(\IE) \simeq -11$.

We fit the \ac{LF} using a combination of parametric functions, specifically a  \citet{1976ApJ...203..297S} function for faint objects and a Gaussian to reproduce the distribution of luminous galaxies:
\begin{align}
\Phi(M) {\rm d} M = &\, \Phi^*_{\rm S}\,  \nonumber \\
\times & \Biggl\{ 0.4\,\ln (10)\,\,10^{-0.4\,(M-M^*_{\rm S})\, (\alpha_{\rm S}+1)} \times \exp{\left[-10^{-0.4\, (M-M^*_{\rm S})}\right]} \nonumber \\
& + A_{\rm G} \,\exp{\brackets{-0.5\, \paren{\frac{M-M^*_{\rm G}}{\sigma_{\rm G}}}^2}} \Biggr\}\, {\rm d} M\, , 
\label{EqLF}
\end{align}
where $\Phi^*_{\rm S}$, $M^*_{\rm S}$, $\alpha_{\rm S}$ are the normalisation, the characteristic absolute magnitude and the faint-end slope of the Schechter function, while $A_{\rm G}$, $M^*_{\rm G}$, and $\sigma_{\rm G}$ are the normalisation relative to the Schechter, the mean absolute magnitude and the width of the Gaussian component.
To derive the parameters of \cref{EqLF} we adopted the Maximum Likelihood Estimator, firstly described in \citeauthor{1979ApJ...232..352S} (\citeyear{1979ApJ...232..352S}, see also \citealp{2005A&A...439..863I,2011A&ARv..19...41J}). 
No binning is used here, instead a log-likelihood is computed for each galaxy magnitude and summed to return a joint log-likelihood. Following the formalism of \citet{2015ApJ...811..141M} and \citet{2021MNRAS.503.3044F}, the fit is carried out by deriving the posterior distribution and the best-fit parameters using the \texttt{MULTINEST} Bayesian algorithm \citep{2008MNRAS.384..449F,2019OJAp....2E..10F}. \Cref{sc:statparams} expends on the robustness of the derived parameters and associated errors.
We limit the fit to galaxies above the completeness magnitude limit of $M(\IE) \lesssim -11.33$. The fit is performed using the full set of galaxies above this magnitude limit, or excluding the central cD (NGC\,1275) and/or the two other cDs (NGC\,1272 and NGC\,1265), as often done in the literature \citep[e.g.,][]{2004ApJ...610..745L,2020A&A...640A..30M}. For comparison with previous results in the literature we also fit the data using a single Schechter function. The best fit to the data are shown in  \cref{LF}, and the best-fit parameters are given in \cref{table:TabLF}. The $\Phi^*$ parameters give characteristic numbers of objects of the fitted function within the surveyed region. The accurate volume of this region is hardly quantifiable given the complex structure of the cluster. Nevertheless, assuming a spherical distribution of size $r_{200} = 2.2\, \mathrm{Mpc}$ (see \cref{sc:Perseus}), and considering that the surveyed region is of approximately $0.7\, {\rm deg}^2$, we can estimate that the corresponding volume is $V \simeq 4.33\, {\rm Mpc}^3$.

\begin{table*}[htbp!]
\caption{Best-fit parameters of the \ac{LF} .}
\begin{center}
\label{table:TabLF}
\begin{tabular}{lcccccc}
\hline\hline
\noalign{\vskip 2pt}
Sample  & $\Phi^*_{\rm S}$           & $M^*_{\rm S}$            & $\alpha_{\rm S}$              & $A_{\rm G}$                   & $M^*_{\rm G}$                & $\sigma_{\rm G}$\\
\noalign{\vskip 2pt}
\hline
\noalign{\vskip 2pt}
All galaxies& $134.63^{+33.18}_{-27.14}$ & $-17.95^{+0.44}_{-0.47}$& $-1.22^{+0.04}_{-0.03}$  &$0.33^{+0.09}_{-0.06}$ & $-20.04^{+0.05}_{-0.37}$   & $1.45^{+0.27}_{-0.23}$\\
\noalign{\vskip 2pt}
\hline
\noalign{\vskip 2pt}
Excluding NGC\,1275& $125.88^{+30.34}_{-25.46}$  & $-18.15^{+0.44}_{-0.44}$& $-1.22^{+0.03}_{-0.03}$  & $0.36^{+0.10}_{-0.07}$ & $-20.22^{+0.43}_{-0.31}$   &  $1.28^{+0.25}_{-0.20}$\\
\noalign{\vskip 2pt}
\hline
\noalign{\vskip 2pt}
Excluding NGC\,1275, NGC\,1272, NGC\,1265& $114.75^{+27.43}_{-24.17}$ &  $-18.38^{+0.41}_{-0.46}$&  $-1.23^{+0.03}_{-0.03}$  &   $0.42^{+0.12}_{-0.09}$ &  $-20.39^{+0.33}_{-0.22}$   &  $1.06^{+0.22}_{-0.15}$\\
\noalign{\vskip 2pt}
\hline
\noalign{\vskip 2pt}
Excluding NGC\,1275, NGC\,1272, NGC\,1265& $26.18^{+4.20}_{-3.72}$ &  $-22.41^{+0.25}_{-0.28}$&  $-1.30^{+0.01}_{-0.02}$  &  --  &  --  &  -- \\
\noalign{\vskip 2pt}
\hline
\end{tabular}
Notes: the parameters correspond the 16, 50, and 84\% percentiles of the distribution. $\Phi^*_{\rm S}$ is in units of N dex$^{-1}$.
\end{center}
\end{table*}


\section{\label{sc:SMF}The stellar mass function}

As described in \cref{sc:Stellarmass}, the available multifrequency data allow us to measure stellar masses for all the detected sources, and thus derive the \ac{SMF} of the cluster. 
The \ac{SMF} has a shape quite different to that of the \ac{LF}. It has a steep rise at the bright end down to ${\mathcal M} \simeq 10^{10}\,M_{\odot}$, then the number of objects moderately increases down to ${\mathcal M}\simeq 10^{7}\,M_{\odot}$ and rapidly decreases below this limit. In this range of stellar masses the increase of number of objects is not as smooth as observed in the \ac{LF}, but quite noisy: a dip might be present at ${\mathcal M}\simeq 10^{9.5-10}\,M_{\odot}$, and the slope of the distribution changes significantly at lower stellar mass.  We fit the  \ac{SMF} using a single Schechter function and show the non parametric function in bins of 0.4 dex. To avoid any complex completeness correction (as mentioned in \cref{sc:Stellarmass}), we fit the data down to stellar masses ${\mathcal M} = 10^7\, M_{\odot}$ (679 objects) where the completeness is near 100\,\%.

\begin{figure*}
\centering
\includegraphics[width=0.8\textwidth]{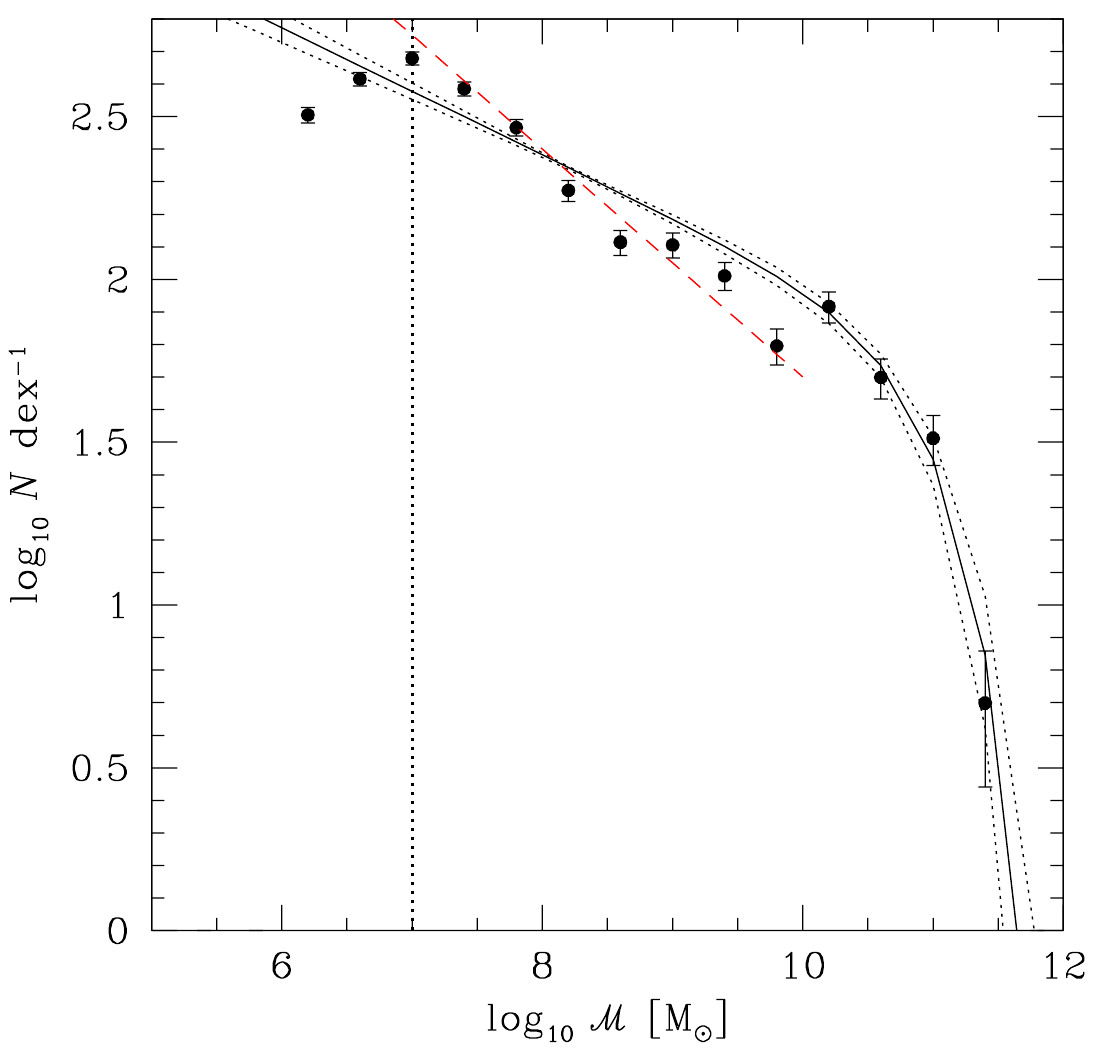}
\caption{\ac{SMF} derived excluding the three cD galaxies NGC\,1275, NGC\,1272, and NGC\,1265. The solid black and dotted black lines show the best-fit and $\pm 1\,\sigma$ confidence regions of the Schechter \ac{SMF} parametrisation. The red dashed line, which is a reasonably good fit to the slope of the faint end of the distribution, has $\alpha = -1.35$. The vertical grey dashed line shows the stellar mass limit used in the fit. The best-fit parameters are given in \cref{table:TabSMF}.}
\label{SMF}%
\end{figure*}

\begin{table*}[htbp!]
\caption{Best-fit parameters of the \ac{SMF}. }
\begin{center}
\label{table:TabSMF}
\begin{tabular}{lccc}
\hline\hline
\noalign{\vskip 1pt}
Sample  & $\Phi^*_{\rm S}$           & $\logten{\mathcal M^*_{\rm S}/M_{\odot}}$            & $\alpha_{\rm S}$  \\
\noalign{\vskip 2pt}
\hline
\noalign{\vskip 2pt}
All galaxies & $51.32^{+9.80}_{-8.35}$ & $11.28^{+0.13}_{-0.11}$ & $-1.21^{+0.02}_{-0.01}$ \\
Excluding NGC\,1275 & $54.66^{+10.25}_{-9.03}$ & $11.21^{+0.12}_{-0.11}$ & $-1.20^{+0.02}_{-0.02}$ \\
Excluding NGC\,1272, NGC\,1275, and NGC\,1265& $60.24^{+10.10}_{-9.62}$ & $11.10^{+0.12}_{-0.12}$&  $-1.19^{+0.02}_{-0.02}$   \\
\noalign{\vskip 2pt}
\hline
\end{tabular}\\
Notes: the parameters correspond the 16, 50, and 84\,\% percentiles of the distribution. $\Phi^*_{\rm S}$ is in units of N dex$^{-1}$.
\end{center}
\end{table*}


The Schechter parametrisation has a slope $\alpha_{\rm S} \simeq -1.2$ and flattens more rapidly than the data at ${\mathcal M} \simeq 10^7\,M_{\odot}$. In the stellar mass range $10^7 \lesssim {\mathcal M} \lesssim 10^{9.5}\,M_{\odot}$ the distribution has a slope close to $\alpha_{\rm S} = -1.35$.
Below this stellar mass limit, the observed distribution might be not complete. The inclusion of the massive cD galaxies NGC\,1272, NGC\,1275, and NGC\,1265, does not have any effect on the determination of the faint-end slope of the distribution. 


\section{\label{sc:Discussion}Discussion}

\subsection{\label{sc:Compact}Compact sources}

Studies of nearby clusters such as Virgo and Fornax indicate that, at the sensitivity of the \Euclid observations, compact sources such as \ac{BCDs} and \ac{UCDs} can be detected \citep[]{1985AJ.....90.1759S,2016A&A...585A...2B,2021MNRAS.504.3580S}. 
If reached, these particular populations of compact sources should thus be accounted for in the determination of the \ac{LF}. The methods described above used to identify members of the Perseus cluster, and in particular those based on the positions of galaxies along several scaling relations, have been mainly defined to identify extended, \ac{LSB} objects \citep[e.g.,][]{1998A&A...333...17B}. 
They might thus fail to recognise \ac{BCDs} and \ac{UCDs}, the former generally characterised by a relatively low luminosity but with a high surface brightness, which make them similar to background luminous star-forming systems, the latter by a point-like morphology similar to that of stars or globular clusters, which are generally poorly resolved in ground-based observations. 
To check whether these objects can be identified in the selection criteria mentioned in \cref{sc:Catalogues} we first derived the mean properties of approximately 40 \ac{BCDs} in the Virgo cluster given in the \ac{VGC} of \citet{1985AJ.....90.1681B}. 
Using the full compilation of photometric data analysed in \citet{2014A&A...570A..69B}, we estimated that the mean $B$-band isophotal diameter of \ac{BCDs} is $D_{25.5}(B) = \ang{;;40} \pm \ang{;;16}$, while the $H$-band effective radius $r_{\rm {e}}(H) = \ang{;;7.4}\pm \ang{;;4.1}$. Scaled to the distance of the Perseus cluster \citep[72\,Mpc versus 16.5\,Mpc for Virgo;][]{1999A&A...343...86G,2007ApJ...655..144M,2018ApJ...856..126C} they are $D_{25.5}(b) = \ang{;;10}\pm \ang{;;4}$ and $r_{\rm e}(H)= \ang{;;1.8} \pm \ang{;;1}$. \ac{BCDs} are thus fully resolved in the \Euclid data, even in the ring-filtered images which remove only star-like sources. They have $B$-, $H$-, and $K$-band absolute magnitudes $M_B\geq -17$ \citep{1985AJ.....90.1759S} and $M_H\geq -17.7 $ \citep{2000A&AS..142...73B}, as well as $M_K\geq -16.6$ \citep{1997A&AS..121..507B}, and have stellar masses ${\mathcal M} \lesssim 10^8\, M_{\odot}$ \citep{2014A&A...570A..69B}. They are, however, rare in rich environments where star formation is generally quenched due to the interaction of galaxies with their surroundings \citep[e.g.,][]{2006PASP..118..517B}. 
Those catalogued in the \ac{VGC} and selected in the $B$ band are 49/2096 (2.3\,\%), and they are principally located at the periphery of the cluster. We checked the nature of all line emitters in the continuum-subtracted H$\alpha$ narrow-band image to see whether we detected compact star-forming sources with morphological properties similar to those of local \ac{BCDs}, but we did not detect any of them. 
\ac{BCDs} are high-surface brightness objects with a very strong star-formation activity, and are thus easily detectable in narrow-band imaging if present \citep{2002A&A...386..134B,2022A&A...659A..46B}. 
Given that the \Euclid observations concern only the inner $\lesssim 0.2\, r_{200}$, that Perseus is a rich and massive cluster, and their very blue colour, we expect that the fraction of \ac{BCDs} in the targeted region is significantly lower than the one observed in Virgo, as suggested by the analysis of the H$\alpha$ data. We thus consider as negligible their possible contribution to our \IE-band \ac{LF}.

\ac{UCDs} are much more compact than \ac{BCDs} and for this reason are more difficult to be identified and distinguished from other point-like sources such as stars and globular clusters. Those identified in the Virgo cluster by \citet{2020ApJS..250...17L} have effective radii in the $g$ and $i$ bands of $10\textrm{pc} \lesssim r_{\rm e}(g,i)\lesssim 100\,\textrm{pc}$, which at the distance of Perseus correspond to $\ang{;;0.03}\lesssim r_{\rm e}(g,i)\lesssim \ang{;;0.3}$, but most of them having $r_{\rm e}(g,i) \lesssim \ang{;;0.1}$ \citep[see][]{2012MNRAS.422..885P}. They are thus hardly resolved even at the excellent angular resolution of the \Euclid \ac{VIS} imaging data (${\rm FWHM}= \ang{;;0.16}$). Their stellar mass is ${\mathcal M} \lesssim 10^7 \, M_{\odot}$, and their luminosity $M_g\geq -12$ and $M_K\geq -12.4$ \citep[]{2020ApJS..250...17L,2021MNRAS.504.3580S}. 
They are often associated with the central massive ellipticals \citep[]{2020ApJS..250...17L,2021MNRAS.504.3580S}, and are thus potentially present in the Perseus field. 
The extraordinary quality of the \Euclid data, however, allows the detection of these and other compact sources (globular clusters) associated with many massive galaxies. 
Their observed \ac{LF} begins at $\IE \simeq 23$ [$M(\IE)\sim -11.3$] and steeply rises up to $\IE \simeq 25.5$ or 26 [$M(\IE)\simeq -8.8$ to $ -8.3$], and drops below this. 
Being compact sources, we expect that the selection criteria mentioned in \cref{sc:Catalogues}, and in particular the one used to discriminate stars from galaxies, have removed them from the sample of possible members. 
In any case, their contribution to the \ac{LF} derived in this work, which is limited to galaxies with $M(\IE)> -11.3$, is certainly negligible.

\subsection{\label{sc:statparams}Robustness of the derived parameters}

\begin{figure*}
\centering
\includegraphics[width=0.8\textwidth]{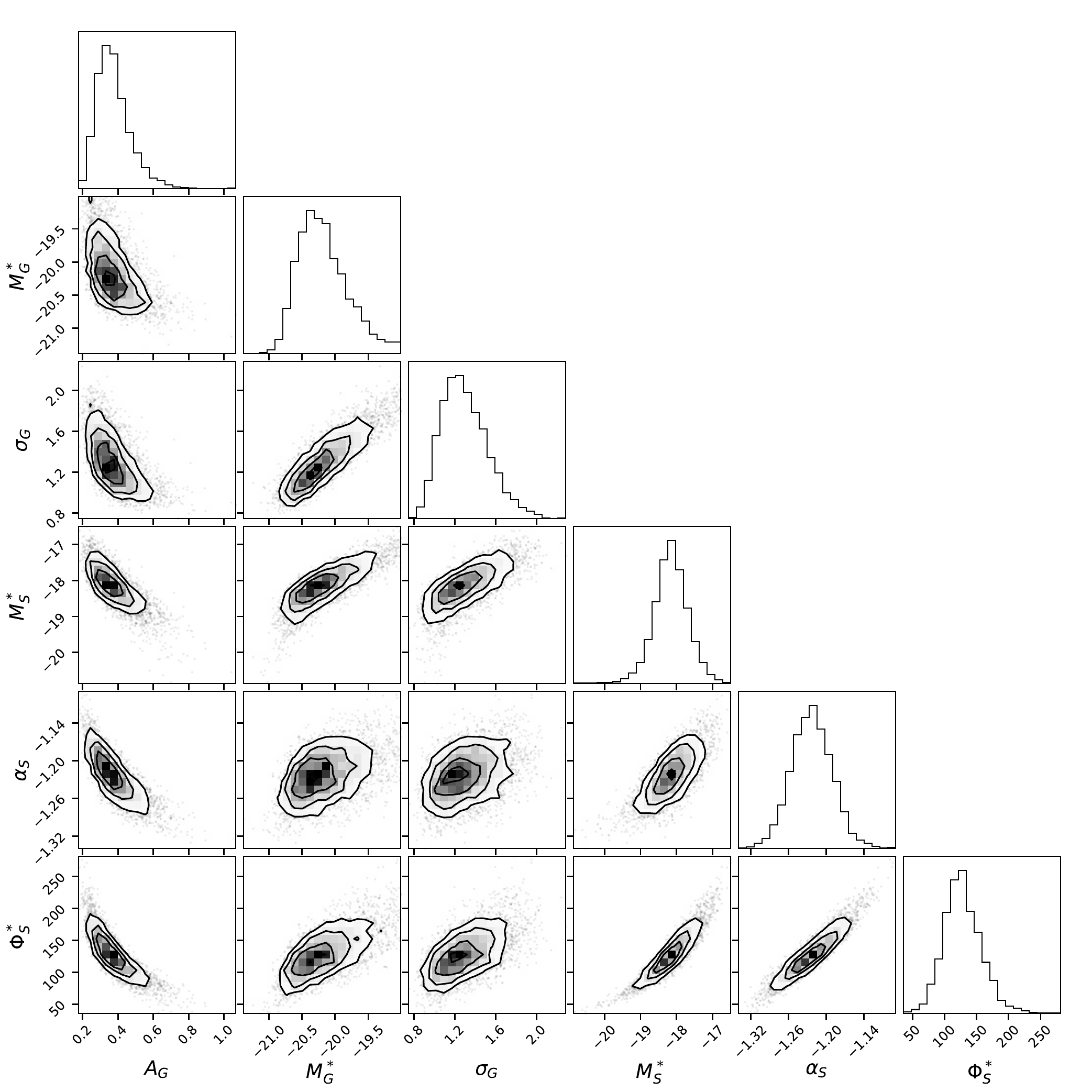}
\caption{Probability distribution function of the output parameters of the fit of the luminosity function for a Schechter $+$ Gauss parametric function when the galaxies NGC\,1275, NGC\,1272, and NGC\,1265 are excluded.}
\label{pdfLF}%
\end{figure*}

The three cD galaxies in the core of the cluster (NGC\,1275, NGC\,1272, and NGC\,1265) introduce a discontinuity in the \ac{LF} that cannot be traced by our simple parametric function given in Eq.\,(\ref{EqLF}). 
We thus fit the data including all galaxies with an absolute magnitude $M(\IE)\leq -11.33$, or excluding either NGC\,1275, or the three brightest objects NGC\,1275, NGC\,1272, and NGC\,1265. 
The best fit to the data derived using these three different samples are given in Table~\ref{table:TabLF}. 
\Cref{pdfLF} shows the probability distribution functions of the output parameters derived when the three cD galaxies are excluded.

As expected, the output parameters of the Gaussian function strongly depend on the adopted sample, i.e., on the inclusion or exclusion of the brightest cD galaxies. This is due to the fact that the number of bright objects is very limited, thus any output parameter in this magnitude regime is statistically uncertain.
More interesting, in the framework of galaxy formation and evolution, is the faint end of the \ac{LF}. The faint end slope of the \ac{LF}, $\alpha_{\rm S}$, is robustly determined, as depicted in \cref{pdfLF}, and around $\alpha_{\rm S} = - 1.2$ to $-1.3$. 
This is significantly flatter than the slope of the dark matter halo mass function predicted by cosmological simulations \citep[$\alpha\simeq -2$, e.g.,][]{2015ARA&A..53...51S}. 
This is also the case if the fit is limited to the steep rise of the distribution [$M(\IE)\lesssim -15$], where $\alpha_{\rm S}\simeq -1.5$, and drops to $\alpha_{\rm S}\simeq -1.1$ in the $ -15 \lesssim M(\IE) \lesssim -11.3$ magnitude range.  
Any possible missed satellite, if it exists, must have a surface brightness at $r_{\rm e}\,(\mu_{I_{\rm E},{\rm e}}) \gtrsim 29\,{\rm mag\,arcsec}^{-2}$.

\subsection{Comparison with other work}

\begin{figure}
\centering
\includegraphics[width=0.49\textwidth]{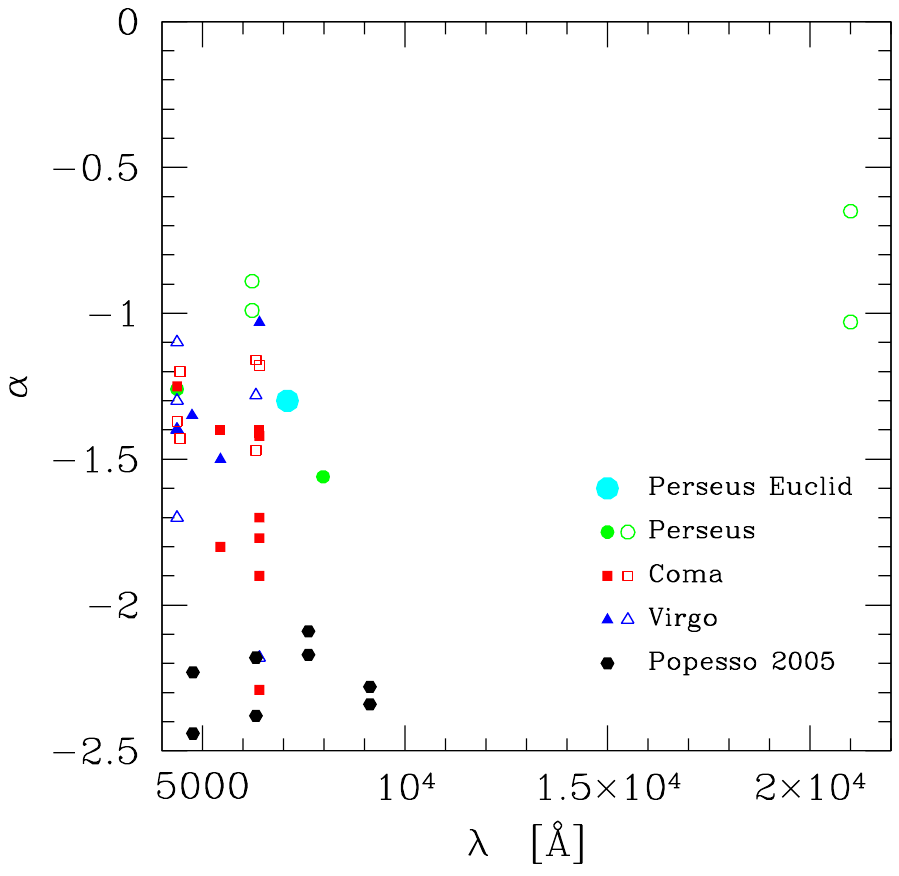}
\caption{The faint-end slope $\alpha_{\rm S}$ of the LF derived in this work [$M(\IE)\leq -12$; big filled cyan circle] is compared to the values of the analogous Schechter parameters $\alpha$ available in the literature in different photometric bands for the Perseus cluster (green circles, references given in the text), for the Coma cluster (red boxes, taken from the compilation of \citealp{2007AJ....133..177M}), the Virgo cluster (blue triangles, from the compilation of \citealp{2016ApJ...824...10F}), and for the composite LF of clusters of \citet[][black hexagons, double Schechter functions, measured using two different background subtraction techniques within 0.5\,Mpc]{2005A&A...433..415P}. Empty symbols are for \ac{LF} of the whole cluster, filled symbols for the core region as in the Euclid field. 
}
\label{fig:comparison}%
\end{figure}

The \IE-band \ac{LF} of the Perseus cluster measured within $R\lesssim0.25\,r_{200}$ is compared to those derived in other bands in \cref{LF}. The faint-end slope of the fitted function is compared to those derived in these other works at different wavelengths in \cref{fig:comparison}. These have been determined in the $K$ band using WISE 2.4-$\mu{\rm m}$ data by \citet{2020A&A...640A..30M} and concern the inner $R\sim r_{200}$, and in the $r$ band by \citet[$R\sim 1.4\,r_{200}$]{2020MNRAS.494.1681A}. The \IE-band \ac{LF} is also compared with that of \citet{2008MNRAS.383..247P} measured in the $B$ band for a limited number of dwarfs in the very inner regions of the cluster.  
To compare \ac{LF} derived in different photometric bands, we cross-matched the \Euclid catalogue with those used in previous studies. We estimated the mean colour in the different bands, which we then used as a systematic correction on the X-axis. Since the Y-axis normalisation depends on the sampled volume, which varies across these studies, we arbitrarily chose the normalisation on this axis to match the distribution observed in the \IE band.

The \IE-band \ac{LF} derived in this work is significantly deeper than those of \citet{2020A&A...640A..30M} and \citet{2020MNRAS.494.1681A}, which are both limited to the bright end, $M(\IE)\simeq  -18.3$ and $M(\IE) \simeq  -16.2$, respectively, when these limiting magnitudes are derived using the mean colour of common galaxies in the different bands. It is also deeper than that of \citet[$M(\IE)\simeq -13.9$]{2008MNRAS.383..247P}, although that does not sample the bright end given the very limited sky region covered during the observations (173\,arcmin$^2$).

For a fair comparison, three things should be remembered.  Firstly, the \IE-band \ac{LF} measured in this work concerns only the inner $R\lesssim 0.25\,r_{200}$ which, for the well known morphology-segregation effect \citep[]{1980ApJS...42..565D,1993ApJ...407..489W}, is highly dominated by early-type systems. Indeed, the number of star-forming galaxies is limited to a handful of objects, as depicted in the colour-magnitude diagram shown in \cref{fig:GUVICS_coloriMAGgH}. Secondly, the overall shape of the \ac{LF} is known to change at different distances from the cluster centre \citep[e.g.,][]{2002MNRAS.329..385B,2006A&A...445...29P,2011MNRAS.414.2771D}. Finally, the shape also changes as a function of the photometric band \citep[e.g.,][]{2002ApJ...569..144A,2005A&A...433..415P,2016A&A...585A...2B}. 

The \IE-band \ac{LF} measured in this paper agrees with the  decrease observed at $M(\IE)\simeq -19$ by \citet{2020A&A...640A..30M} and \citet{2020MNRAS.494.1681A} when limited to the early-type galaxy population. This decrease is due to the presence of the dip, which corresponds to the relative decrease of objects at $\IE \sim 15$ in the scaling relations shown in \cref{Ferrarese2avg} (see also figure~11 of \citealt{EROPerseusDGs}). The faint end of the \IE-band \ac{LF} is comparable to the one observed in the $B$ band by \citet{2008MNRAS.383..247P} in the same magnitude range.

The \IE-band \ac{LF} of Perseus can also be compared to those derived for other nearby clusters of galaxies. \Cref{LF} shows the comparison with the $i$-band \ac{LF} of composite SDSS clusters measured within $R<0.5\,h^{-1}$\,Mpc by \citet{2005A&A...433..415P}. It shows a discontinuity at $M(\IE)\simeq -19$, where the \IE-band \ac{LF} of Perseus has a dip, and is significantly steeper at fainter luminosities ($\alpha = -2.17$; see also \cref{fig:comparison}). The presence of the dip with properties similar to the one presented here has been observed in other very massive nearby clusters, such as Coma \citep{1995A&A...297..610B,1997A&A...317..385L,1998ApJ...503L..45D,2007A&A...462..411A,2007AJ....133..177M} and Abell 85 \citep{1999A&A...343..760D}. The slope of the LF in the $-19 \lesssim M(\IE) \lesssim -14$ absolute magnitude range is $\alpha \simeq -1.5$, and is significantly flatter than the one derived by \citet[$\alpha \simeq -2.17$]{2005A&A...433..415P}. However, this difference should be considered with caution for several reasons. First of all, this luminosity range is generally lacking  spectroscopic data, thus cluster membership is often derived using statistical arguments, such as a statistical subtraction of the background galaxy counts, as done in \citet{2005A&A...433..415P}. This technique is obviously very sensitive to sampling variance, since in this luminosity regime the number of background galaxies largely exceeds that of cluster members. For the region analysed in this work, for instance, the number of galaxies detected by \texttt{SourceExtractor} with $\IE < 23.6$ is about $33\,000$, most of which have faint luminosities (see \cref{sc:Catalogues}),with only 1220 finally being identified as cluster members. As a consequence, the statistical subtraction of the background galaxy counts introduces large uncertainties at the faint end, as suggested by the very discrepant results obtained even on the same cluster when different techniques are used (see \cref{fig:comparison}). In the Coma cluster, by far the most studied local massive cluster of galaxies, the results obtained so far give slopes ranging from $\alpha = -1.0$ to $\alpha = -2.0$ \citep[see table~1 in][]{2007AJ....133..177M,2012AJ....144...40Y}, suggesting that the faint end slope of the \ac{LF} is still poorly constrained whenever the identification of cluster members suffers from large uncertainties. This is also the case in other nearby clusters where different values of $\alpha$ are often derived when using different techniques to account for the background contamination \citep[e.g.][]{2008A&A...479..335B}. We will be able to test this statistical correction and compare the results to those found in this work once a robust estimate of the galaxy number counts will be measured thanks to the \Euclid mission.

A particular case is the Virgo cluster. Located at only 16.5\,Mpc \citep{1999A&A...343...86G,2007ApJ...655..144M,2018ApJ...856..126C}, this is the closest cluster of galaxies and for this reason it has been the target of several multifrequency surveys. Thanks to its proximity, deep ground-based imaging surveys can easily sample the dwarf galaxy regime over a luminosity range comparable to the one reached by \Euclid in Perseus. NGVS \citep{2012ApJS..200....4F} managed to detect and identify cluster members down to a stellar mass limit of ${\mathcal M} \simeq 10^6\,M_{\odot}$. The typical angular resolution of these data, \ang{;;0.6} in the $i$ band (49\,pc), is comparable to the physical scales probed by \ac{VIS} in the \IE band at the distance of Perseus (\ang{;;0.16}, 56\,pc, see \cref{sc:EuclidData}). \citet{2016ApJ...824...10F} derived the $i$-band \ac{LF} of the cluster within $R \simeq 0.2\, r_{200}$ down to $M_i \simeq -10$, where the 336 cluster members have been identified with criteria very similar to those used in this work. The two sets of data can thus be directly compared. The bright part of $i$-band Virgo cluster \ac{LF} is poorly constrained because of the very limited number of massive objects; it is thus impossible to see whether a dip is present also in this cluster. The faint end, however, is well represented by a Schechter function of slope $\alpha = -1.35$, a value slightly steeper than the one found for Perseus ($-1.30\lesssim\alpha \lesssim -1.20$). This difference is not surprising given that Perseus is a very massive spiral-poor relaxed cluster ($M_{200} = 1.2 \times 10^{15}\, M_{\odot}$), while Virgo is a spiral-rich, intermediate mass unrelaxed system \citep[$M_{200} = 1$--$4 \times 10^{14}\, M_{\odot}$,][]{2012ApJS..200....4F,2022A&ARv..30....3B}. A flattening of the $B$-band LF measured in selected regions out to $R \simeq 1\, r_{200}$ in the magnitude range $-14 \lesssim$ $M_B \lesssim -11$  has been claimed by \citet[$\alpha = -1.3$ to $-1.0$]{2002MNRAS.333..423T} with shallower INT data, and only partly confirmed by the deeper NGVS data in the inner region \citep[$\alpha = -1.39$ for $-17.4 \lesssim M_g \lesssim -14.4$  versus $\alpha = -1.26$ for $-14.4\lesssim M_g \lesssim -9.13$,][]{2016ApJ...824...10F}.

The results obtained in this work are consistent with those available in the literature for Virgo, and extend those derived for other very massive clusters such as Perseus to a luminosity range never explored before. They provide a unique sample for comparison with the prediction of cosmological simulations to constrain galaxy evolution in a very rich environment. 

Finally, we compare the observed \ac{LF} to that predicted for cluster galaxies using the Cluster-EAGLE cosmological simulations in the \ac{SDSS} $i$ band \citep{2022MNRAS.515.2121N}. The two \ac{LF}s are fairly different: the cosmological simulations do not show an obvious dip at $M(\IE)\simeq -19$, but just a flattening of the curve, although they have a faint-end slope in the sampled magnitude range [$M(\IE)< -15$] close ($\alpha = -1.3$) to the observed one. The \ac{LF} of \citet{2022MNRAS.515.2121N}  has been derived combining 30 clusters of mass $10^{14}\lesssim M_{200}/M_{\odot}\lesssim 10^{15.4}$ with galaxies located within $R=r_{200}$. The difference with the observed one might partially be due to the fact that in Perseus we are sampling only the very innermost regions ($R\lesssim 0.25\,r_{200}$) and that Perseus is a very massive cluster, where a dip at intermediate luminosity is generally observed.  

\subsection{Implications of the results in an evolutionary picture}

The field galaxy \ac{LF} and \ac{SMF} are generally considered to be composed of two different components, one for the quiescent population and one for the star-forming systems, dominant in low-density regimes \citep[][]{2003ApJS..149..289B,2012MNRAS.421..621B,2013ApJ...777...18M}. They can both be represented by two functions. The former, generally characterised by a Gaussian or a Schechter function declining at faint luminosities ($\alpha > -1$), is dominated in the high-luminosity, high-mass regime. The latter, mainly fitted with a Schechter function, is dominant at the faint end. In the cluster environment the relative weight of the two populations is significantly changed, in particular at the faint end where the dominant population consists of dwarf ellipticals \citep[e.g.,][]{1985AJ.....90.1759S}. Given the tight positive relation between the fraction of slow/fast rotators with stellar mass \citep{2016ARA&A..54..597C} observed also in clusters such as Virgo \citep{2014A&A...570A..69B}, it is likely that the Gaussian part of the luminosity and mass functions are dominated by massive ellipticals formed during major merging events, and those of lower luminosities/stellar masses by gentler mechanisms. 

The steep rise of the dwarf elliptical galaxy population in the luminosity range $-19\lesssim M(\IE) \lesssim -14$ observed here and in other nearby clusters \citep[e.g.,][]{1985AJ.....90.1759S} can be explained if the galaxies infalling into the cluster since its formation, which are preferentially star-forming objects, are quenched after their interaction with the hostile surrounding environment \citep[e.g.,][]{2014A&ARv..22...74B}. We recall that the field \ac{LF}, when measured including the dominant \ac{LSB} population, has a slope in the visible bands close to $\alpha = -1.5$ \citep{2005ApJ...631..208B}. A simple fading of the star-formation activity, which might follow ram-pressure-stripping events \citep[e.g.,][]{2022A&ARv..30....3B}, can transform blue galaxies into red systems \citep[e.g.,][]{2010A&A...517A..73G} without affecting their structural \citep[e.g.,][]{2008A&A...489.1015B,2008ApJ...674..742B} and kinematic properties \citep[e.g.,][]{2011A&A...526A.114T,2015ApJ...799..172T}, and more important the shape of the luminosity and mass functions. This evolutionary picture is consistent with the observed variation of the slope of the \ac{LF} of the Virgo cluster derived using tracers sensitive to the youngest stellar populations \citep[H$\alpha$, UV,][]{2016A&A...585A...2B,2023A&A...675A.123B}  or to older populations \citep[visible bands,][]{1985AJ.....90.1759S,2016ApJ...824...10F}. It is also consistent with the observed increase of the fraction of red/blue galaxies with decreasing redshift at the faint end of the \ac{LF} in clusters of galaxies \citep{2007MNRAS.374..809D,2009MNRAS.400...68D,2007ApJ...661...95S,2009MNRAS.394.2098S,2008ApJ...673..742G,2014A&ARv..22...74B}.

The flattening at lower luminosities should still be confirmed with more accurate spectroscopic observations. If borne out, lowest-mass objects gravitationally perturbed during their flyby encounters with other cluster members \citep[galaxy harassment,][]{1996Natur.379..613M,2005ASPC..331...89M} might explain this flattening. In the most extreme cases a large fraction of the mass is removed during the interaction, leaving just the galaxy nucleus \cite[][]{2003MNRAS.344..399B,2003Natur.423..519D,2008MNRAS.385.2136G,2013MNRAS.433.1997P}. This mechanism, which has been often proposed to explain the formation of \ac{UCDs}, might reduce the number of objects in this luminosity/mass regime, flattening the observed distribution. Interestingly, \cite{EROPerseusICL} have shown a significant contribution of dwarf galaxies to the intracluster globular cluster \ac{LF}. These dwarf galaxies would be entirely disrupted to form the ICL so this could partly explain the observed difference between the steep faint-end slope of the dark matter halo distribution predicted by simulations with the flatter slope observed in the \ac{LF} and \ac{SMF}. This speculative interpretation should be tested by comparison with hydrodynamic cosmological simulations once they reach this very low-mass regime with sufficient statistics. 

\subsection{Perspectives}

The analysis and results presented in this work demonstrate the extraordinary capabilities of \Euclid in the study of galaxy evolution in rich environments. Although not reaching the spectacular angular resolution of \ac{HST}, \Euclid will provide a homogeneous set of optical (and \ac{NIR}) images with a typical ${\rm FWHM} \sim \ang{;;0.16}$   ($\sim \ang{;;0.49}$) over more than 14\,000\,deg$^2$. This resolution will be crucial for the identification of galaxies belonging to different local environments, from massive clusters such as Perseus down to groups and filaments, hundreds of which will be covered during the mission.
Furthermore, the extraordinary quality of the images in terms of sensitivity to \ac{LSB} regions will be crucial to study different characteristic features associated with galaxies in rich environments, such as tails, filaments, shells of tidally perturbed objects \citep{2015MNRAS.446..120D}, or the intracluster light present in the central regions of massive clusters \citep{EROPerseusICL}. These particular stellar features are formed only during gravitational perturbations that are able to displace the stellar component, while not in hydrodynamic interactions affecting only the gas \citep[e.g.,][]{2022A&ARv..30....3B}. The \Euclid observations will thus be of prime importance for the identification of the dominant perturbing mechanism in different environments, from filaments and groups, to massive clusters. It will also be extremely important to identify \ac{UDGs}, a population seen in nearby clusters as well as in the field \citep{EROPerseusDGs, 2015ApJ...807L...2K,2015ApJ...809L..21M,2015ApJ...804L..26V, 2024ApJS..271...52Z}, whose origin is still under debate \citep{2017MNRAS.466L...1D,2019MNRAS.485..382C,2020MNRAS.497.2786T,2022A&A...667A..76J}.

\section{\label{sc:Conclusions}Conclusions}

The \Euclid \ac{ERO} programme gathered deep optical and \ac{NIR} imaging data of the inner regions ($R \leq 0.25\,r_{200}$) of the nearby Perseus cluster. 
We used this unique set of imaging data, which reached a point-source depth of $\IE = 27.3$  and surface brightnesses $\mu_{I_{\rm E}} = 30.1 {\rm\,mag\,arcsec}^{-2}$, to measure the \ac{LF} and \ac{SMF} of this cluster. 
We identified cluster members out of the roughly half-million detected sources using a combination of photometric redshifts, scaling relations, cross-matches with catalogues available in the literature, and visual inspection of the images. We identified 1220 cluster members. Using the observed distribution of their structural properties (total magnitude, surface brightness, extension, S\'ersic parameter, and inclination) we accurately measured the completeness function of the sample, which reaches 50\,\% at $M(\IE) = -11.3$. We derived the \IE-band \ac{LF} of the inner region of the cluster and fitted it using a parametric Schechter plus Gaussian function, which reproduces well the observed distribution, in particular whenever the three brightest cD galaxies (NGC\,1275, NGC\,1272, and NGC\,1265) are excluded. The fitted function, which is significantly deeper than those derived so far using ground-based data, has a dip at $M(\IE) \simeq -19$ and a faint-end slope $\alpha_{\rm S} \simeq -1.2$ to $ -1.3$. We used the unique set of multifrequency data in hand, which also include deep, unpublished $ugriz$ and narrow-band H$\alpha$ imaging data gathered at the \ac{CFHT}, to measure the stellar mass of all the cluster members, and thus derive the \ac{SMF} of the cluster. The \ac{SMF} can be represented by a Schechter function with a faint-end slope $\alpha_{\rm S}\simeq -1.2$ to $-1.35$. 

Despite the exceptional sensitivity of \Euclid to \ac{LSB} objects ($\sim$ 30.1\,mag\,arcsec$^{-2}$), which are expected to dominate at faint luminosities, the observed faint-end slope of the \ac{LF} and \ac{SMF} of Perseus are significantly flatter than those predicted for the dark matter halo distribution by cosmological simulations. The unique set of data that the \Euclid mission will soon provide will be crucial to extend this analysis to a much larger sample of local galaxies in different environments, from the field to groups and rich clusters. These data will offer new, stringent constrains to cosmological simulations on galaxy evolution, and thus provide a unique opportunity to understand the processes that shape the \ac{LF} and \ac{SMF} in different environments.  

\begin{acknowledgements}
\AckERO  
\AckEC
Based on observations obtained with MegaPrime/MegaCam, a joint project of CFHT and CEA/DAPNIA, at the Canada-France-Hawaii Telescope (CFHT) which is operated by the National Research Council (NRC) of Canada, the Institut National des Science de l'Univers of the Centre National de la Recherche Scientifique (CNRS) of France, and the University of Hawaii. The observations at the Canada-France-Hawaii Telescope were performed with care and respect from the summit of Maunakea which is a significant cultural and historic site.
This work presents results from the European Space Agency (ESA) space mission Gaia. Gaia data are being processed by the Gaia Data Processing and Analysis Consortium (DPAC). Funding for the DPAC is provided by national institutions, in particular the institutions participating in the Gaia MultiLateral Agreement (MLA).
Funding for the Sloan Digital Sky Survey V has been provided by the Alfred P. Sloan Foundation, the Heising-Simons Foundation, the National Science Foundation, and the Participating Institutions. SDSS acknowledges support and resources from the Center for High-Performance Computing at the University of Utah. SDSS telescopes are located at Apache Point Observatory, funded by the Astrophysical Research Consortium and operated by New Mexico State University, and at Las Campanas Observatory, operated by the Carnegie Institution for Science. The SDSS web site is \url{www.sdss.org}. SDSS is managed by the Astrophysical Research Consortium for the Participating Institutions of the SDSS Collaboration, including Caltech, The Carnegie Institution for Science, Chilean National Time Allocation Committee (CNTAC) ratified researchers, The Flatiron Institute, the Gotham Participation Group, Harvard University, Heidelberg University, The Johns Hopkins University, L’Ecole polytechnique f\'{e}d\'{e}rale de Lausanne (EPFL), Leibniz-Institut f\"{u}r Astrophysik Potsdam (AIP), Max-Planck-Institut f\"{u}r Astronomie (MPIA Heidelberg), Max-Planck-Institut f\"{u}r Extraterrestrische Physik (MPE), Nanjing University, National Astronomical Observatories of China (NAOC), New Mexico State University, The Ohio State University, Pennsylvania State University, Smithsonian Astrophysical Observatory, Space Telescope Science Institute (STScI), the Stellar Astrophysics Participation Group, Universidad Nacional Aut\'{o}noma de M\'{e}xico, University of Arizona, University of Colorado Boulder, University of Illinois at Urbana-Champaign, University of Toronto, University of Utah, University of Virginia, Yale University, and Yunnan University.
CS acknowledges the support of the Natural Sciences and Engineering Research Council of Canada (NSERC). Cette recherche a été financée par le Conseil de recherches en sciences naturelles et en génie du Canada (CRSNG). CS also acknowledges support from the Canadian Institute for Theoretical Astrophysics (CITA) National Fellowship program.
FB acknowledges support from the project PID2020-116188GA-I00, funded by MICIU/AEI /10.13039/501100011033.
LQ acknowledges funding from CNES postdoctoral fellowship program.
\end{acknowledgements}

\bibliography{Perseus,Euclid,EROpipelineRefs,EROplus}

\begin{appendix}

\section{\label{appendix:ScienceOverview}Overview of the science programme}

The extraordinary data quality and richness of this \ac{ERO} Perseus cluster programme opens the door to other exciting analyses. In the following subsections, we briefly present soon-to-be published research activities by the \ac{ERO} Perseus science team beyond the primary focus of this paper.


\subsection{Dwarf galaxies}

Dwarf galaxies are the oldest and most numerous galaxy type in the Universe \citep{1990A&A...228...42B,1994A&ARv...6...67F}, yet the driving mechanism(s) in their evolution and the transformation into the different types we see today remains elusive. The two main types are the dwarf irregular (dI) galaxies and the dwarf elliptical/spheroidal (dE/dSph) galaxies, but many other subclasses make up this population of galaxies, such as \ac{UCDs} and \ac{UDGs}. The defining distinction between dwarf and giant galaxies is that dwarfs are extremely faint, with magnitudes fainter than $-18$ in the $V$ band \citep{1985ApJ...296L...7A}, although this magnitude limit varies greatly in the literature (see, for example, \citealt{1971ARA&A...9...35H}). They are also defined by their typically low stellar masses \citep[$\leq 10^9\,M_{\odot}$,][]{2017ARA&A..55..343B} and small physical sizes, with sizes ranging from a few hundred parsecs to a few kiloparsecs. Dwarf galaxies often host globular clusters (GCs), which can be used to probe the underlying mechanisms driving galaxy assembly, the dynamics of stellar populations, and the influence of environment on star-formation processes.

In \citet{EROPerseusDGs}, we conduct a comprehensive census of the dwarf galaxies, their nuclei, and their globular cluster populations in the Perseus \ac{ERO} programme. Our study maps their spatial distribution and visual morphologies, while also extracting their photometric and structural properties. Our analysis reveals a total of 1100 dwarf galaxy candidates, 630 of which are found to be new discoveries. The morphological mix consists of 96\,\% dE, 53\,\% nucleated, 26\,\% GC-rich, and 6\,\% with disturbed morphology. We classify a relatively high fraction of galaxies, 9\,\%, as \ac{UDGs}. The majority of the dwarf candidates follow the scaling relations of dwarf galaxies in the literature. However, our dwarf sample significantly extends the known scaling relation towards diffuse galaxies. In agreement with studies of nucleated populations, we find fewer nuclei in faint, small, and high-ellipticity dwarfs, with flatter surface brightness profiles. Globally, the GC specific frequency, $S_{N}$, of the Perseus dwarf candidates is intermediate between those measured in the Virgo and Coma clusters. While the dwarf candidates with the largest GC counts are found throughout the \Euclid \ac{FOV}, the dwarfs located around the east-west strip, where most of the brightest cluster members are found, exhibit larger $S_{N}$, on average.


\subsection{Asymmetries in dwarf galaxies for studying environmental processes}

Dwarf galaxies are excellent tracers of the effects of environment, due to their \ac{LSB} signature. We analyse the morphology of dwarf galaxies in Perseus to study the role of the environment in this massive cluster, where strong gravitational forces shape the evolution of its galaxies. We concentrate on the asymmetries of early-type dwarfs (dE and dSph), which are a solid indicator of tidal forces \citep[see][]{2011MNRAS.410.1076P}. We conducted such an analysis for the 564 dwarfs catalogued in the Fornax Deep Survey \citep{2018A&A...620A.165V}. \citet{2011MNRAS.410.1076P} found strong asymmetries in 11 galaxies in HST data of the outer parts of Perseus. With the \Euclid data on the Perseus cluster, strong of 1100 dwarf galaxies, we now have a 100-fold increase in the number of dwarf galaxies available, permitting a comprehensive analysis of their morphological evolution within the cluster.


\subsection{Environmental dependence of the nucleation fraction in dwarf galaxies}

\ac{NSCs} are compact stellar systems with sizes (half-light radii) of 1--$50\,{\rm pc}$ and stellar masses spanning several decades, from as low as $10^4\, M_{\odot}$ to as high as $10^{8}\, M_{\odot}$. This makes them, on average, larger and more massive than a typical GC, and their central stellar surface densities can be even more extreme. Sitting at the bottom of their host galaxy's potential well, NSC stellar populations and dynamics encode a record of the formation and evolution of galactic inner regions \citep[see][and references therein]{2020A&ARv..28....4N} and inform us about the efficiency of high-pressure star formation at early times. 

\ac{NSCs} are found in galaxies spanning a wide range of masses, morphological types and environments, and nucleation seems to be a complex function of all these parameters. Recent studies in the nearby Universe have cemented the role of galaxy mass/luminosity as the main driver of NSC occurrence \citep{2019ApJ...878...18S}. Remarkably, they have also uncovered an unambiguous secondary dependence on the environment at low galaxy masses [$6 < {\rm log_{10}}({\cal M}/M_{\odot}) < 9$], whereby more massive haloes feature higher nucleation fractions than lower density environments \citep{2021MNRAS.507.3246H,2021MNRAS.506.5494P,2021MNRAS.508..986Z,2024MNRAS.tmp..961Z}.  With a virial mass $M_{200}\,{\sim}\,10^{15}\,M_{\odot}$ Perseus is an ideal laboratory to study nucleation at extreme environmental densities, with over a thousand dwarfs already identified as candidate cluster satellites within the \Euclid footprint \citep{EROPerseusDGs}. We will take advantage of \Euclid's exquisite sensitivity for both extended and compact sources to carry out a comprehensive study of the phenomenon of nucleation down to the luminosity regime of the `classical' Galactic satellites, as well as provide a thorough comparison with other relevant environments in the nearby Universe.


\subsection{Globular clusters of dwarf galaxies}

In this project, we use \Euclid \ac{ERO} data on the Perseus galaxy cluster and identify GCs around dwarf galaxies in this cluster. Given the observations of a number of GC-rich dwarf galaxies in nearby galaxy clusters (\citealp{amorisco2018,prole2019,2020ApJ...899...69L,saifollahi2022}) and galaxy groups (\citealp{muller2021,carlsten2022}), the GC content of dwarf galaxies has become a topic of interest in the past few years. The GC-rich dwarfs host a few times more GCs compared to the general population of dwarf galaxies of the same stellar mass. It appears that the high-density environments within galaxy clusters, environments similar to the environment of the Perseus cluster, play a role in the formation of GC-rich dwarf galaxies. However, until now, most of the observations have focused on a sub-class of dwarf galaxies with large size and very low surface brightness, known as \ac{UDGs} \citep{vandokkum2015}. In the meantime, the GC properties of the general population of dwarf galaxies have not been studied in detail, which potentially can lead to inconsistencies in interpretations of GC observations of \ac{UDGs}. 

However, now, with the \ac{ERO} data of the Perseus cluster, we are able to have a uniform analysis and in-depth study of the GC properties of more than 1000 dwarf galaxies, including several \ac{UDGs}. Given the spatial resolution, as well as the depth of the \Euclid $\IE$-band image of this \ac{ERO} programme, we are able to identify GCs brighter than the turn-over magnitude of the GC luminosity function (GCLF) at the distance of the Perseus galaxy cluster (\citealp{EROPerseusICL,EROPerseusDGs}). For the GCs, the combination of $\IE$ with the three \Euclid \ac{NISP} bands, $\YE$, $\JE$, and $\HE$, help to further clean the sample of GC candidates from possible contaminants (\citealp{EROFornaxGCs}, Euclid Collaboration: Voggel et al. in prep.). As a result of our search for GCs, we produce catalogues of GC candidates for each dwarf galaxy identified within the \ac{ERO} data. Then, using these catalogues, we examine the properties of GCs of dwarf galaxies in the Perseus cluster for a wide range of galaxy luminosities and sizes, and study the galaxy-GC scaling relations. Furthermore, given the correlation between GC numbers and galaxy halo mass (\citealp{spitler2009,harris2013}), we convert the estimated GC numbers to dwarf halo mass and further investigate their stellar-to-halo mass ratios. 


\subsection{Measuring distances to early-type and dwarf galaxies using surface brightness fluctuations}

The \ac{SBF} method is a robust technique for determining distances to early-type galaxies out to about $200\,{\rm Mpc}$ \citep[and beyond, with JWST data, see][and references therein]{CantielloBlakeslee23}. By measuring  intrinsic variance in surface brightness distribution, it estimates distance with up to 5\,\% accuracy for individual targets \citep{2007ApJ...655..144M,Blakeslee21,Jensen21}. The \ac{SBF} signal arises from the stochastic stellar counts and luminosities in the host stellar population and corresponds to a quantity closely related to the luminosity-weighted mean brightness of the stars within the target galaxy \citep{ts88,tal90}. 

To derive the distances from \ac{SBF}, accurate knowledge of fluctuations of magnitude ($\overline{m}$) and absolute magnitude ($\overline{M}$) in the same passband is crucial. Typically, $\overline{M}$ is calibrated in a two-step procedure: first using galaxies with known distances to establish the zero point, and then examining the dependence of $\overline{M}$ on stellar population properties \citep[see, e.g.,][]{tonry01,2018ApJ...856..126C}. This is often done using galaxies at a common distance (e.g., in groups or clusters) and colour measurements to standardize the stellar population dependence of $\overline{M}$.

\Euclid's  high-resolution imaging, depth, and wide sky survey area offer potential for precise \ac{SBF} measurements for thousands of galaxies in the local Universe. This makes the Perseus \ac{ERO} programme invaluable for testing and calibrating \ac{SBF} measurement procedures using \Euclid's specific data set.

\ac{SBF} magnitudes (and colours) offer insights into  unresolved stellar populations in targeted galaxies \citep{jensen15}. Despite  lower angular resolution in \ac{NIR} observations, \ac{NISP} instrument will provide ample data for \ac{NIR} \ac{SBF} magnitude calibration, particularly beneficial for elliptical galaxies where red-giant-branch stars dominate \ac{NIR} flux. Contemporary \ac{VIS} and \ac{NISP} fluctuation amplitudes aid in understanding potential sources of scatter, such as the impact of intermediate-age asymptotic giant branch stars on \ac{NIR} \ac{SBF} \citep{cantiello03,raimondo09}.

The Perseus cluster,  at $72\,{\rm Mpc}$, is an ideal environment for \ac{SBF} measurements. Its minimal depth effects enable precise characterization of the \ac{SBF} amplitude relative to galaxy colour, available from \Euclid's \ac{VIS} and \ac{NISP} imaging. We aim to conduct \ac{SBF} measurements  on all Perseus cluster galaxies with smooth morphology using \ac{VIS} and \ac{NISP} imaging data. Challenges include standardising $\overline{M}$ in the \IE band (due to its wide wavelength range), and addressing undersampling of the \ac{PSF} in \ac{NISP} data. Preliminary tests are currently underway.


\subsection{\label{sc:SpatiallyResolved}Spatially-resolved stellar populations}

The deep images and highly resolved, precise photometry obtained in the optical and \ac{NIR} by \textit{Euclid} on the Perseus cluster enable the mapping of spatially-resolved stellar mass density of galaxies in the field via resolved \ac{SED} fitting. Combining these data sets with UV and optical imaging from GALEX \citep{Morrissey2007} and \ac{CFHT} offers sufficient constraints on other important properties, including stellar age, metallicity, and even star-formation history. By deriving spatially resolved maps of stellar mass and other properties of the stellar populations, we can delve deeper into the understanding of the assembly of galaxy mass within the cluster, disentangling the roles of endogenous galaxy processes, environmental influences, and galaxy mergers. The mass profiles derived will be the benchmark for future kinematic and dynamical follow-up studies and high-resolution \ion{H}{i} and H$_2$ gas maps and masses to constrain the interplay of the small-scale physics of gas and star formation with galaxy-scale properties.

\begin{figure*}
\centering
\includegraphics[width=0.99\textwidth]{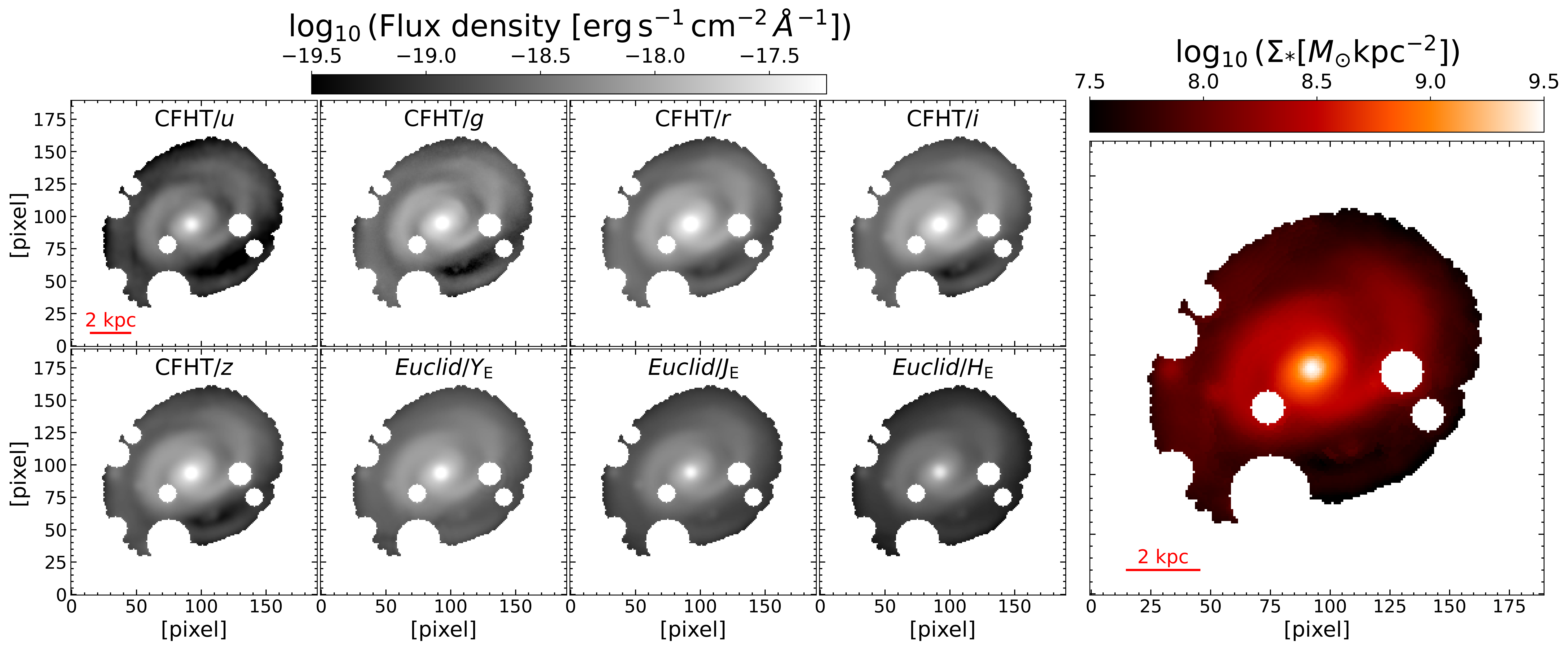}
\caption{{\it Left}: Flux maps for NGC\,1268 in various \ac{CFHT} and \Euclid bands. {\it Right}: Stellar mass density map obtained after SED fitting. }
\label{fig:ngc1268}
\end{figure*}

We will use the \texttt{piXedfit} \citep{Abdurrouf2021,Abdurrouf2022a,Abdurrouf2022b_pixedfitcode} pipeline for performing spatially resolved \ac{SED} fitting on the combined data from \Euclid and \ac{CFHT}.  This code has been rigorously tested on galaxies simulated by the TNG collaboration (Abdurro'uf et al., in prep.). Leveraging synthetic spectral models (e.g., \citealt{2003MNRAS.344.1000B}), we will determine the best-fitting template and its characteristic stellar population parameters for each pixel, thereby constructing stellar population maps for the target galaxies extending to their outer galactic regions. For NGC\,1268, in Fig.~\ref{fig:ngc1268} we show the 2D maps of the fluxes in each band and mass density map obtained after \ac{SED} fitting. 

\subsection{The mysterious origin of \texorpdfstring{NGC\,1277}{NGC1277}}
 
NGC\,1277 is the most extreme case of relic galaxy, i.e., a local counterpart of the massive ultra-compact galaxy population also known as `red nuggets' common at $z > 2$ \citep{Ferre-Mateu17,Spiniello24}. How this object remains similar to those in the primeval Universe is a mystery. In addition, it has also been found that this galaxy is consistent with being dark matter deficient \citep[$f_{\rm dm} (5r_{\rm e}) < 0.05$ at the $2\,\sigma$ confidence level, see][]{Comeron23}, being the first such galaxy found with a high stellar mass (${\mathcal M} \sim 1.6\times 10^{11}\, M_{\odot}$). Our project aims at shedding light into these conundrums by looking for interactions of this galaxy, combining both HST and \Euclid photometry, in the optical but crucially also in the \ac{NIR} that better tracks stellar mass. \Euclid's \ac{LSB} capabilities and unique spatial resolution is an asset for this kind of study. Finding asymmetries in the galaxy surface brightness profiles favours a scenario where it has lost some of its baryonic mass (and thus the lower-bound dark matter) via tidal interactions with the cluster or its neighbour giant elliptical NGC\,1278. Conversely, the dearth of these features allows us to detect well-defined galaxy edges or truncations, a novel physically-motivated galaxy size indicator related with the gas density threshold for enabling efficient star formation \citep[e.g.,][]{Buitrago24,Fernandez-Iglesias24}. In fact, because of NGC\,1277's very high formation redshift \citep{Trujillo14,Buitrago18}, our \Euclid \ac{ERO} data are in a position to prove that relic galaxies are those for which these truncations appear the sharpest.


\subsection{Euclid view of known jellyfish galaxies in the Perseus cluster}

Two star-forming perturbed galaxies, UGC\,2665 (49:51:50.7,    41:38:07) and MCG\,+07-07-070 (50:05:30.9, 41:38:27) are included within the \Euclid \ac{ERO} images of the Perseus cluster. These objects are characterised by filamentary dusty and stellar structures escaping from their stellar discs, like those observed in a few cluster objects with available \ac{HST} data \citep{2015AJ....150...59K,2016AJ....152...32A,2019ApJ...870...63C}. The two galaxies also have extended radio continuum tails of cometary shape \citep{2022ApJ...941...77R}. All this observational evidence suggests that the two galaxies are undergoing a ram-pressure-stripping event \citep{2022A&ARv..30....3B}. We use the optical and \ac{NIR} imaging data of extraordinary quality gathered during the \textit{Euclid} \ac{ERO} programme of the Perseus cluster to study the impact of the perturbing process on the galaxy properties down to scales of 50\,--\,100\,pc. The data are first used to identify the dominant perturbing mechanism, hydrodynamical versus gravitational, since the two processes have different effects on the stellar component at very low surface brightness levels, such as those reached by \Euclid. We then compare the observed colours of the filamentary structures formed during the interaction to date their epoch of formation with the purpose of posing strong observational constraints on models and simulations. Finally, we can discuss the possible impact that \Euclid can have on local studies of galaxy evolution in rich environments thanks to its extraordinary image quality in terms of sensitivity to \ac{LSB} emission and angular resolution. 
 

\subsection{The density and colour profiles of the far outskirts of galaxies}

The high spatial resolution of \Euclid offers an unprecedented view of the Perseus cluster with a \ac{FOV} large enough to encompass a wide range of environments, from field galaxies to groups and galaxies in the cluster itself, each harbouring massive disc galaxies with stellar masses exceeding $10^{10}\,M_{\odot}$. \Euclid's advanced \ac{VIS} and \ac{NISP} cameras provide us with a unique opportunity to explore the detailed stellar structures of the galactic outer discs, thanks to their depth in the optical and \ac{NIR} domains. The depth and uniformity of the imaging coverage is unequalled for a survey covering both a cluster and field galaxies. Traditionally, the luminosity profiles of disc galaxies have been described by a single exponential \citep{deVaucouleurs1958}, but over the years observations have been challenging this first model, revealing a more complex reality characterised (at the next level of approximation) by double exponentials \citep{Erwin2005,Pohlen_Trujillo2006}. Simulations suggest that truncated profiles may stem from internal dynamical processes within the discs, such as Lindblad resonance or star-formation thresholds \citep{Kennicutt2001}, while the origins of anti-truncations remain elusive. These outer disc profiles serve as cosmic fingerprints, offering insights into their late-stage assembly processes \citep{Guiterrez2011} and the influence of dark matter halos, sculpted by the surrounding cosmic environment. 

The analysis of \ac{ERO} images of the Perseus cluster -- which lies within the gigantic Perseus-Pisces filament -- unveils a spectrum of profile types, from truncated to anti-truncated, each providing clues about the disc's composition and evolution. Preliminary studies conducted on clusters like Coma \citep{Head2015} and Virgo \citep{Erwin2012}, where evolved discs abound, have yielded ambiguous conclusions regarding the environmental influences. Within the framework of the \ac{ERO}, \Euclid's resolution and its large field of view offer a promising avenue for pilot studies, providing limited yet sufficient statistics to delve into the intricate interplay between galaxies and their cosmic surroundings. 


\begin{figure*}[htbp!]
\centering
\includegraphics[width=1.0\textwidth]{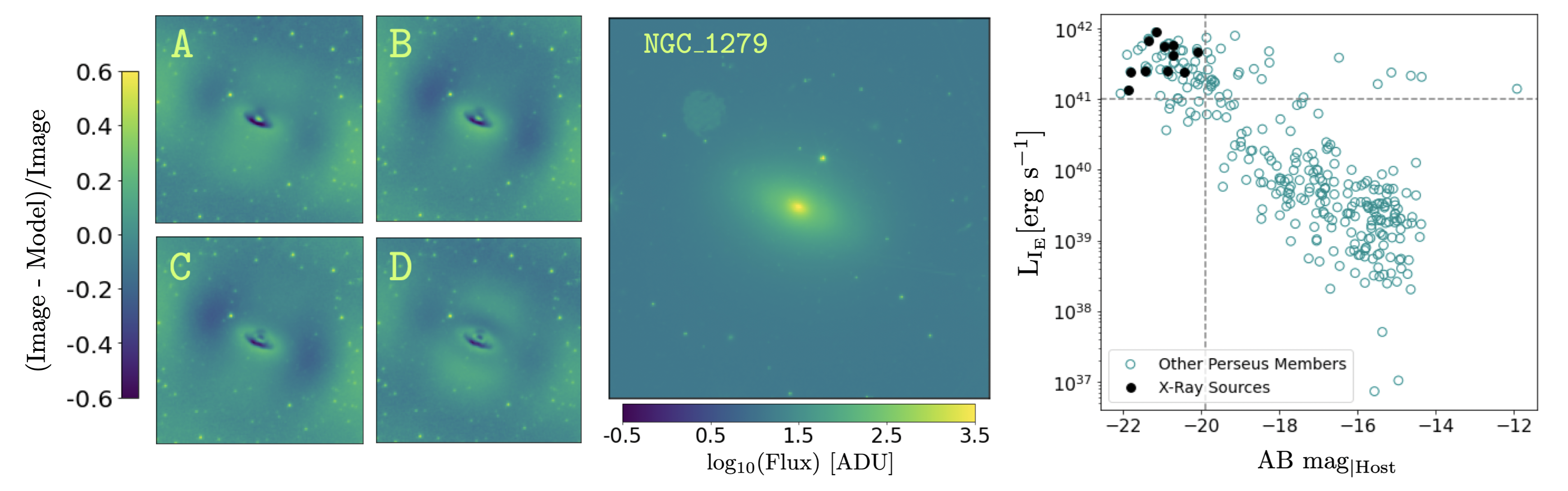}
\caption{\textit{Left and middle}: example of AGN decomposition, performed with the new package \texttt{GingaFit}, featuring the NGC\,1279 X-ray source galaxy. Fitting residuals from single S\'ersic, double S\'ersic (bulge+disc), PSF+S\'ersic, and PSF+double S\'ersic models are labelled from A to D. The residuals, shown in linear scale, are normalised to the original image and zoomed-in to show the central area of the galaxy, enclosed in half the Kron Radius. \textit{Right}: \IE AGN luminosity versus total AB magnitude of the host. The sub-populations of objects identified with properties that are similar to the X-ray sources are identified within the boundaries motivated by the faintest of the X-ray sources.} 
\label{fig:ngc1279}
\end{figure*}

\subsection{Analysis of the intracluster light and intracluster globular clusters}

\ac{ICL} is a ubiquitous feature in galaxy clusters \citep[e.g.,][]{Feldmeier2004, Kluge2020, golden-marx2023}. Its assembly history is central to understanding the global evolution of the cluster galaxy population, while it can also be a tool to infer the radius of the cluster and even its dark matter distribution \citep[e.g.,][]{MT19, Alonso-Asensio2020, Gonzalez2021,  Yoo2022, ContrerasSantos2024}. 

In \citet{EROPerseusICL}, we make a comprehensive study of the \ac{ICL} and \ac{ICGC} in the Perseus \ac{ERO} programme, mapping their distribution and properties out to a radius of 600\,kpc from the \ac{BCG}. We estimate that the central 500\,kpc of the Perseus cluster is host to $70\,000\pm2800$ globular clusters (\ac{BCG}+\acp{ICGC}), with half of them situated at a distance greater than $(239\pm16)$\,kpc, as well as $1.6\times10^{12}\,L_{\odot}$ of \ac{BCG}+\ac{ICL} in \HE. We use various properties of the \ac{ICL} and \acp{ICGC} to determine that these intracluster stellar components are most likely to have been tidally stripped from the outskirts of massive elliptical galaxies and from the disruption of dwarf galaxies.
The \ac{ICL} and \acp{ICGC} share a coherent spatial distribution, and we find that the \ac{ICL} and \ac{ICGC} contours on the largest scales are not centred on the \ac{BCG}, but rather are offset westwards of the \ac{BCG} core by approximately 60\,kpc, in the direction of several bright cluster members. 


\subsection{Galaxy morphology from bulge and disc decomposition of galaxies at z<1}

The Hubble sequence has been observed to be in place as early as a billion years after the Big Bang \citep{Ferreira-2023-hubble-seq-JWST}. Thanks to its high spatial resolution and wide field of view, \Euclid will enable the morphological analysis of billions of galaxies out to redshifts $z \simeq 3$ \citep{Euclid-2022-morpho-forecast}, and the reconstruction of the history of the Hubble sequence over most of the age of the Universe. 

The \ac{ERO} Perseus data allow us to test the ability of \Euclid images to characterise galaxy morphology for galaxies, mostly in the cluster background, out to redshifts of $z \sim 1$. In this study, we perform multi-band luminosity model-fitting of the \ac{VIS} and \ac{NISP} images with the \texttt{SourceXtractor++} software \citep{Bertin-2020-SourceXtractor-plus-plus, Kummel-2022-SE-use},
 whose efficiency was demonstrated in the \Euclid Morphology Challenge \citep{Euclid-2023-morpho-challenge-morpho}. We perform and analyse the results of several model-fitting runs, using either a single S\'ersic profile \citep{Sersic-1963-sersic-model}, or decomposing galaxies as the sum of a S\'ersic bulge and an exponential disc. Examining multi-variate distributions of galaxy, bulge and disc properties, we examine the consistencies and biases between the various modelling configurations, and we assess the reliability of the derived structural parameters in these first \Euclid data, in the light of the results obtained by the \Euclid Morphology Challenge for synthetic galaxy images \citep{Euclid-2023-morpho-challenge-morpho}. We are then able to provide robust model-fitting photometry and measure biases between adaptive aperture, single S\'ersic, and bulge-disc model photometry, depending on the morphology of the galaxies.

Finally, to build upon the results of \citet{Quilley-2022-bimodality} that highlighted the Hubble sequence as an inverse evolutionary sequence, along which bulge growth and disc reddening are indicators of the quenching of galaxies, we derive preliminary variations of bulge and disc fluxes, colours, and sizes as a function of redshift out to $z\sim1$.


\subsection{Probing \texorpdfstring{\ac{AGN}}{AGN} activity and host galaxy morphology}
 
The Perseus cluster provided the first and most spectacular example to date of radio-mode \ac{AGN} feedback from a \ac{BCG} on the surrounding intracluster medium \citep{fabian2003, fabian2006, sanders2020, reynolds2021, veilleux2023}. Such feedback is expected to be ubiquitous among galaxy clusters throughout the Universe, but is challenging to identify in large numbers. The high-resolution imaging of the \ac{EWS} opens up the possibility of teasing out point-source \ac{AGN} emission from host galaxies, and thereby building a census of \ac{AGN} activity in the \Euclid survey. We explore the reliability of this \ac{AGN}-host decomposition on the members of the Perseus cluster, including the hosts of 13 known X-ray point sources \citep{Santra2007}, which serve as a reference sample.
For the structural decomposition \citep{kim2008, bruce2016, 2019ApJS..245...10W, peng2010} we carefully account for sources of bias, making sure that our measurements are robust against degeneracies in the multi-dimensional parameter space of galaxy structural properties \citep{tarsitano2018}. We perform multiple fits of increasing complexity, combining one or multiple S\'ersic profiles and \ac{PSF} components, using our newly developed python package named \texttt{GingaFit}.\footnote{\url{https://github.com/AstroFederica/GingaFit/tree/main}}

In the fitting residuals (as those shown in the left panels of \cref{fig:ngc1279}), both the single S\'ersic (panel A) and the bulge+disc (panel B) models leave a central concentration of light and structure at large radii, arising from the fact that galaxies do not have smooth, axi-symmetric light profiles. In each case, the addition of a point-source component significantly improves the fitting model, whilst also accounting for the point-source residual. In the right panel of \cref{fig:ngc1279}, we show the implied point-source luminosity for each cluster member galaxy, integrated over the \IE band, against their host's absolute magnitudes from the bulge and disc components. We are able to isolate two sub-samples, one within the same luminosity-magnitude range of the X-ray sources, and another low-mass sample with high luminosity. The measurement reliability and nature of the emission of all luminous point sources will be determined in a future paper.


\begin{figure}
\centering
\includegraphics[width=0.9\columnwidth]{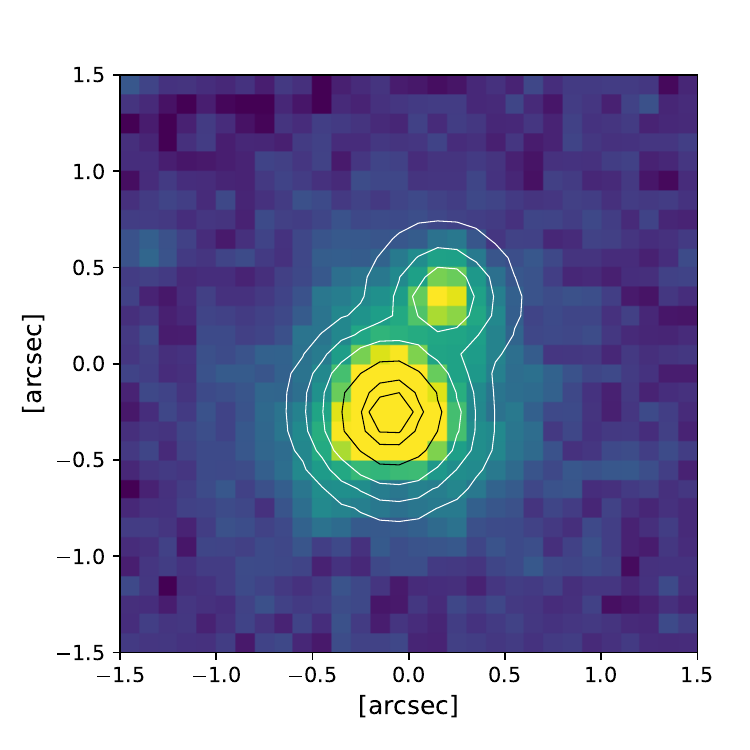}
\caption{VIS image of the dual lensed AGN candidate SDSS~J032023.93+412131.5 at $z=2.811$, showing the nearby companion at \ang{;;0.65} of separation. The contours show two fitted PSFs.} 
\label{fig:dual}
\end{figure}

\subsection{Dual/lensed QSO candidates at sub-arcsec separations}

During their life, galaxies experience a number of major and minor mergers with other galaxies. The supermassive black holes (SMBHs) present in the centres of most massive galaxies merge together with their hosts, creating a SMBH pair that slowly reduces its separation and eventually merges, creating a burst of gravitational waves 
\cite[GW; e.g.,][]{Colpi2014}.
During the in-spiralling phase, which can last gigayears \citep[e.g.,][]{Volonteri2022}, these pairs can be detected as dual AGN at kiloparsec separations.

The appearance of two or more QSOs at the same redshift and close separations can also be due to strong gravitational lensing of a single QSO due a foreground galaxy, producing multiple images of the background source. These are systems or great relevance in many fields of the astrophysics \citep[e.g.,][]{Shajib2022}.

Detecting these rare systems of dual or lensed QSOs is challenging because it requires high resolution imaging over large parts of the sky, and only a small number of confirmed such systems have been discovered, especially outside the local universe \citep{DeRosa2020,Mannucci2023}. Recently, several techniques based on the ESA \textit{Gaia} satellite have been proposed to select homogeneous samples of dual/lensed AGN candidates at sub-arcsec separations \citep[e.g.,][]{Lemon2017,Hwang2020,Mannucci2022}.
With its large FoV, superb spatial resolution, and high sensitivity, the \Euclid VIS camera can outperform the previous techniques based on optical data and provide large samples of dual/lensed AGN candidates, i.e., AGN with nearby companions. 

One such system have already been discovered by examining the eight known QSO with $z>0.1$ covered by the VIS image of the Perseus cluster. One of them, SDSS~J032023.93+412131.5, a QSO at $z=2.811$
discovered by the \ac{SDSS} \citep{Lyke2020},
shows a nearby companion at $\ang{;;0.65}$ of separation and 3.1\,magnitude fainter than the primary component, see \cref{fig:dual}.
This system in unresolved in the \Euclid NISP image, characterised by a much lower spatial resolution. Preliminary high-resolution \ac{NIR} imaging with LBT (Ulivi et al., in prep), has revealed that the two objects have very similar optical-to-nearIR colours, indicating that the faint component is likely to be a secondary QSO in either a dual or a strongly lensed system. In the former case, the projected separation between the two component would be $\sim5.2$\,kpc.


\subsection{Properties of the interstellar medium: albedo derivation from the optical to the NIR}

The \ac{ERO} Perseus data allow us to test the idea of using the \Euclid images to study the structure of the Galactic interstellar medium (ISM) as well as the properties of interstellar dust. These micron-size solid particles absorb starlight, producing absorption, $A_V$, and reddening, $E(B-V)$, in the visible (and \ac{NIR}), as well as a wide emission spectrum extending from the mid-IR to  millimetre wavelengths. These signatures have been extensively used to understand the nature of interstellar dust, and also as tracers of the ISM structure. The underlying idea is that dust and gas are well mixed, in every phase of the \ac{ISM} \citep{1988ApJ...330..964B}. 

In addition to absorption, a significant fraction of the starlight impinging on micron-size dust grains is scattered off their surfaces. Given their size and shape, the scattering is observed in the UV through the \ac{NIR}, with an expected maximum around $\lambda=1\,\mu$m \citep{2004ASPC..309...77G,2017A&A...602A..46J}. This scattered light, historically dubbed diffuse Galactic light (DGL) has been known for several decades \citep{1937ApJ....85..213E,1976AJ.....81..954S}, but it is only recently that it has been used, like absorption and emission, as a tracer of interstellar structure \citep{2015A&A...579A..29B,2016A&A...593A...4M,2018A&A...617A..42M,2020arXiv201201465R,Liu2023}. An important difference with emission and absorption is that the scattered light amplitude depends on the 3D structure of the radiation field because stellar photons are predominantly scattered in the forward direction. Therefore, in regions with very anisotropic radiation fields, the scattered light image will give a distorted picture of the interstellar density field. Fortunately most of the sky at high Galactic latitude is composed of diffuse interstellar clouds that are not forming stars and are illuminated by the general interstellar radiation field that can be considered to be close to isotropic. In this case, large area mapping of diffuse areas of the sky in the optical and \ac{NIR} with very high angular resolution offers the opportunity to map the ISM structure down to AU scales over hundreds of parsecs. 

The Perseus cluster field is a good test case for this. It is at intermediate latitude ($b=-13^\circ$) with a moderate column density of $N_{\rm H} \simeq 1.4 \times 10^{21}\, {\rm cm}^{-2}$ using the $E(B-V) \simeq 0.16$ estimate of \citet[][]{2014A&A...571A..11P} in this area and the $E(B-V)/N_{\rm H}$ conversion factor of \citet{2017ApJ...846...38L}. This corresponds to the translucent part of the interstellar medium ($A_V \simeq 0.5$) where matter is expected to be mostly atomic. According to the THEMIS dust model \citep{2017A&A...602A..46J}, the dust scattering should be maximum in the \Euclid \JE band. However, these theoretical estimates are based on very little data \citep[see the review by][]{2004ASPC..309...77G} and the exact shape of the scattered light spectrum should vary locally, since it contains information about dust shape, composition, and sizes. The combination of \Euclid's four bands, coupled with deep complementary optical $u$, $g$, $r$, $i$, and $z$ ground-based \ac{CFHT} data, provides a powerful wavelength range to study the DGL. In particular, combining scattering and emission offers a great opportunity to improve current dust models \citep[e.g.,][]{Zhang2023}.



\section{\label{appendix:Completeness}Completeness of the main catalogue}

\begin{figure}[htbp!]
\centering
\includegraphics[width=0.49\textwidth]{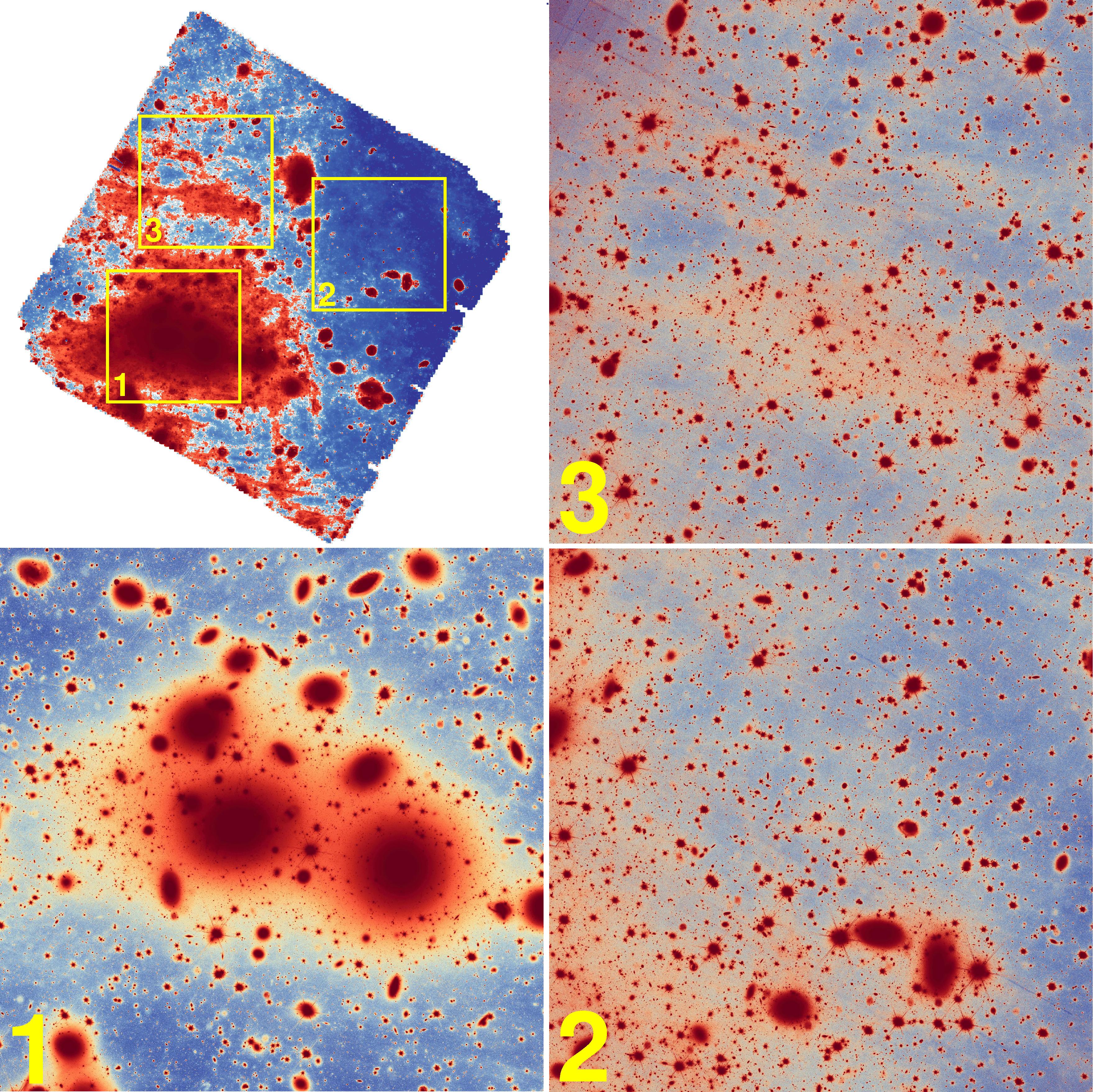}
\caption{The Perseus field is divided into three regions 10\,000\,pixels\,$\times$\,10\,000\,pixels each ($\ang{;;1000}$ on a side) to investigate completeness under various conditions: Region~1, large galaxies, Region~2, regular field density composed of background galaxies and stars; and Region~3, Galactic cirrus. The Galactic cirrus manifests itself as a faint orange haze in frame 3, with the top-left thumbnail displaying an enhanced version of the image that highlights extended diffuse emission, cirrus, and intracluster light. Collectively, these three representative regions cover $44\,\%$ of the total area. In all images, north is oriented upwards and east to the left.}
\label{fig:3CompletenessRegions}%
\end{figure}

We derived the main catalogue of bright and dwarf galaxies through exhaustive visual scrutiny of the entire data set, initially detecting galaxies from the high-resolution \ac{VIS} \IE-band image and using the VIS+NISP colour images for additional verification \citep{EROPerseusDGs}. The manageable size of our \Euclid \ac{FOV} for the Perseus cluster allows this approach over a fully automated method \citep[e.g.,][]{2020ApJ...890..128F}. This method ensures a comprehensive review of cluster galaxy candidates, yielding a nearly complete sample, although mastering the completeness function is essential to correct the LF. 

Completeness loss occurs due to the finite depth of the data set or the complex nature of the observed field, where high star density and dense interstellar matter, along with large, bright cluster galaxies, can obscure fainter galaxies. About $10\,\%$ of the \Euclid Perseus field is occupied by bright sources, complicating the detection process. 

To analyse completeness, we simulated the injection of dwarf galaxies into the original \Euclid \IE image. We selected three square areas, each \ang{;1000;}\,$\times$\,\ang{;1000;} (or 10\,000\,pixels across), representing a significant $44\,\%$ of the whole image, while encompassing the three most diverse environments: one centred on the cluster core rich in large galaxies (Region~1); one in an area heavily affected by Galactic cirrus (Region~3); and one relatively free of these complexities, but with similar stellar density (Region~2). \Cref{fig:3CompletenessRegions} illustrates these regions, highlighting the variability in environmental conditions that affect light detection.

\begin{figure}[htbp!]
\centering
\includegraphics[width=0.49\textwidth]{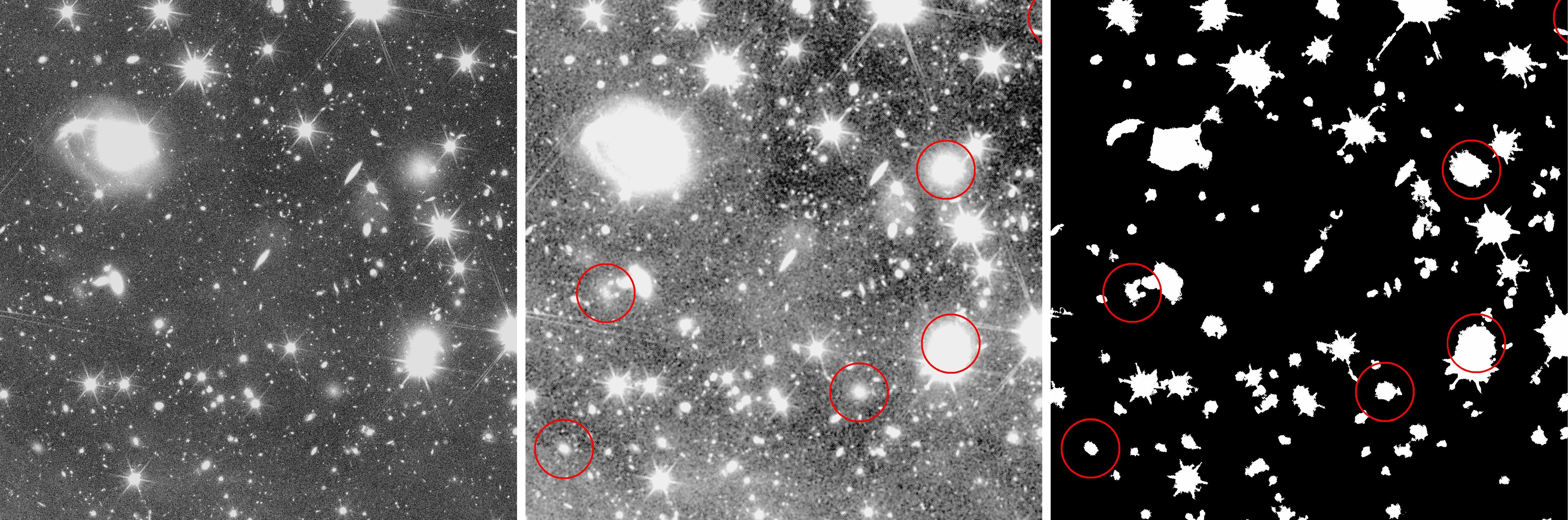}
\caption{\textit{Left}: Section of the original VIS image (900\,pixels\,$\times$\, 900\,pixels, \ang{;;90} on the side) displaying standard features, specifically extended cluster galaxies, stars, background galaxies, and optical ghosts near bright stars appearing as faint round structures. \textit{Middle}: Ring-filtered version enhancing extended sources and removes compact ones, with red circles (\ang{;;10} in diameter) marking dwarfs from our main catalogue. \textit{Right}: \texttt{SourceExtractor} segmentation map confirming detection of all dwarfs, maintaining a low count of other detected sources.} 
\label{fig:DwarfsRingFilterSegmentation}%
\end{figure}

The initial step involved selecting an automated detection method capable of accurately identifying the majority of galaxies in the main catalogue while excluding most sources that do not share their morphological characteristics. This strategy ensures that the detection method can be confidently used to identify injected simulated galaxies without mistaking them for other compact sources.

Testing different image-filtering techniques revealed that the ring filter, recommended by \citet{Secker1995} and \citet{2020ApJ...890..128F} for this specific task, is near to optimal.\footnote{This is effectively a low-pass spatial filter for removing \ac{PSF}-sized sources.}  We utilised the background map created by the \ac{ICL} team \citep{EROPerseusICL} aimed at detecting compact sources. The parameters for the ring filter were set with an inner radius of 2\,pixels and an outer radius of 4\,pixels, with the \ac{VIS} \ac{FWHM} at 1.6\,pixels. The resulting smoothed background map effectively meets our requirements for detecting faint extended structures.

\begin{figure*}[htbp!]
\centering
\includegraphics[width=0.7\textwidth]{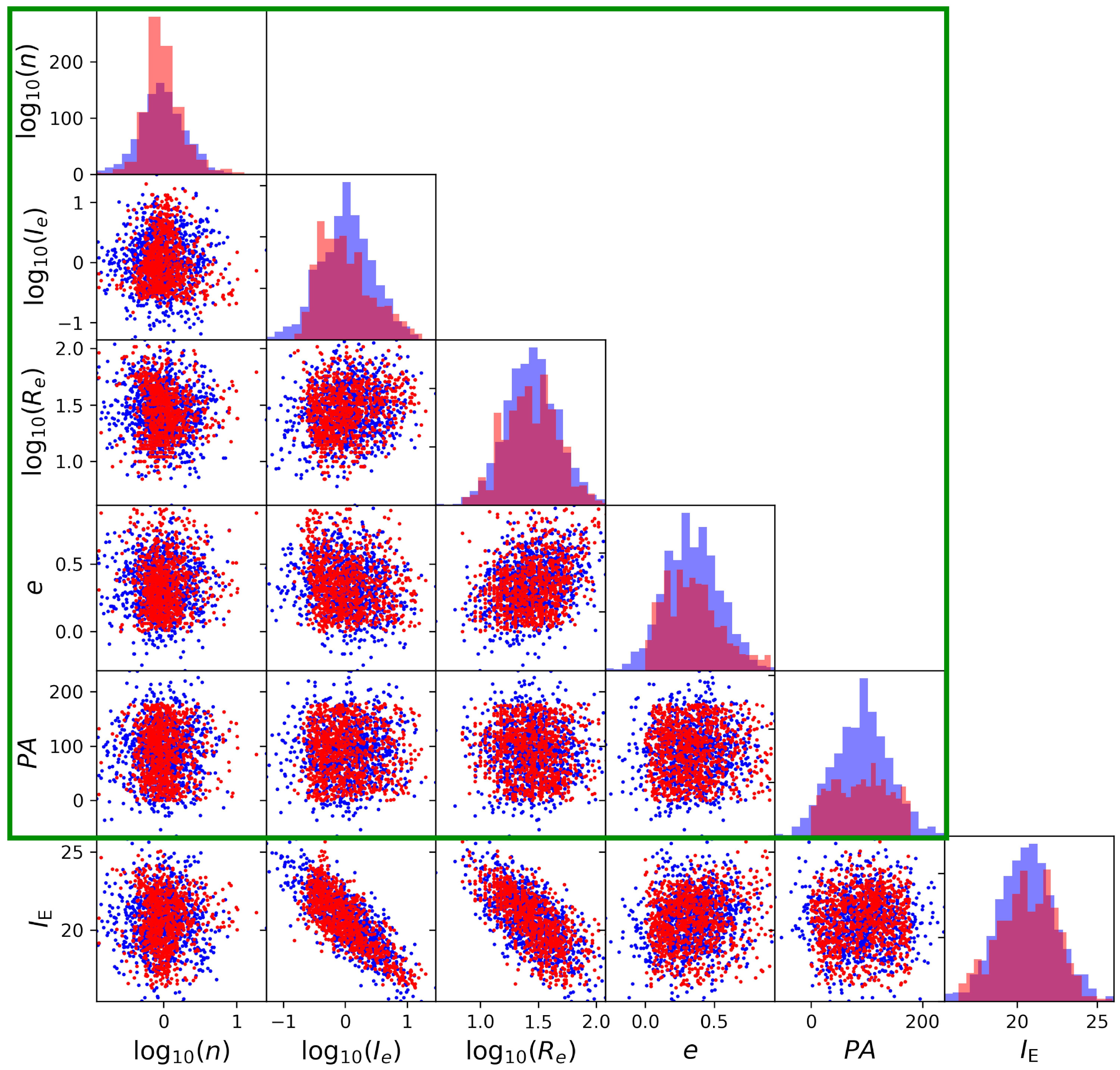}
\caption{Fits to model parameters for dwarf galaxies.  A multivariate Gaussian fit (blue dots) was applied to the dwarf sample from the main catalogue (red dots), ensuring the five defining parameters per galaxy -- S\'ersic index $n$, surface brightness $I_\mathrm{e}$ at the effective radius $R_\mathrm{e}$, ellipticity $e$, and position angle ${\rm PA}$, indicated in the green box -- are weighted proportionally to their occurrence in the real data. The total magnitude \IE, dependent on S\'ersic index, $I_\mathrm{e}$, and $R_\mathrm{e}$, is not used in the simulation.}
\label{fig:SimulationParameters}
\end{figure*}

Visual inspection showed that all faint galaxies ($\IE\,{>}\,16$) were significantly enhanced. \texttt{SourceExtractor} \citep{1996A&AS..117..393B} was then configured to detect only objects exceeding a size of 500 pixels (DETECT$\_$MINAREA), with a detection threshold above $2\,\sigma$ (DETECT$\_$THRESH), and using a 64-pixel mesh size (BACK$\_$SIZE), along with a moderate smoothing filter ($\mathrm{BACK\_FILTERSIZE}=3$) for internal background subtraction.

This approach achieved a recovery rate of 93\,\% for all faint galaxies in our catalogue, which have a limited angular size on the sky and remain unaffected by the internal sky background subtraction -- ensuring they are not partially erased. \Cref{fig:DwarfsRingFilterSegmentation} demonstrates these steps, from the original image through the ring-filtered version to the \texttt{SourceExtractor} segmentation map, confirming the effective detection of all dwarf galaxies.

While the \texttt{SourceExtractor} measurements are not as precise as those from \texttt{AutoProf}/\texttt{AstroPhot} for determining the total magnitude of these peculiar objects -- being within 0.5\,magnitude of accuracy -- they still enable a fully automated method for testing completeness. This is done by injecting simulated galaxies and verifying how many are successfully identified in the extracted catalogue.

To ensure comprehensive statistics, we inject each simulated galaxy onto a grid every \ang{;;25} in both the $x$ and $y$ directions across our observational regions, resulting in 400 injections per region, totaling 1200 injections for each simulated galaxy. This process was replicated for 817 galaxies spanning nine magnitude bins from $\IE=16$ to $\IE=24$. Initially, we analyse the primary physical parameters of the dwarfs in our main catalogue on the \IE image: S\'ersic index; effective radius $R_{\rm e}$; surface brightness at $R_{\rm e}$ ($I_{\rm e}$); ellipticity; position angle; and total magnitude \IE \citep{EROPerseusDGs}. An average of 90 galaxies per magnitude bin was sampled using a multivariate Gaussian fit to reflect the actual distribution of these parameters among the dwarf population.

\Cref{fig:SimulationParameters} illustrates the correlations between these six parameters in red. While the total magnitude \IE, which is closely linked to the S\'ersic index $n$, $I_{\rm e}$, and $R_{\rm e}$, was included for demonstration only, and was not directly used in the simulations. Instead, it was derived from $n$, $I_{\rm e}$, $R_{\rm e}$, and the ellipticity. The blue dots in \cref{fig:SimulationParameters} represent the randomised parameters drawn to accurately depict the actual dwarf population in the Perseus cluster.

\begin{figure}[htbp!]
\centering
\includegraphics[width=0.49\textwidth]{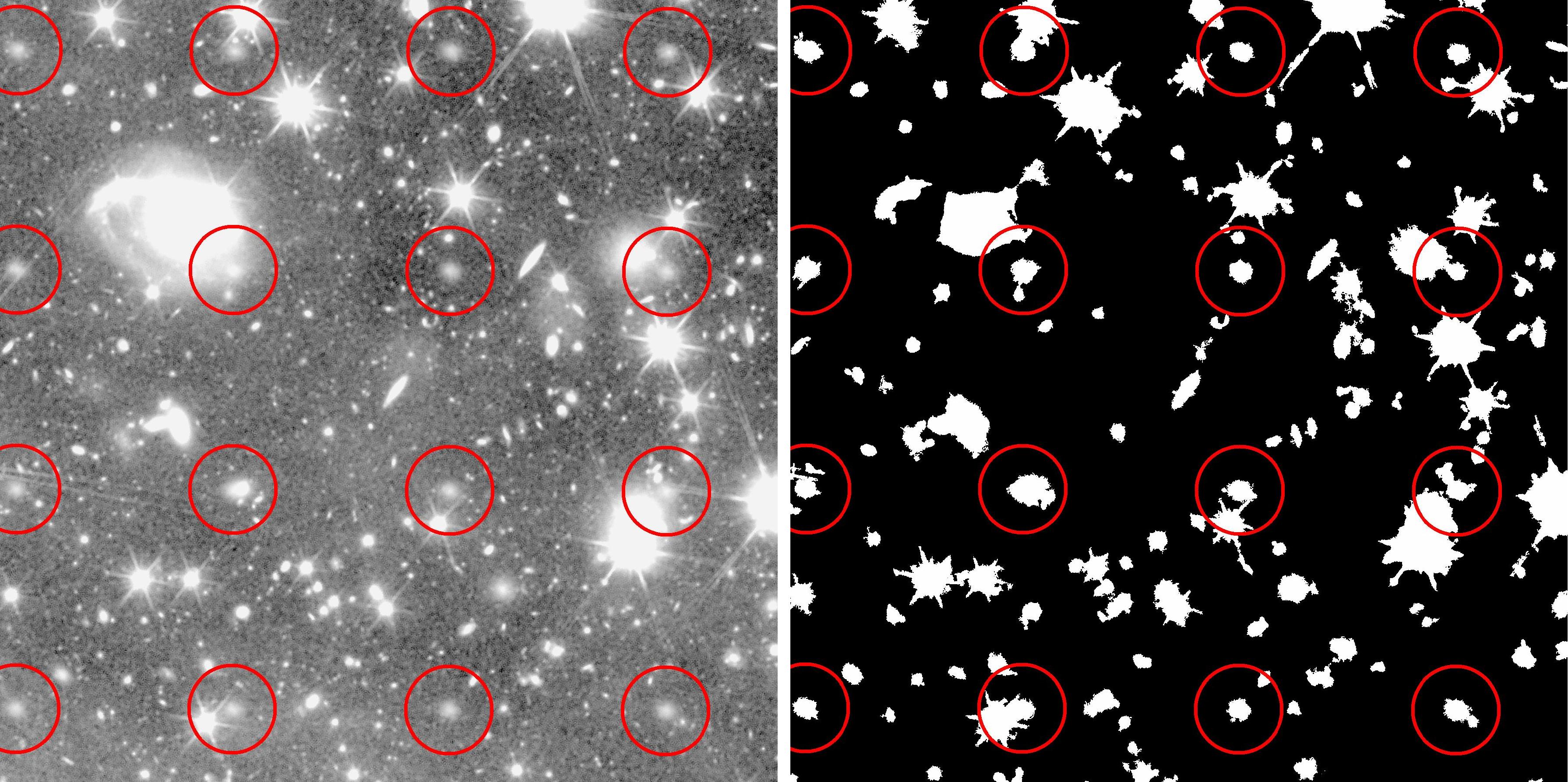}
\caption{Injection of simulated galaxies. \textit{Left}: A simulated $\IE=22.4$ dwarf galaxy with specific physical parameters (S\'ersic index = 0.61, $I_{\rm e} = 0.49\, \mathrm{ADU\, per\, pixel}$, $R_{\rm e} = 18.10\, \mathrm{pixels}$, ellipticity = 0.14, and position angle = \ang{72;;} is injected into a 250\,pixel\,$\times$\,250\,pixel grid (marked by red circles). \textit{Right}: The \texttt{SourceExtractor} segmentation map shows successful recovery of most injected sources. This example clearly demonstrates how completeness is compromised by crowding from nearby stars and galaxies. The image area is the same as in \cref{fig:DwarfsRingFilterSegmentation}.}
\label{fig:SimulationsRingFilterSegmentation}%
\end{figure}

\begin{figure}[htbp!]
\centering
\includegraphics[width=0.49\textwidth]{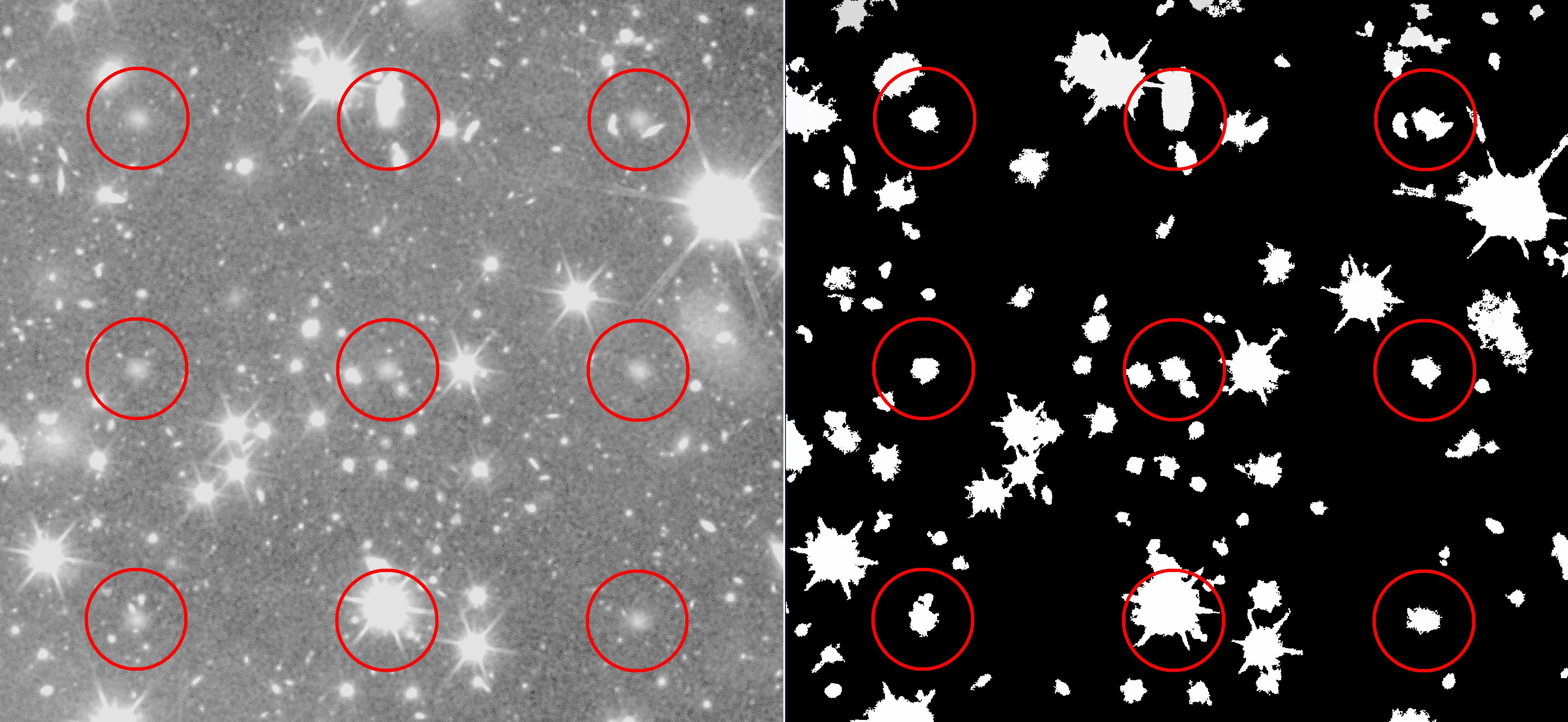}
\caption{In this \ang{;;75} wide area featuring nine injected dwarf galaxies (the same as shown in \cref{fig:SimulationsRingFilterSegmentation}), two are not recovered: the centre top red circle overlaps with a background galaxy, and the centre bottom overlaps with a bright star. The segmentation map on the right displays a single polygon at the location of each simulated dwarf, highlighting where recoveries succeeded and where they were hindered by overlapping objects.}
\label{fig:FailedRecovery}%
\end{figure}

We utilise the image simulation tool \texttt{makeimage} from the \texttt{Imfit} package \citep{2015ApJ...799..226E} to generate 817 250\,pixel$\times$250\,pixel stamps based on our five drawn parameters. The largest dwarfs in our main catalogue do not exceed \ang{;;20} in diameter, as highlighted in the sub-panel for $R_{\rm e}$ in \cref{fig:SimulationParameters}. 

Each stamp is replicated 400 times in a $20\times20$ grid and inserted into each of the three ring-filtered regions to streamline the process and conserve computing resources. This method is effective since these stamps, featuring flat and featureless profiles, are minimally impacted by the tight ring filter, with differences at the percent level.

There is no need to randomise the injection positions due to the automated detection scheme employed. \Cref{fig:SimulationsRingFilterSegmentation} demonstrates this process, where the \texttt{SourceExtractor} segmentation map shows that the simulated galaxies are typically successfully recovered (indicated by red circles). \Cref{fig:FailedRecovery} displays two instances where recovery failed due to overlaps with a bright star and a background galaxy.

The recovery level, which assesses completeness for each galaxy, is determined by automatically comparing the extracted catalogue with the values in a catalogue generated by \texttt{SourceExtractor} from simulations using the same parameters. We use a matching radius of \ang{;;1}, and a galaxy is considered lost if its recovered magnitude deviates by $5\,\sigma$ from the expected value, based on the dispersion observed across 400 measurements without injections.

Overall, 980\,400 simulated galaxies were processed by \texttt{SourceExtractor} to generate three completeness measurements per galaxy, corresponding to each of the three regions. 
To validate the method, we ran a matching code for each galaxy against the \texttt{SourceExtractor} catalogue produced for the three regions without any injected galaxies, ensuring that our tests were not influenced by existing sky data. 
The negligible matching level of $0.4\,\%$ confirms that the completeness values accurately reflect the ability to recover the injected simulated galaxies, free from background contamination.

\begin{figure}
\centering
\includegraphics[width=0.49\textwidth]{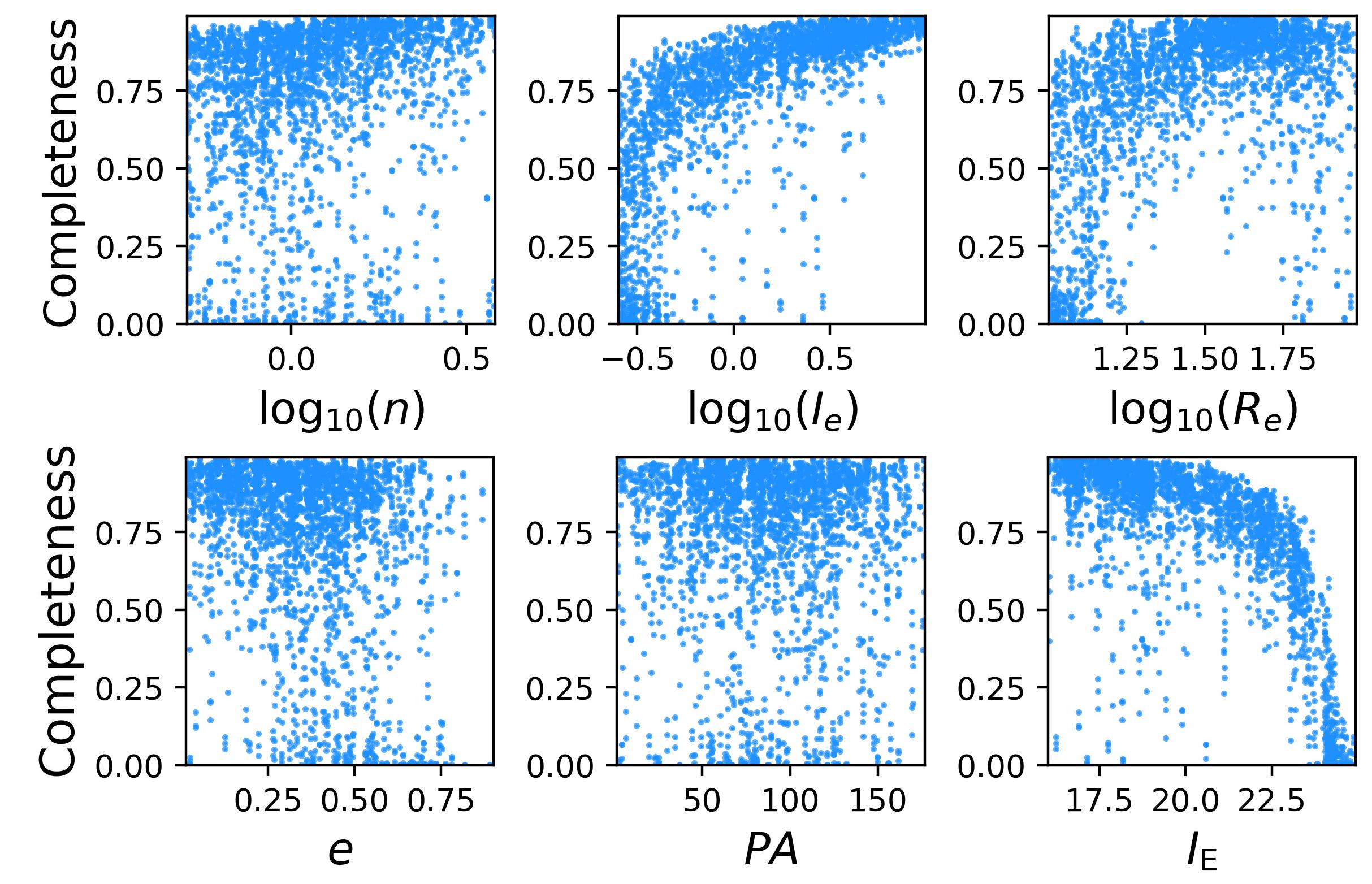}
\caption{Completeness covariance with respect to the main physical parameters. The total magnitude \IE, which derives from the S\'ersic index $n$, the effective radius $R_{\rm e}$, $I_{\rm e}$ the surface brightness at $R_{\rm e}$, and the ellipticity $e$, shows the tightest relation to completeness.}
\label{fig:CompletenessCovariance}%
\end{figure}

\begin{figure}[htbp!]
\centering
\includegraphics[width=0.49\textwidth]{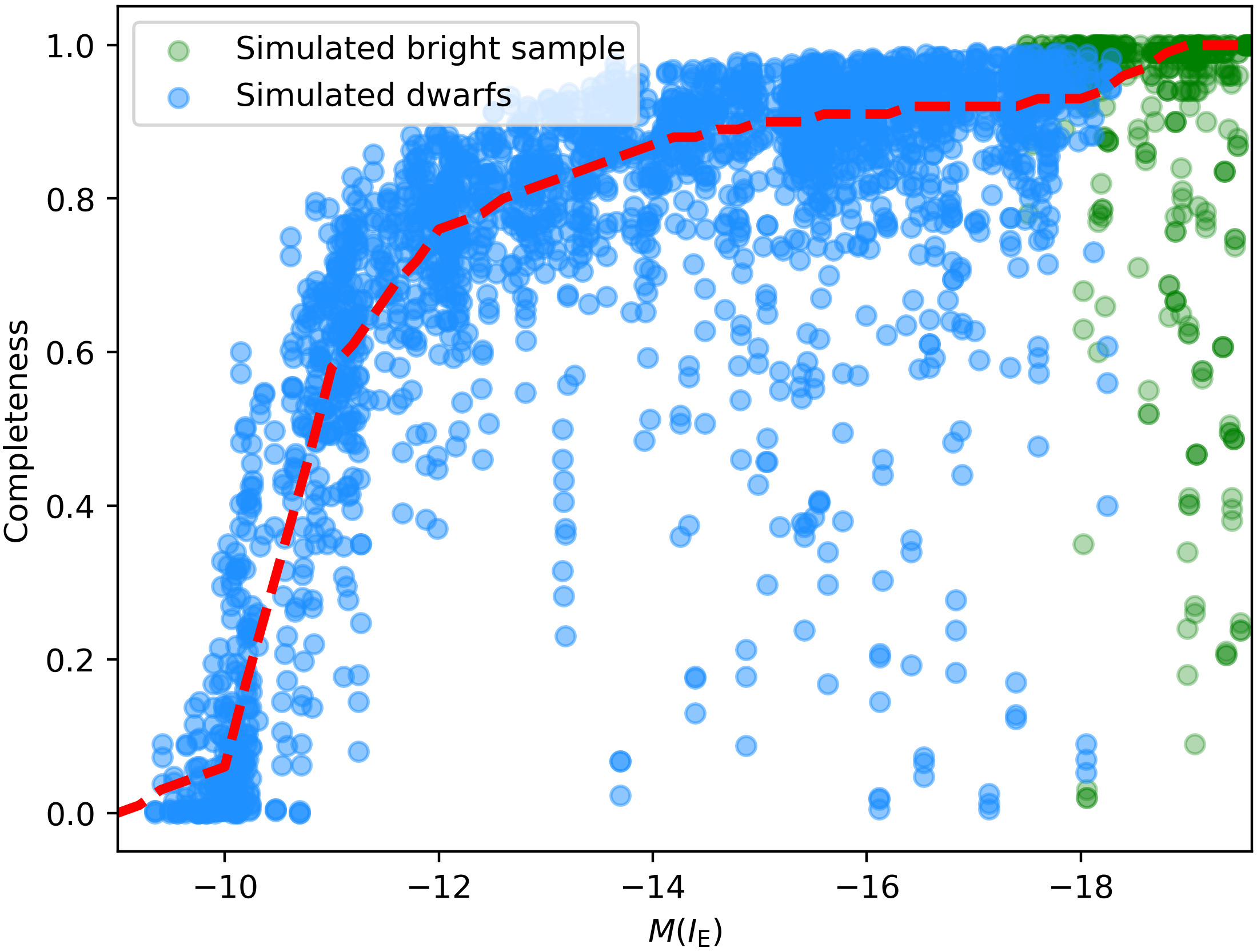}
\caption{Completeness estimates.  The completeness of the Perseus dwarf galaxies is depicted in blue, based on 980\,400 individual injections across three distinct Perseus regions, each containing 400 samples from each of the 817 simulated galaxies, resulting in 2451 completeness measurements shown on this plot. The dashed red median line shows that completeness begins to decline at $M(\IE)=-12$, dropping to 50\,\% at $-11$. The green data points illustrate that even the faintest of our bright galaxy samples achieve 96\,\% completeness when simulated similarly to the dwarfs, with completeness approaching 100\,\% at $M(\IE)=-19$. For brighter and larger galaxies, the testing method used here is not ideally suited, leading to an apparent drop in completeness (green dots on the right). However, our catalogue achieves 100\,\% completeness for all bright galaxies.}
\label{fig:Completeness}%
\end{figure}

The correlation of the 2451 measurements with our key parameters is detailed in \cref{fig:CompletenessCovariance}. Clear expected trends are evident with respect to the S\'ersic index, $R_{\rm e}$, and $I_{\rm e}$, yet it is the total magnitude that shows the strongest correlation. We delve deeper into this completeness relationship in \cref{fig:Completeness}, now focusing on absolute magnitudes.

A dashed red median line illustrates a slight decline in completeness from $M(\IE)=-18$ to $M(\IE)=-12$, beyond which completeness rapidly decreases, hitting 50\,\% at $M(\IE)=-11$ and nearly 12\,\% at $M(\IE)=-10$. Comparisons across the three regions (\cref{fig:3CompletenessRegions}) reveal that only the presence of large galaxies in Region\,1 slightly impacts completeness across all magnitude bins, reducing it by 7\%. In consequence we adopt a unique completeness correction across the entire \IE range.

The polynomial fit in \cref{fig:Completeness} incorporates all simulated dwarf measurements and thus represents the median completeness across the entire image.

The faintest members of the bright galaxies in our catalogue were tested using the same method as the dwarfs, with results depicted in green on \cref{fig:Completeness}. This approach is effective primarily for galaxies in the faintest magnitude bin at $M(\IE)=-18$, where galaxies are also more compact. To accommodate their size and prevent truncation, the simulation and injection grid was expanded from $250\times250$ to 1000\,pixels\,$\times$\,1000\,pixels.

This testing shows that completeness stands at 96\,\% for non-dwarf galaxies at $M(\IE)=-18$. However, completeness quickly decreases for brighter and more extended galaxies because our dwarf-focused measurement method unintentionally suppresses flux due to \texttt{SourceExtractor}'s background subtraction being influenced by the galaxy itself. This was visually confirmed, as even the faintest bright galaxies are distinguishable when near a star or adjacent to a larger galaxy.
Given the few faint non-dwarf galaxies in the $M(\IE)=-18$ bin, it is evident that a completeness correction for the luminosity function (LF) is unnecessary for the bright galaxies ($M(\IE) > -18$).




\section{Photometry pipeline comparisons}

In this paper, the photometry is carried out with \texttt{AutoProf} for the bright galaxies and a combination of \texttt{AstroPhot} and \texttt{Galfit} for the dwarf galaxies \citep{EROPerseusDGs}. Other photometry software codes are available for comparison. In \citet{ConnorAP}, \texttt{AutoProf} has been compared to {\texttt{PHOTUTILS}}, {\texttt{XVISTA}}, and {\texttt{Galfit}}. In this appendix, an additional comparison is made with the {\texttt{Archangel}} pipeline \citep{2007astro.ph..3646S,2012PASA...29..174S}  using the same specific \Euclid\ \ac{ERO} data. Previously, the {\texttt{Archangel}} software had been adapted for {\it Spitzer} space observations as described in detail in \citet{2012AJ....144..133S}. For the comparisons in this appendix, it was further modified to be applied to \Euclid\ data. Briefly, {\texttt{Archangel}} performs the masking of stars and flaws. Subsequently, it replaces masked regions by mean isophote values. Ellipses are then fit to isophotes with increasing radii. The 2D information is compressed into uni-dimensional surface brightness and magnitude growth curves. Other parameters such as extrapolated magnitudes, are finally derived. {\texttt{Archangel}} was run once on each galaxy of our test sample. We did not modify the parameters between elliptical, spiral, and dwarf galaxies so as not to bias the comparisons. In any case, \citet{2012AJ....144..133S} showed that the major contribution to the magnitude uncertainties for space-based observations is the setting of the sky level. Furthermore, \citet{2012AJ....144..133S} and \citet{2014MNRAS.444..527S} demonstrated the agreement between magnitudes obtained with the {\it Spitzer}-adapted version of {\texttt{Archangel}} and other alternative pipelines, such as the software developed for the GALEX Large Galaxy Atlas\footnote{https://sites.astro.caltech.edu/~neill/GALEX/glga/} \citep[GLGA,][]{2007galx.prop...91S} or for the Spitzer Survey of Stellar Structure in Galaxies \citep[S$^4$G,][]{2010PASP..122.1397S}. The sky setting was identified as the cause of any difference between a few outliers. 

\begin{figure}[htbp!]
\centering
\includegraphics[width=\linewidth]{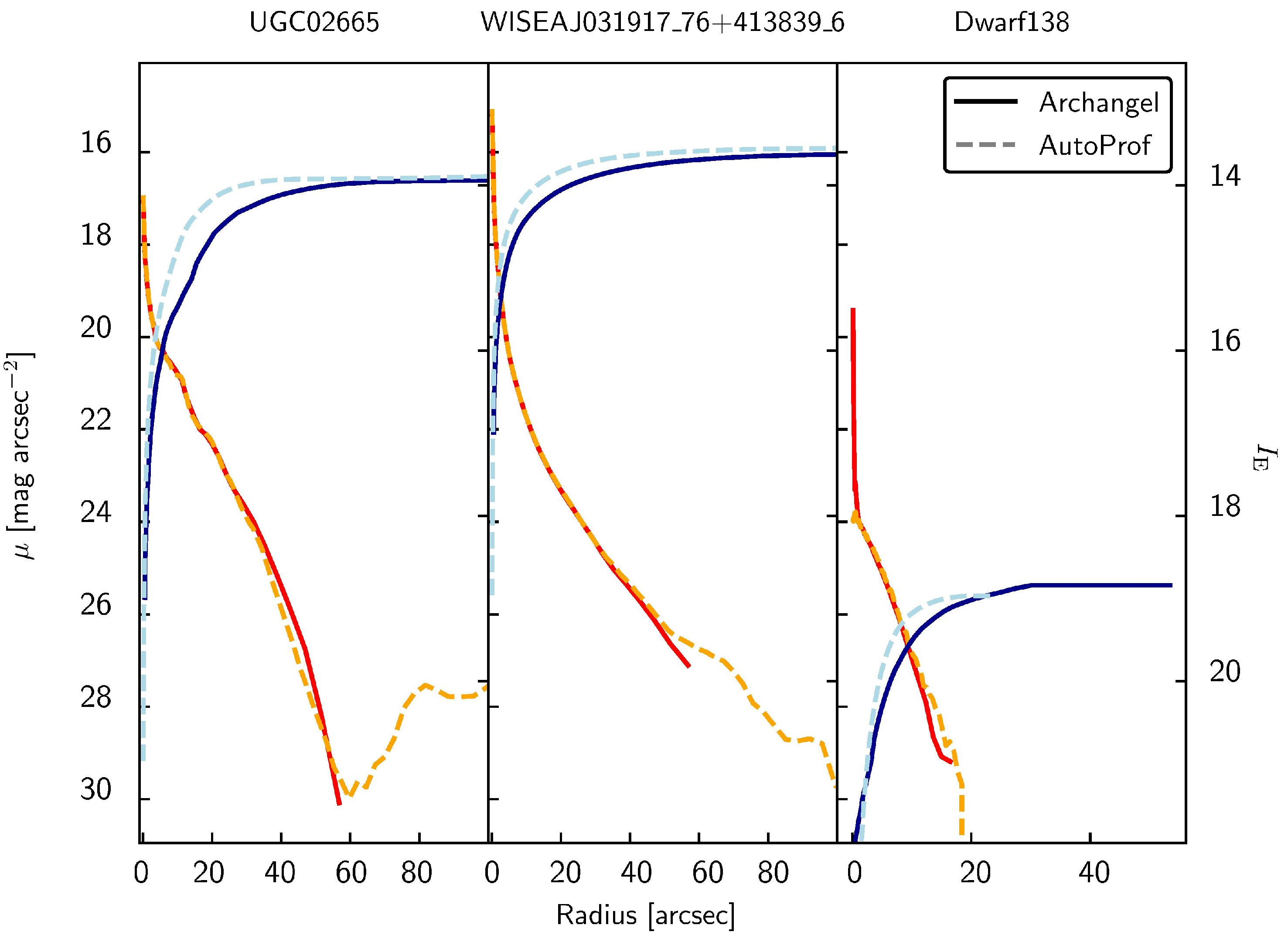}
\vspace{-0.4cm}
\caption{Surface-brightness profiles (red and orange lines, and left axis) and curves of growth (blue lines and right axis) obtained for three galaxies whose names are given at the top of each panel. The photometry has been performed with \texttt{AutoProf} (dashed lines) and \texttt{Archangel} (solid lines). There is a good agreement between both software codes.}
\label{fig:Arch-AutoP}%
\end{figure}

Following the main analysis of the paper, the zero point is set to 30.13 (30) for the $\IE$ ($\YE$, $\JE$, $\HE$) bands, the pixel size is \ang{;;0.1} and magnitudes are given in the AB system. In this appendix, detailed comparisons are presented for an elliptical, a spiral, and a dwarf galaxy, to emphasise the agreement between both software codes. \Cref{fig:Arch-AutoP} shows the surface brightness profile (blue solid and dashed lines) and the curve of growth (red solid and orange dashed lines) obtained for the three galaxies using \texttt{Archangel} (solid lines) and \texttt{AutoProf} (dashed lines). There is an agreement between the surface brightness profiles out to the edges of the galaxies. The curves of growth tend to increase faster with \texttt{AutoProf} but converge asymptotically to the same magnitude as those obtained with \texttt{Archangel}. Axis ratios and position angles derived with both codes are similar though. The curves of growth thus differ only by the radius definition. Table~\ref{tbl:Arch-AutoP} summarises the magnitudes in the four \Euclid\ bands obtained with both codes for the three galaxies, confirming the agreement.

\begin{table*}[htbp!]
\centering
\newcommand{\pd}{\phantom{1}}
\setlength{\tabcolsep}{3.25pt}
\caption{Magnitudes in the four \Euclid bands. Uncertainties on magnitudes derived by \texttt{Archangel} and \texttt{AutoProf} are held below $0.05$ but for the dwarf galaxy. Magnitudes obtained with both software codes are very similar, if not equal, and differences are held below a few percent.}
\label{tbl:Arch-AutoP}
\small
\begin{tabular}{l@{\hspace{4ex}}cccc@{\hspace{4ex}}cccc}
\hline\hline
\noalign{\vskip 1pt}
\omit & \multicolumn{4}{c}{{\texttt{Archangel}}}  & \multicolumn{4}{c}{\texttt{AutoProf}} \\
\hline
\noalign{\vskip 1pt}
 Name  &   \IE  &   \YE  &   \JE  &  \HE &   \IE &  \YE &   \JE  &   \HE \\
\hline
\noalign{\vskip 1pt}
UGC02665 (spiral) & 13.92  & 13.12 &  12.97  & 12.79 & 13.92  & 13.12   & 12.97  & 12.79 \\
WISEAJ031917\_76+413839\_6 (elliptical) & 13.61  & 12.78  & 12.65 & 12.47 & 13.57  & 12.97 & 12.59 & 12.41\\
Dwarf138 (dwarf) & 18.88  &  18.35 &  18.26  & 18.23 & 18.85 & 17.97 & 17.52 & 17.81 \\
\hline
\end{tabular}
\end{table*}





\section{\label{appendix:tables}Photometric tables of bright galaxies}

This section presents nine tables offering a complete overview of the catalog which will be available as supplementary material at the Strasbourg astronomical Data Centre (CDS):\\
\cref{tab:parameter_summary}: Complete list of parameters extracted for Tables~\ref{table:1} to \ref{table:2}, along with their descriptions and associated table indexes.\\
\cref{tab:parameter_summary2}: Complete list of parameters extracted for Tables~\ref{table:3} to \ref{table:4err}, along with their descriptions and associated table indexes.\\
\cref{table:1}: Primary data from VIS photometry.\\
\cref{table:1err}: Associated errors for the primary data catalogue.\\
\cref{table:2}: Redshift and extinction corrections.\\
\cref{table:3}: Magnitudes within one effective radius ($R_{\rm e}$) in VIS and NISP bands.\\
\cref{table:3}: Associated errors of the 1\,$R_{\rm e}$ magnitudes catalogue.\\
\cref{table:4}: Magnitudes using VIS photometry to constrain NISP photometry.\\
\cref{table:4err}: Associated errors of the forced magnitudes.\\

\begin{table*}
\caption{Summary table of parameters extracted from Tables~\ref{table:1} to \ref{table:2}, along with their descriptions and associated table numbers.}
\label{tab:parameter_summary}
\begin{tabular}{lll}
\hline\hline
\noalign{\vskip 1pt}
Parameter                     & Description                                                           & Table numbers     \\
\hline
\noalign{\vskip 1pt}
ID                           & Object identifier                                                     & \cref{table:1} to \cref{table:4err}\\
RA                           & Right ascension in degrees                                            &\cref{table:1} to \cref{table:4err} \\
Dec                           & Declination in degrees                                                & \cref{table:1} to \cref{table:4err} \\
x$_\mathrm{pix}$              & $x$-coordinate in pixels for AutoProf run                                                & \cref{table:1}                            \\
y$_\mathrm{pix}$              & $y$-coordinate in pixels for AutoProf run                                               & \cref{table:1}                            \\
\IE                           & Apparent magnitude in VIS band                                                    & \cref{table:1}                            \\
$M(\IE)$          & Absolute magnitude in VIS band                                                   & \cref{table:1}                            \\
$R_{\rm e}$            & Effective radius in pixels                                            & \cref{table:1}                            \\
$R_{\rm e}$         & Effective radius in arcseconds                                        & \cref{table:1}                            \\
$R_{\rm e}$            & Effective radius in kpc                                               & \cref{table:1}                            \\
$n$                       & Sersic index                                                          & \cref{table:1}                            \\
AR                    & Axis ratio                                                            & \cref{table:1}                            \\
PA                     & Position angle in degrees                                                       & \cref{table:1}                            \\
$\mu_{\IE,\rm{0,GD}}$            & Central surface brightness (Graham and Driver)                                       & \cref{table:1}                            \\
$\mu_{\IE,\rm{e,GD}}$           & Surface brightness at effective radius (Graham and Driver)                           & \cref{table:1}                            \\
$\langle \mu_{\IE,\rm{e,GD}}\rangle $        & Average surface brightness at effective radius (Graham and Driver)                   & \cref{table:1}                            \\
$\mu_{\IE,\rm{0}}$                          & Central surface brightness                                            & \cref{table:1}                            \\
$\mu_{\IE,\rm{e}}$                        & Surface brightness at effective radius                                & \cref{table:1}                            \\
$\langle\mu_{\IE,\rm{e}}\rangle$                 & Average surface brightness at effective radius                        & \cref{table:1}                            \\
$\sigma_{\rm x}$          & Error on $x$-coordinate in pixels                                       & \cref{table:1err}                         \\
$\sigma_{\rm y}$               & Error on $y$-coordinate in pixels                                       & \cref{table:1err}                         \\
$\sigma_{I_{\rm E}}$                & Error on apparent magnitude                                           & \cref{table:1err}                         \\
$\sigma_{M(\IE)}$        & Error on absolute magnitude                                            & \cref{table:1err}                         \\
$\sigma_{R_{\rm e}}$         & Error on effective radius in pixels                                   & \cref{table:1err}                         \\
$\sigma_{R_{\rm e}}$        & Error on effective radius in arcseconds                               & \cref{table:1err}                         \\
$\sigma_{R_{\rm e}}$         & Error on effective radius in kpc                                      & \cref{table:1err}                         \\
$\sigma_n$         & Error on Sersic index                                                 & \cref{table:1err}                         \\
$\sigma_{\rm AR}$      & Error on axis ratio                                                   & \cref{table:1err}                         \\
$\sigma_{\rm PA}$           & Error on position angle in degrees                                              & \cref{table:1err}                         \\
$\sigma_{\mu_{\IE,\rm{0,GD}}}$        & Error on central surface brightness (Graham and Driver)                              & \cref{table:1err}                         \\
$\sigma_{\mu_{\IE,\rm{e,GD}}}$        & Error on surface brightness at effective radius (Graham and Driver)                  & \cref{table:1err}                         \\
$\sigma_{\langle \mu_{\IE,\rm{0,GD}}\rangle}$     & Error on average surface brightness at effective radius (Graham and Driver)          & \cref{table:1err}                         \\
$\sigma_{\mu_{\IE,\rm{0}}}$             & Error on central surface brightness                                    & \cref{table:1err}                         \\
$\sigma_{\mu_{\IE,\rm{e}}}$           & Error on surface brightness at effective radius                       & \cref{table:1err}                         \\
$\sigma_{\langle\mu_{\IE,\rm{e}}\rangle}$     & Error on average surface brightness at effective radius               & \cref{table:1err}                         \\
$z_{\rm spec}$                         & Spectroscopic redshift from literature                                               & \cref{table:2}                            \\
$z_{\rm phot}$                            & SDSS or NED Photometric redshift                                                  & \cref{table:2}                            \\
$\sigma_{z_{\rm spec}}$         & Error on spectroscopic redshift                                       & \cref{table:2}                            \\
$\sigma_{z_{\rm phot}}$             & Error on photometric redshift                                         & \cref{table:2}                            \\
flagtool                      & Tool indicator (AP for AutoProf, APh for AstroPhot)                                                       &  \cref{table:2}                            \\
EC(\IE)                   & VIS extinction correction $c_{I_{\rm E}}\, E(B-V)$ from \cref{eq:MWext} to be used for all VIS magnitudes                                             &  \cref{table:2}                            \\
EC(\YE)                     & Y extinction correction $c_{Y_{\rm E}}\, E(B-V)$ from \cref{eq:MWext} to be used for all \YE magnitudes                                            &  \cref{table:2}                            \\
EC(\JE)                    & J extinction correction $c_{J_{\rm E}}\, E(B-V)$ from \cref{eq:MWext} to be used for all \JE magnitudes                                              &  \cref{table:2}                            \\
EC(\HE)                    & H extinction correction $c_{H_{\rm E}}\, E(B-V)$ from \cref{eq:MWext} to be used for all \HE magnitudes                                              &  \cref{table:2}                            \\
$\sigma_{\rm EC}(\IE)$           & Error on VIS extinction correction                                    & \cref{table:2}                            \\
$\sigma_{\rm EC}(\YE)$              & Error on \YE extinction correction                                      & \cref{table:2}                            \\
$\sigma_{\rm EC}(\JE)$              & Error on \JE extinction correction                                      & \cref{table:2}                            \\
$\sigma_{\rm EC}(\HE)$               & Error on \HE extinction correction                                      & \cref{table:2}                            \\ 
\noalign{\vskip 1pt}
\hline
\end{tabular}
\end{table*}

\clearpage

\begin{table*}
\caption{Summary table of parameters extracted from Tables~\ref{table:3} to \ref{table:4err}, along with their descriptions and associated table numbers.}
\label{tab:parameter_summary2}
\begin{tabular}{lll}
\hline\hline
\noalign{\vskip 1pt}
Parameter& Description & Table numbers \\
\hline
\noalign{\vskip 1pt}

ID & Object identifier & \cref{table:3} to \cref{table:4} \\
RA & Right Ascension in degrees & \cref{table:3} to \cref{table:4}\\
Dec & Declination in degrees & \cref{table:3} to \cref{table:4} \\
\IE & Magnitude within 1\,$R_{\rm e}$ aperture in the VIS band & \cref{table:3} \\
$M(\IE)$ & Absolute magnitude within 1\,$R_{\rm e}$ aperture in the VIS band & \cref{table:3} \\
\YE & Magnitude within 1\,$R_{\rm e}$ aperture in the \YE band & \cref{table:3} \\
$M(\YE)$ & Absolute magnitude within 1\,$R_{\rm e}$ aperture in the \YE band & \cref{table:3} \\
\JE & Magnitude within 1\,$R_{\rm e}$ aperture in the \JE band & \cref{table:3} \\
$M(\JE)$ & Absolute magnitude within 1\,$R_{\rm e}$ aperture in the \JE band & \cref{table:3} \\
\HE & Magnitude within 1\,$R_{\rm e}$ aperture in the \HE band & \cref{table:3} \\
$M(\HE)$ & Absolute magnitude within 1\,$R_{\rm e}$ aperture in the \HE band & \cref{table:3err}\\
$\sigma_{I_{\rm E}}$ & Error on the magnitude within 1\,$R_{\rm e}$ aperture in the VIS band & \cref{table:3err} \\
$\sigma_{M(\IE)}$  & Error on the absolute magnitude within 1\,$R_{\rm e}$ aperture in the VIS band & \cref{table:3err} \\
$\sigma_{Y_{\rm E}}$  & Error on the magnitude within 1\,$R_{\rm e}$ aperture in the \YE band & \cref{table:3err} \\
$\sigma_{M(\IE)}$ & Error on the absolute magnitude within 1\,$R_{\rm e}$ aperture in the \YE band & \cref{table:3err} \\
$\sigma_{J_{\rm E}}$ & Error on the magnitude within 1\,$R_{\rm e}$ aperture in the \JE band & \cref{table:3err} \\
$\sigma_{M(\JE)}$ & Error on the absolute magnitude within 1\,$R_{\rm e}$ aperture in the \JE band & \cref{table:3err} \\
$\sigma_{H_{\rm E}}$  & Error on the magnitude within 1\,$R_{\rm e}$ aperture in the \HE band & \cref{table:3err} \\
$\sigma_{M(\HE)}$ & Error on the absolute magnitude within 1\,$R_{\rm e}$ aperture in the \HE band & \cref{table:3err} \\
\IE & Total forced magnitude in the VIS band & \cref{table:4} \\
$M(\IE)$& Total absolute forced magnitude in the VIS band & \cref{table:4} \\
\YE & Total forced magnitude in the \YE band & \cref{table:4} \\
$M(\YE)$& Total absolute forced magnitude in the \YE band & \cref{table:4} \\
\JE & Total forced magnitude in the \JE band & \cref{table:4} \\
$M(\JE)$ & Total absolute forced magnitude in the \JE band & \cref{table:4} \\
\HE & Total forced magnitude in the \HE band & \cref{table:4} \\
$M(\HE)$& Total absolute forced magnitude in the \HE band & \cref{table:4} \\
$\sigma_{I_{\rm E}}$& Error on the forced magnitude in the VIS band & \cref{table:4err} \\
$\sigma_{M(\IE)}$& Error on the absolute forced magnitude in the VIS band & \cref{table:4err} \\
$\sigma_{{I_{\rm E}}}$& Error on the forced magnitude in the \YE band & \cref{table:4err} \\
$\sigma_{M(\YE)}$ & Error on the absolute forced magnitude in the \YE band & \cref{table:4err} \\
$\sigma_{J_{\rm E}}$& Error on the forced magnitude in the \JE band & \cref{table:4err} \\
$\sigma_{M(\JE)}$& Error on the absolute forced magnitude in the \JE band & \cref{table:4err} \\
$\sigma_{H_{\rm E}}$ & Error on the forced magnitude in the \HE band & \cref{table:4err} \\
$\sigma_{M(\HE)}$& Error on the absolute forced magnitude in the \HE band & \cref{table:4err} \\
\noalign{\vskip 1pt}
\hline
\end{tabular}
\end{table*}

\clearpage

\begin{sidewaystable}[htbp!]
\newcommand{\pd}{\phantom{1}}
\begin{minipage}[t]{0.48\textwidth}
\setlength{\tabcolsep}{3.25pt}
\caption{Primary data from VIS photometry.}
\label{table:1}
\small
\resizebox{25.1cm}{!}{
\begin{tabular}{lrrrrrrrrrrrrrrrrrr}
\hline\hline
\noalign{\vskip 2pt}
\omit\hfil ID \hfil & \omit\hfil RA \hfil & \omit\hfil Dec \hfil & \omit\hfil x$_\mathrm{pix}$ \hfil & \omit\hfil y$_\mathrm{pix}$ \hfil & \omit\hfil \IE \hfil & \omit\hfil $M(\IE)$ \hfil & \omit\hfil $R_{\rm e}$  \hfil & \omit\hfil $R_{\rm e}$ \hfil & \omit\hfil $R_{\rm e}$ \hfil & \omit\hfil $n$ \hfil & \omit\hfil AR \hfil & \omit\hfil PA \hfil & \omit\hfil $\mu_{\IE,\rm{0, GD}}$ \hfil & \omit\hfil $\mu_{\IE,\rm{e}}$ \hfil & \omit\hfil $\langle \mu_{\IE,\rm{e, GD}}\rangle$ \hfil & \omit\hfil $\mu_{\IE,\rm{0}}$ \hfil & \omit\hfil $\mu_{\IE,\rm{e}}$ \hfil & \omit\hfil $\langle \mu_{\IE,\rm{e}} \rangle$\hfil\\
\noalign{\vskip 1pt}
\hline
\noalign{\vskip 1pt}
CGCG540-074 & 49.115 & 41.627 & 1001.880 & 1002.000 & 13.402 & $-20.885$ & 96.487 & 9.649 & 3.368 & 3.273 & 0.744 & 29.697 & 16.781 & 25.108 & 23.715 & 15.540 & 21.357 & 16.711 \\
WISEAJ031637\_12+414721\_3 & 49.155 & 41.789 & 2008.470 & 2009.570 & 15.110 & $-19.177$ & 82.883 & 8.288 & 2.893 & 4.017 & 0.929 & 152.641 & 18.394 & 26.721 & 25.328 & 17.169 & 22.840 & 17.745 \\
2MASXJ03165143+4127342 & 49.214 & 41.459 & 987.590 & 999.130 & 14.942 & $-19.345$ & 148.537 & 14.854 & 5.185 & 2.817 & 0.754 & 76.206 & 19.162 & 27.489 & 26.096 & 18.518 & 23.503 & 18.619 \\
... &... &... &... &... &... &... &... &... &... &... &... &... &... &... &... &... &... &... \\
 GAIADR3239372962793322368 & 49.301 & 41.381 & 499.500 & 499.020 & 16.650 & $-17.637$ & 35.246 & 3.525 & 1.230 & 4.880 & 0.560 & 17.690 & 17.690 & 26.020 & 24.620 & 17.080 & 22.090 & 18.060 \\
 7488 & 49.495 & 41.163 & 499.300 & 499.180 & 17.390 & $-16.897$ & 42.342 & 4.234 & 1.478 & 1.280 & 0.480 & 158.050 & 18.510 & 26.840 & 25.440 & 19.870 & 22.630 & 20.590 \\
\hline
\end{tabular}
}
\hfill \\

\setlength{\tabcolsep}{3.25pt}
\caption{Associated errors for the primary data catalogue.}
\label{table:1err}
\small
\resizebox{25.1cm}{!}{
\begin{tabular}{lrrrrrrrrrrrrrrrrrr}
\hline\hline
\noalign{\vskip 2pt}
\omit\hfil ID \hfil & \omit\hfil RA \hfil & \omit\hfil Dec \hfil & \omit\hfil $\sigma_{x_\mathrm{pix}}$ \hfil & \omit\hfil $\sigma_{y_\mathrm{pix}}$\hfil & \omit\hfil $\sigma_{I_\mathrm{E}}$ \hfil & \omit\hfil $\sigma_{M(\IE)}$ \hfil & \omit\hfil $\sigma_{{\rm R_e}}$ \hfil & \omit\hfil $\sigma_{{\rm R_e}}$ \hfil & \omit\hfil $\sigma_{{\rm R_e}}$ \hfil & \omit\hfil $\sigma_n$ \hfil & \omit\hfil $\sigma_{\rm AR}$\hfil & \omit\hfil $\sigma_{\rm PA}$ \hfil & \omit\hfil $\sigma_{\mu_{\IE,\rm{0,GD}}}$ \hfil & \omit\hfil $\sigma_{\mu_{\IE,\rm{e,GD}}}$  \hfil & \omit\hfil $\sigma_{\langle\mu_{\IE,\rm{e,GD}}\rangle}$  \hfil & \omit\hfil $\sigma_{\mu_{\IE,\rm{0}}}$ \hfil & \omit\hfil $\sigma_{\mu_{\IE,\rm{e}}}$  \hfil & \omit\hfil $\sigma_{\langle\mu_{\IE,\rm{e}}\rangle}$  \hfil\\
\noalign{\vskip 1pt}
\hline
\noalign{\vskip 1pt}
 CGCG540-074 & 49.115 & 41.627 & 0.010 & 0.010 & 0.001 & 0.001 & 1.892 & 0.189 & 0.066 & 0.118 & 0.018 & 1.851 & 0.043 & 0.043 & 0.043 & 0.000 & 0.043 & 0.000 \\
WISEAJ031637\_12+414721\_3 & 49.155 & 41.789 & 0.010 & 0.010 & 0.003 & 0.003 & 1.625 & 0.163 & 0.057 & 0.056 & 0.002 & 1.128 & 0.029 & 0.029 & 0.029 & 0.000 & 0.029 & 0.000 \\
2MASXJ03165143+4127342 & 49.214 & 41.459 & 0.010 & 0.010 & 0.001 & 0.001 & 2.913 & 0.291 & 0.102 & 0.033 & 0.039 & 1.158 & 0.023 & 0.023 & 0.023 & 0.000 & 0.023 & 0.000 \\
... &... &... &... &... &... &... &... &... &... &... &... &... &... &... &... &... &... &... \\
 GAIADR3239372962793322368 & 49.301 & 41.381 & 0.010 & 0.010 & 0.000 & 0.000 & 0.691 & 0.069 & 0.024 & 0.150 & 0.040 & 0.280 & 0.030 & 0.030 & 0.030 & 0.000 & 0.030 & 0.000 \\
 7488 & 49.495 & 41.163 & 0.010 & 0.010 & 0.000 & 0.000 & 0.830 & 0.083 & 0.029 & 0.010 & 0.050 & 1.440 & 0.050 & 0.050 & 0.050 & 0.000 & 0.050 & 0.000 \\
\hline
\end{tabular}
}
\end{minipage}
\end{sidewaystable}

\begin{sidewaystable}[htbp!]
\newcommand{\pd}{\phantom{1}}
\setlength{\tabcolsep}{3.25pt}
\caption{Redshift and extinction corrections.}
\label{table:2}
\begin{tabular}{lrrrrrrrrrrrrrrr}
\hline\hline
\noalign{\vskip 2pt}
\omit\hfil ID \hfil & \omit\hfil RA \hfil & \omit\hfil Dec \hfil & \omit\hfil $z_{\rm spec}$ \hfil & \omit\hfil $z_{\rm phot}$ \hfil & \omit\hfil $\sigma_{z{\rm spec}}$ \hfil & \omit\hfil $\sigma_{z{\rm phot}}$ \hfil & \omit\hfil flagtool \hfil  & \omit\hfil EC(\IE) \hfil & \omit\hfil EC(\YE) \hfil & \omit\hfil EC(\JE) \hfil & \omit\hfil EC(\HE) \hfil & \omit\hfil $\sigma_{EC(\IE)}$  \hfil & \omit\hfil $\sigma_{EC(\YE)}$ \hfil & \omit\hfil $\sigma_{EC(\JE)}$  \hfil & \omit\hfil $\sigma_{EC(\HE)}$  \hfil\\
\noalign{\vskip 1pt}
\hline
\noalign{\vskip 1pt}
CGCG540-074 & 49.115 & 41.627 & 0.017 & 0.024 & 0.000 & 0.007 & AP & 0.276 & 0.138 & 0.094 & 0.061 & 0.015 & 0.008 & 0.005 & 0.003 \\
WISEAJ031637\_12+414721\_3 & 49.155 & 41.789 & 0.018 & 0.046 & 0.000 & 0.013 & AP & 0.302 & 0.152 & 0.103 & 0.067 & 0.016 & 0.008 & 0.006 & 0.004 \\
2MASXJ03165143+4127342 & 49.214 & 41.459 & $-99.000$ & 0.018 & $-99.000$ & 0.049 & AP & 0.285 & 0.143 & 0.098 & 0.063 & 0.016 & 0.008 & 0.006 & 0.004 \\
... &... &... &... &... &... &... &... &... &... &... &... &... &... &... &...  \\
GAIADR3239372962793322368 & 49.301 & 41.381 & $-99.990$ & $-99.990$ & $-99.990$ & $-99.990$ & AP & 0.309 & 0.155 & 0.106 & 0.069 & 0.018 & 0.009 & 0.006 & 0.004 \\
 7488 & 49.495 & 41.163 & $-99.990$ & $-99.990$ & $-99.990$ & $-99.990$ & AP & 0.406 & 0.204 & 0.139 & 0.090 & 0.025 & 0.013 & 0.009 & 0.006 \\
\hline
\end{tabular}
\end{sidewaystable}

\begin{sidewaystable}[htbp!]
\newcommand{\pd}{\phantom{1}}
\begin{minipage}[t]{0.48\textwidth}
\setlength{\tabcolsep}{3.25pt}
\caption{Magnitudes within 1 effective radius in VIS and NISP bands.}
\label{table:3}
\begin{tabular}{lrrrrrrrrrr}
\hline\hline
\noalign{\vskip 2pt}
\omit\hfil ID \hfil & \omit\hfil RA \hfil & \omit\hfil Dec \hfil & \omit\hfil \IE \hfil & \omit\hfil $M(\IE)$ \hfil & \omit\hfil \YE \hfil & \omit\hfil $M(\YE)$  \hfil & \omit\hfil \JE \hfil & \omit\hfil $M(\JE)$  \hfil & \omit\hfil \HE \hfil & \omit\hfil $M(\HE)$  \hfil \\
\noalign{\vskip 1pt}
\hline
\noalign{\vskip 1pt}
 CGCG540-074 & 49.115 & 41.627 & 14.119 & -20.168 & 13.346 & -20.941 & 13.174 & $-21.112$ & 13.035 & $-21.252$ \\
WISEAJ031637\_12+414721\_3 & 49.155 & 41.789 & 15.827 & $-18.460$ & 15.039 & $-19.248$ & 14.885 & $-19.402$ & 14.740 & $-19.547$ \\
2MASXJ03165143+4127342 & 49.214 & 41.459 & 15.667 & $-18.620$ & 15.029 & $-19.257$ & 14.903 & $-19.383$ & 14.776 & $-19.510$ \\
... &... &... &... &... &... &... &... &... &... &... \\
GAIADR3239372962793322368 & 49.301 & 41.381 & 17.370 & $-16.917$ & 16.546 & $-17.741$ & 16.399 & $-17.888$ & 16.189 & $-18.098$ \\
7488 & 49.495 & 41.163 & 18.100 & $-16.187$ & 17.464 & $-16.822$ & 17.383 & $-16.904$ & 17.266 & $-17.021$ \\
\hline
\end{tabular}
\hfill \\

\setlength{\tabcolsep}{3.25pt}
\caption{Associated errors of the 1\,$R_{\rm e}$ magnitudes catalogue.}
\label{table:3err}
\begin{tabular}{lrrrrrrrrrr}
\hline\hline
\noalign{\vskip 2pt}
\omit\hfil ID \hfil & \omit\hfil RA \hfil & \omit\hfil Dec \hfil & \omit\hfil $\sigma_{{I_{\rm E}}}$  \hfil & \omit\hfil $\sigma_{M(\IE)}$    \hfil & \omit\hfil $\sigma_{{Y_{\rm E}}}$   \hfil & \omit\hfil $\sigma_{M(\YE)}$    \hfil & \omit\hfil $\sigma_{{J_{\rm E}}}$    \hfil & \omit\hfil $\sigma_{M(\JE)}$    \hfil & \omit\hfil $\sigma_{{H_{\rm E}}}$   \hfil & \omit\hfil $\sigma_{M(\HE)}$    \hfil \\
\noalign{\vskip 1pt}
\hline
\noalign{\vskip 2pt}
 CGCG540-074 & 49.115 & 41.627 & 0.001 & 0.001 & 0.002 & 0.002 & 0.001 & 0.001 & 0.002 & 0.002 \\
WISEAJ031637\_12+414721\_3 & 49.155 & 41.789 & 0.013 & 0.013 & 0.004 & 0.004 & 0.004 & 0.004 & 0.004 & 0.004 \\
 2MASXJ03165143+4127342 & 49.214 & 41.459 & 0.015 & 0.015 & 0.001 & 0.001 & 0.001 & 0.001 & 0.001 & 0.001 \\
... &... &... &... &... &... &... &... &... &... &... \\
 GAIADR3239372962793322368 & 49.301 & 41.381 & 0.010 & 0.010 & 0.005 & 0.005 & 0.006 & 0.005 & 0.004 & 0.004 \\
 7488 & 49.495 & 41.163 & 0.020 & 0.020 & 0.001 & 0.001 & 0.001 & 0.001 & 0.001 & 0.001 \\
\noalign{\vskip 1pt}
\hline
\end{tabular}
\end{minipage}
\end{sidewaystable}

\begin{sidewaystable}[htbp!]
\newcommand{\pd}{\phantom{1}}
\begin{minipage}[t]{0.48\textwidth}
\setlength{\tabcolsep}{3.25pt}
\caption{Magnitudes using VIS photometry to constrain NISP photometry.}
\label{table:4}
\begin{tabular}{lrrrrrrrrrr}
\hline\hline
\noalign{\vskip 2pt}
\omit\hfil ID \hfil & \omit\hfil RA \hfil & \omit\hfil Dec \hfil & \omit\hfil \IE   \hfil & \omit\hfil ${M(\IE)}$    \hfil & \omit\hfil \YE \hfil & \omit\hfil ${M(\YE)}$  \hfil & \omit\hfil \JE \hfil & \omit\hfil ${M(\JE)}$    \hfil & \omit\hfil \HE \hfil & \omit\hfil ${M(\HE)}$   \hfil\\
\noalign{\vskip 1pt}
\hline
\noalign{\vskip 2pt}
CGCG540-074 & 49.115 & 41.627 & 13.402 & $-20.885$ & 12.952 & $-21.335$ & 12.447 & $-21.839$ & 12.300 & $-21.986$ \\
WISEAJ031637\_12+414721\_3 & 49.155 & 41.789 & 15.110 & $-19.177$ & 14.529 & $-19.758$ & 14.277 & $-20.009$ & 14.112 & $-20.174$ \\
2MASXJ03165143+4127342 & 49.214 & 41.459 & 14.942 & $-19.345$ & 14.354 & $-19.933$ & 14.234 & $-20.053$ & 14.066 & $-20.221$ \\
... &... &... &... &... &... &... &... &... &... &... \\
GAIADR3239372962793322368 & 49.301 & 41.381 & 16.650 & $-17.637$ & 15.948 & $-18.339$ & 15.776 & $-18.511$ & 15.602 & $-18.685$ \\
7488 & 49.495 & 41.163 & 17.390 & $-16.897$ & 16.812 & $-17.474$ & 16.747 & $-17.540$ & 16.628 & $-17.659$ \\
\noalign{\vskip 1pt}
\hline
\end{tabular}
\hfill \\

\setlength{\tabcolsep}{3.25pt}
\caption{Associated errors of the forced magnitudes.}
\label{table:4err}
\begin{tabular}{lrrrrrrrrrr}
\hline\hline
\noalign{\vskip 2pt}
\omit\hfil ID \hfil & \omit\hfil RA \hfil & \omit\hfil Dec \hfil & \omit\hfil $\sigma_{{I_{\rm E}}}$    \hfil & \omit\hfil $\sigma_{M(\IE)}$    \hfil & \omit\hfil $\sigma_{{Y_{\rm E}}}$    \hfil & \omit\hfil $\sigma_{M(\YE)}$    \hfil & \omit\hfil $\sigma_{{J_{\rm E}}}$    \hfil & \omit\hfil $\sigma_{M(\JE)}$   \hfil & \omit\hfil $\sigma_{{H_{\rm E}}}$    \hfil & \omit\hfil $\sigma_{M(\HE)}$    \hfil\\
\noalign{\vskip 1pt}
\hline
\noalign{\vskip 1pt}
CGCG540-074 & 49.115 & 41.627 & 0.001 & 0.001 & 0.002 & 0.002 & 0.001 & 0.001 & 0.002 & 0.002 \\
 WISEAJ031637\_12+414721\_3 & 49.155 & 41.789 & 0.003 & 0.003 & 0.003 & 0.003 & 0.003 & 0.003 & 0.003 & 0.003 \\
2MASXJ03165143+4127342 & 49.214 & 41.459 & 0.001 & 0.001 & 0.002 & 0.002 & 0.002 & 0.002 & 0.002 & 0.002 \\
... &... &... &... &... &... &... &... &... &... &... \\
 GAIADR3239372962793322368 & 49.301 & 41.381 & 0.000 & 0.000 & 0.004 & 0.004 & 0.004 & 0.004 & 0.003 & 0.003 \\
 7488 & 49.495 & 41.163 & 0.000 & 0.000 & 0.002 & 0.002 & 0.001 & 0.001 & 0.001 & 0.001 \\
\hline
\end{tabular}
\end{minipage}
\end{sidewaystable}

\newpage

\end{appendix}

\label{LastPage}
\end{document}